\documentclass{article}

\usepackage{arxiv}
\usepackage[utf8]{inputenc} 
\usepackage[T1]{fontenc}    
\usepackage{hyperref}       
\usepackage{url}            
\usepackage{booktabs}       
\usepackage{amsfonts}       
\usepackage{nicefrac}       
\usepackage{microtype}      
\usepackage{lipsum}
\usepackage{graphicx}
\usepackage{subcaption}
\usepackage{amsmath,amsfonts,amssymb}
\graphicspath{{Figures/}}
\usepackage{color}
\usepackage[ruled,vlined]{algorithm2e}
\usepackage{mdframed}
\usepackage{comment}

\newcommand{\mbf}{\mathbf}
\newcommand{\mbb}{\mathbb}
\newcommand{\mc}{\mathcal}
\newcommand{\f}{\frac}
\newcommand{\bs}{\boldsymbol}
\newcommand{\mb}{\mathbf}
\newcommand{\R}{\mathbb{R}}
\newcommand{\RNum}[1]{\uppercase\expandafter{\romannumeral #1\relax}}

\title{Machine-Learning for Nonintrusive Model Order Reduction of the Parametric Inviscid Transonic Flow past an airfoil }

\author{
  S. Ashwin Renganathan \\
  Mathematics and Computer Science Division\\
  Argonne National Laboratory\\
  Lemont, IL 60439, USA\\
  \texttt{srenganathan@anl.gov} \\
  \AND
  Romit Maulik \\
  Argonne Leadership Computing Facility\\
  Argonne National Laboratory\\
  Lemont, IL 60439, USA\\
  \texttt{rmaulik@anl.gov} \\
  \AND
  Vishwas Rao \\
  Mathematics and Computer Science Division\\
  Argonne National Laboratory\\
  Lemont, IL 60439, USA\\
  \texttt{vhebbur@anl.gov} \\
}

\begin{document}
\maketitle

\begin{abstract}
Fluid flow in the transonic regime finds relevance in aerospace engineering, particularly in the design of commercial air transportation vehicles. Computational fluid dynamics models of transonic flow for aerospace applications are computationally expensive to solve because of the high degrees of freedom as well as the coupled nature of the conservation laws. While these issues pose a bottleneck for the use of such models in aerospace design, computational costs can be significantly minimized by constructing special, structure-preserving surrogate models called reduced-order models. Such models are known to incur huge off-line costs, however, which can sometimes outweigh their potential benefits. Furthermore, their prediction accuracy is known to be poor under transonic flow conditions. In this work, we propose a machine learning method to construct reduced-order models via deep neural networks, and we demonstrate its ability to preserve accuracy with significantly lower offline and online costs. In addition, our machine learning methodology is physics-informed and constrained through the utilization of an interpretable encoding by way of  proper orthogonal decomposition. Application to the inviscid transonic flow past the RAE2822 airfoil under varying freestream Mach numbers and angles of attack, as well as airfoil shape parameters with a deforming mesh, shows that the proposed approach adapts to high-dimensional parameter variation well. Notably, the proposed framework precludes knowledge of numerical operators utilized in the data generation phase, thereby demonstrating its potential utility in fast exploration of design space for diverse engineering applications. 
\end{abstract}


\section{Introduction}
Projection-based model order reduction is a structure-preserving technique to develop computationally cheap but accurate surrogate models of systems of partial differential equations (PDEs). Such models have seen widespread interest in  recent years because of the promise they have shown in efficient compression of spatiotemporal dynamics for a variety of systems \cite{san2018extreme,wang2019non,choi2019space,mohebujjaman2019physically}. These reduced-order models (ROMs) find extensive application in control \cite{proctor2016dynamic}, multifidelity optimization \cite{peherstorfer2016optimal}, and uncertainty quantification \cite{sapsis2013statistically,zahr2018efficient}, among others. We direct the interested reader to \cite{taira2019modal, benner2015survey} for  excellent reviews of the recent advances and opportunities in ROMs. To derive  a ROM, one  reduces the dimensionality of the full-order model (FOM), which comprises the PDEs in their discrete or semi-discrete form. The dimensionality reduction is achieved mainly via a \emph{projection} step where the FOM is projected onto a suitably chosen test basis set. The proper orthogonal decomposition (POD)~\cite{HolmesPhilip.LumleyJohnL.BerkoozGahlandRowley1998} is a suitable choice to extract such bases, specifically for PDEs governing fluid flow, because of its physical interpretability. The resulting model is referred to as POD-ROM in this work. Furthermore, the \emph{state space} dimension of the PDEs is directly reduced via POD by expressing the PDE state as a linear expansion of the POD modes.

To illustrate the mechanics of a POD-ROM, we consider a state variable $\mbf{u} \in \mbb{R}^N$ that is the numerical solution of a PDE on a computational mesh of size $N$. Then the POD-ROM approximates $\mbf{u}$ as the linear expansion on a finite number of $k$ orthonormal basis vectors (i.e., the POD bases) $\bs{\phi}_i \in \mbb{R}^N$. That is,
\begin{equation}
\mbf{u} \approx \sum_{i=1}^{k} \tilde{u}_i \bs{\phi}_i,
\label{e:POD}
\end{equation}
where $\tilde{u}_i \in \mbb{R}$ is the $i$th component of $\tilde{\mbf{u}} \in \mbb{R}^k$, which are the coefficients of the basis expansion, also known as the \emph{reduced} state.  The  $\lbrace \bs{\phi}_i \rbrace,~i=1,\hdots,k$, $~\bs{\phi}_i\in \mbb{R}^N$ are also called  POD \emph{modes}. One can  show that the POD modes in Equation (1) are the left singular vectors of the snapshot matrix (obtained by stacking $M$ snapshots of $\mbf{u}$), $\mbf{U} = [\mbf{u}_1, \hdots, \mbf{u}_{M}]$, extracted by performing a singular value decomposition (SVD) on $\mbf{U}$ \cite{HolmesPhilip.LumleyJohnL.BerkoozGahlandRowley1998,Chatterjee2000}. That is,

\begin{equation}
\mbf{U}  \underset{\text{svd}}{=} \mbf{V} \bs{\Sigma} \mbf{W}^\top ,
\label{e:svd}
\end{equation}
where $\mbf{V} \in \mbb{R}^{N \times M}$ and $\mbf{\Phi}_k$ represent the first $k$ columns of $\mbf{V}$ after truncating the last $M-k$ columns based on the relative magnitudes of the cumulative sum of their singular values. The total $L_2$ error in approximating the snapshots via the truncated POD basis is then given as
\begin{align}
\sum_{j=1}^M \left\Vert \mbf{u}_j - (\mbf{\Phi}_k \mbf{\Phi}_k^\top) \mbf{u}_j \right \Vert ^2_2 = \sum_{i=k+1}^M \sigma_i^2,
\end{align}
where $\sigma_i$ is the singular value corresponding to the $i$th column of $\mbf{\Phi}_k$ and is also the $i$th diagonal element of $\bs{\Sigma}$. It is well known that the POD bases are $L_2$-optimal and are thus a good choice for an efficient compression of high-dimensional dynamics. In dynamical systems, the reduced state $\tilde{\mbf{u}}$ is assumed to be a function of time with the POD basis set fixed in the time domain. In this work, however, we consider static but parametric systems, in which case the $\tilde{\mbf{u}}$ is assumed to be a function of the parameters while the POD basis set is assumed to be globally fixed across the parameter domain. Therefore, the POD-ROM reduces the unknowns $\mbf{u}$ of the FOM to the $\tilde{\mbf{u}}$ in the ROM, which is significantly cheaper to solve. The dimension reduction of the overall system is achieved via the projection step described as follows.


The projection step projects the original FOM onto the low-dimensional subspace spanned by the POD basis vectors, forming the ROM ~\cite{HolmesPhilip.LumleyJohnL.BerkoozGahlandRowley1998}. Consider the discrete representation of a steady nonlinear parametric system that is the result of the discretization of a PDE,
\begin{equation}
\mb{A}(\bs{\theta})\mb{u} = \mb{f}(\mb{u}),
\label{e:FOM}
\end{equation}
where $\mb{A}(\bs{\theta}) \in \R^{N\times N}$ is the linear differential operator that arises as a result of the discretization of linear terms, $\bs{\theta} \subset \bs{\Theta} \in \R^d$ are the input parameters in some compact domain $\bs{\Theta}$, and $\mb{f}(\mb{u}) \in \mathbb{R}^{N\times 1}$, is the nonlinear term that arises as a result of the discretization of the nonlinear terms, in addition to lumping the boundary condition discretization terms and source terms if present. This represents the full-order system with $N$ unknowns. The projection step begins by realizing that the residual of the FOM is orthogonal to an appropriately chosen test basis, $\mb{\Psi_k} \in \mbb{R}^{N \times k}$. Removing the $\bs{\theta}$ for convenience of notation, this can be stated as
\begin{equation}
\mb{\Psi}_k^\top(\mb{A} \mb{u} - \mb{f}(\mb{u})) = \mb{0}.
\end{equation}

Since the reduced state variable $\tilde{\mb{u}} \approx \mb{\Phi}_k^\top \mb{u}$, this equation can be written as 
\begin{equation}
\mb{\Psi}_k^\top \mb{A} \mb{\Phi}_k \tilde{\mb{u}} = \mb{\Psi}_k^\top \mb{f}(\mb{\Phi}_k \tilde{\mb{u}}),
\end{equation}
defining the reduced matrix $\tilde{\mb{A}} = \mb{\Psi}_k^\top \mb{A} \mb{\Phi}_k \in \mathbb{R}^{k \times k}$, 
\begin{equation}
\tilde{\mb{A}} \tilde{\mb{u}} = \mb{\Psi}_k^\top \mb{f}(\mb{\Phi}_k \tilde{\mb{u}}).
\end{equation}

Equation 7 represents a reduced system with $k << N$ unknowns and therefore can be solved efficiently. The computation of the nonlinear term still involves operations that scale as $\mathcal{O}(N)$, making its computation inefficient with iterative solution methods. However, this inefficiency can be potentially overcome with the discrete empirical interpolation method (DEIM) \cite{Chaturantabut2010}. 

A fundamental challenge arises when the FOM is available as a simulation code (i.e., it is a black box). In such a situation, the full-order system matrices $\left( \mb{A}, \mb{f}(\mb{u}) \right)$ are not available for the projection step. To overcome this, Ranganathan et al.~\cite{renganathan2018koopman} showed that the system matrices can be constructed if the system is first lifted via a linearization assumption based on  Koopman theory~\cite{koopman1931hamiltonian}, following which projection-based model reduction proceeds in the traditional fashion; this method is briefly reviewed in Section~\ref{s:PB_ROM}. One of the primary limitations observed with this approach is that in the transonic regime, when the flow encounters sharp gradients due to shocks, the prediction accuracy suffers unless a \emph{densely} populated training dataset is used. Such a limitation is typical of any POD-ROM framework, and the usual workaround is to isolate the regions of the shock via \emph{domain decomposition} and solve the FOM in this region (see \cite{lucia2002domain} and \cite{legresley2003dynamic}). We note, however, that such approaches are \emph{intrusive} and do not apply to black-box simulation codes. A second limitation is that the projection-based ROMs require stable numerical solvers and stability issues can arise (particularly when shocks are involved) despite the FOM's being stable for a given set of initial and boundary conditions \cite{barone2009stable}. In \cite{renganathan2018koopman} the initial conditions as well as a certain hyperparameter of the ROM are tuned to overcome such issues; however, when ROMs are wrapped within other iterative methods (such as derivative-free optimizers), such issues are inevitable and can lead to failed cases. Moreover, with large-scale computationally expensive simulations, the offline costs of projection-based POD-ROMs can be prohibitive because of (i) the generation of the snapshots themselves, (ii) the POD step (whose algorithmic complexity scales as $\mc{O}(NM^2)$ and (iii) the projection step. Therefore, existing ROMs need to be improved in terms of efficiency particularly when modeling nonlinear flows with parameter-dependent discontinuities, such as flows in the transonic regime. Such improved ROMs will find widespread engineering applications particularly in the design of next-generation aerospace and mechanical systems.


In this work, we hypothesize that with an approach that circumvents the projection step altogether, the heavy offline costs associated with projection-based ROMs can be significantly reduced, in addition to overcoming the stability issues of projection-based ROMs. Furthermore, by leveraging state-of-the-art machine learning models that can learn highly nonlinear embeddings from data, one can construct a potentially cheap, but accurate surrogate model that learns the parameter-to-state mapping well for complex fluid flow in the transonic regime. In this regard, this work focuses  specifically on the use of deep neural networks (DNNs) to approximate $\tilde{\mbf{u}}$ in the parameter space for static-parametric POD-ROMs. Furthermore, to evaluate the effectiveness of our approach, we compare the DNN-based approach with the projection-based approach in \cite{renganathan2018koopman}. We demonstrate our method on the inviscid transonic flow past the RAE-2822 airfoil with up to eight parameters.

The rest of the article is organized as follows. We revisit nonintrusive projection-based model order reduction in Section \ref{s:PB_ROM} and outline the proposed DNN-based approach in Section ~\ref{s:DNN}. The transonic flow problem is introduced in Section~\ref{s:gov_eqns}. The paper concludes with a summary of the key findings and some directions for future work in Sections~\ref{Results} and \ref{Conclusion}, respectively.
\section{Nonintrusive Projection-Based Model Order Reduction}
\label{s:PB_ROM}

\begin{figure}
\centering
\includegraphics[width = 7in]{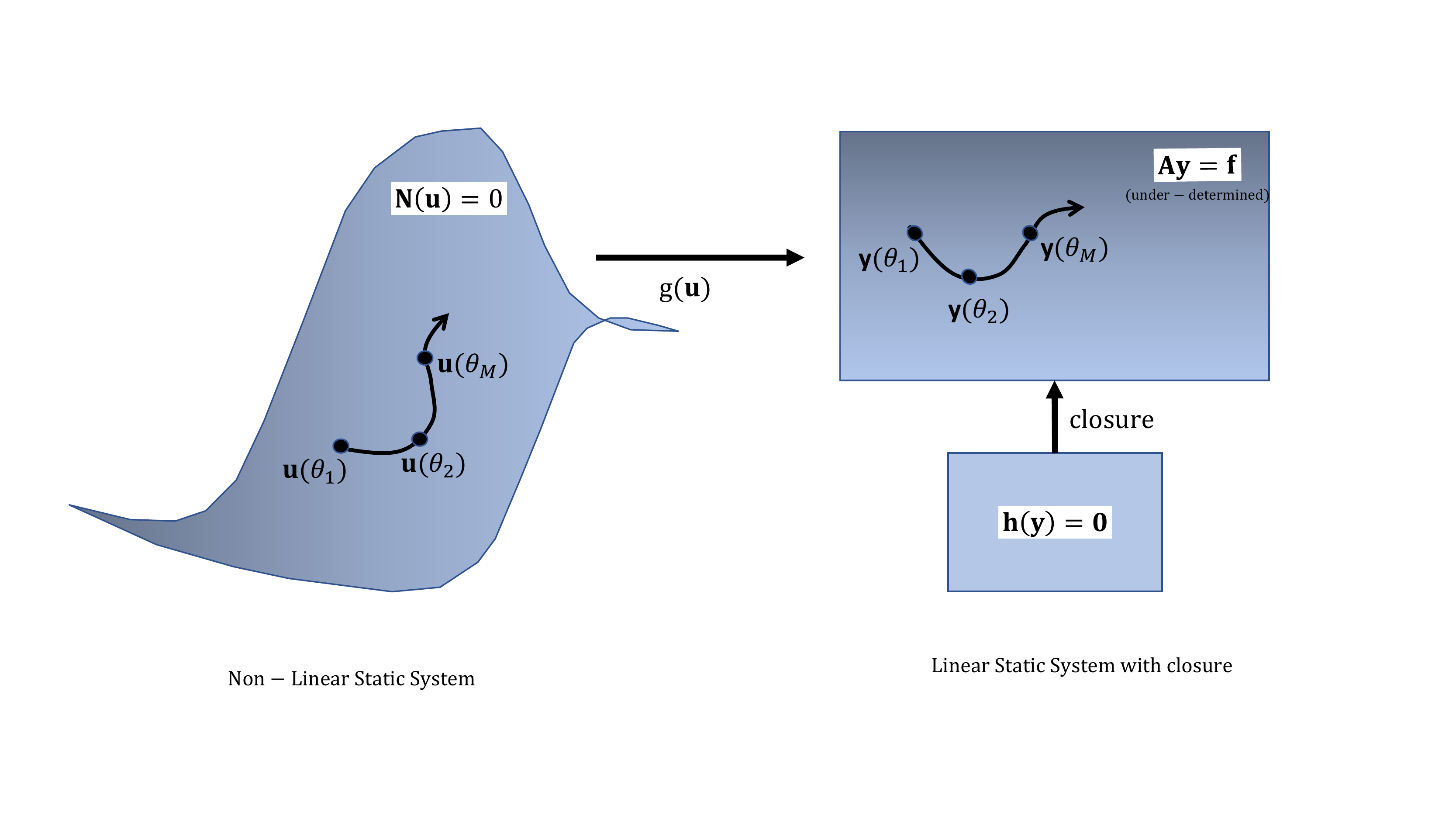}
\caption{Graphical depiction of the present methodology. The original nonlinear static system is transformed to an underdetermined linear system with closure~\cite{renganathan2018methodology}}
\label{f:Koopman_Method}
\end{figure}

We now review the nonintrusive model order reduction originally introduced in \cite{renganathan2018koopman} for freestream boundary parameters and \cite{renganathan2018koopman_applications} for geometry boundary parameters. Consider a static nonlinear system representing the FOM and given by the following discretized form,
\begin{equation}
\mbf{N}(\mbf{u}) = \mbf{0},
\label{e:nl_static_FOM}
\end{equation}
where $\mbf{N}:\R^N \rightarrow \R^N$ represents a nonlinear operator acting on the state variable $\mbf{u} \in \R^N$.  We consider the parametric case where $\mbf{u} = \mbf{u}(\bs{\theta})$; however, we omit the $\bs{\theta}$ in what follows, in order to keep the notation concise. Let $g:\mbb{R}^N \rightarrow \mbb{R}^N$ be a function that operates on the state (such as $g(\mbf{u}) = $ $\mbf{u}\otimes\mbf{u}, \mbf{u}\otimes\mbf{u}\otimes \mbf{u}$, and $e^{\mbf{u}}$, where $\otimes$ denotes the elementwise or Hadamard product). Then we state that 
\begin{equation}
\mbf{N}(\mbf{u}) \equiv \mbf{L} \left( [g_1(\mbf{u})^\top, g_2(\mbf{u})^\top, \hdots, g_O(\mbf{u})^\top]^\top \right), 
\label{e:lifted}
\end{equation}
where $\mbf{L}:\R^{ON} \rightarrow \R^N$ is a linear operator acting on the  \emph{lifted} system with the $g_i(\mbf{u})$'s replacing $\mbf{u}$. We call each $g_i(\mbf{u})$ an \emph{observable} following the convention of other works on the topic (particularly~\cite{Kutz2016}). The number of such observables $O$ in \eqref{e:lifted} is dependent on the system under consideration (as illustrated in Section~\ref{s:gov_eqns}). We then decompose the linear operator in Equation (9) as
\begin{equation}
\mbf{L}[g_1(\mbf{u})^\top, g_2(\mbf{u})^\top, \hdots, g_O(\mbf{u})^\top]^\top \approx \mbf{A}[g_1(\mbf{u})^\top, g_2(\mbf{u})^\top, \hdots, g_O(\mbf{u})^\top]^\top + \mbf{b}_a = \mbf{0},
\label{e:Obs_Form}
\end{equation}
which follows from the discretization of linear PDEs, where $\mbf{b_a} \in \R^{N \times 1}$ is the vector that arises due to the discretization of boundary conditions in addition to lumping any source terms and $\mbf{A}\in \R^{N \times ON}$ is the matrix resulting from the discretization of the linear differential terms. Overall, the parametric changes that deforms the mesh (such as geometry shape) are captured in $\mbf{A}$, whereas the rest (such as free-stream boundary conditions) are captured in $\mbf{b_a}$. In summary, $\mbf{A}_i = \mbf{A}(\bs{\theta}_i)$ and $\mbf{f}_i = \mbf{f}(\bs{\theta}_i)$ are parameter dependent and hence unique for each parameter snapshot, $\bs{\theta}_i$. Note that \eqref{e:Obs_Form} is a linear but underdetermined system and the uniqueness of the solution requires the addition of constraints as discussed in section~\ref{ss:MOR_Cons}. We rewrite \eqref{e:Obs_Form} by modifying the notation as $[g_1(\mbf{u})^\top, g_2(\mbf{u})^\top, \hdots, g_O(\mbf{u})^\top]^\top \rightarrow [\mbf{y}_1^\top,\hdots,\mbf{y}_O^\top]^\top = \mbf{y}$ and $-\mbf{b_a} \rightarrow \mbf{f}$, leading to 
\begin{equation}
\mbf{A}\mbf{y}= \mbf{f},
\label{e:transformed_FOM}
\end{equation}
where \eqref{e:transformed_FOM} is the transformed version of the FOM that we reduce to construct the ROM. Such a transformation enables us to extract $\mbf{A}$~\cite{renganathan2018koopman} nonintrusively and, furthermore, makes the overall methodology amenable to parametric interpolation, as will be illustrated in Section~\ref{ss:ROM_Interp}. The overall idea behind the lifting transformation to the FOM is depicted in Figure~\ref{f:Koopman_Method}~\cite{renganathan2018methodology}, and the model reduction is performed on the transformed equations (the right-hand side of the figure), which is explained in the following subsection.

\subsection{Model Order Reduction}
\label{ss:MOR_Cons}

The total number of observables $O$ is essentially \emph{infinite} if one were seeking a closed linear transformation of the nonlinear FOM~\cite{koopman1931hamiltonian}. However, we set $O$ to be finite, which results in an \emph{unclosed} linear transformation that is then closed with a set of constraints. In this work, $O$ represents the total number of terms in the FOM that are functions of the state and are operated by a linear differential operator. Each of the $g_i(\mbf{u})$'s is picked from knowledge of the FOM in its \emph{continuous} PDE form (as will be illustrated in Section~\ref{s:gov_eqns}). For a FOM that is a system of $S$ coupled PDEs, we note  that $O \geq S$ always; for a linear system $O=S$ and  for a nonlinear system $O > S$. Therefore the observables can be written

\begin{equation}
\mbf{y} = [\mbf{y}^\top_1,...,\mbf{y}^\top_S,\mbf{y}^\top_{S+1},...,\mbf{y}^\top_O]^\top
\end{equation}
To close the underdetermined transformed system, we add algebraic equations that establish the nonlinear consistency relationship between the observables and the state, thereby providing closure;  {the overall concept is depicted in Figure~\ref{f:Koopman_Method}}. These constraints are of the form

\begin{equation}
h_i := \mbf{y}_{S+i} - f_i(\mbf{y}_1,..., \mbf{y}_S) = \mbf{0},~~i=1,\hdots,O-S,
\label{e:nl_cons_generalform}
\end{equation}
where $f_i:\R^N \rightarrow \R^N$ are functions that operate on the observables and $h_i:\R^N \rightarrow \R^N$ are the \emph{equality} constraints. Note that for a system of $S$ coupled PDEs, $O-S$ constraints must be specified in order to achieve closure. Although there is no unique way of specifying these constraints, we provide some guidelines in Section~\ref{a:Euler_MOR}. Equations~\eqref{e:nl_cons_generalform} and \eqref{e:transformed_FOM} form a closed system upon which model reduction is performed. To perform the projection, the truncated basis set for each observable $\mbf{y}_i$ is extracted by performing POD on the corresponding snapshot matrix. Let a design of experiments (DOE) of $M$ points $\lbrace \bs{\theta}_1,\hdots, \bs{\theta}_M \rbrace$ be generated a priori. Then SVD is performed on the matrix $\mbf{Y}_i \in \R^{N \times M}$ defined as

\begin{equation}
\mbf{Y}_i = \begin{bmatrix}
\vdots & \vdots & \vdots & & \vdots \\
\mbf{y}_i^{(1)} & \mbf{y}_i^{(2)} & \mbf{y}_i^{(3)} & \hdots & \mbf{y}_i^{(M)} \\
\vdots & \vdots & \vdots & \vdots & \vdots \\
\end{bmatrix} \underset{\text{thin-svd}}{=} \mbf{V}_i \mbf{\Sigma}_i \mbf{W}_i^\top,
\label{e:thin-svd}
\end{equation}
where $\mbf{y}_i^{(j)}$ is the $j$th snapshot of observable $\mbf{y}_i$ (generated by running the FOM at $\bs{\theta}_j$),  {$\mbf{V}_i \in \R^{N \times M}$ are the left singular vectors, $\mbf{\Sigma}_i \in \R^{M\times M}$ are diagonal matrices of the singular values,  $\mbf{W}_i \in \R^{M\times M}$ are the right singular vectors}, and $\Phi_i \in \R^{N \times k_i}$ are the POD basis corresponding to $\mbf{y}_i$ and are the first $k_i$ columns of $\mbf{V}_i$. This leads to the trial basis matrix for the overall system defined as a block-diagonal matrix of all the $O$ POD basis set given below,
\begin{equation}
    \mbf{\Phi} = \text{blkdiag} \lbrace \Phi_1, \hdots, \Phi_O \rbrace ~ \in \mathbb{R}^{ON \times k},
    \label{e:POD_basis}
\end{equation}
where  $k=k_1 + \hdots + k_O$ and $\text{blkdiag}$ denotes an operator that constructs a block-diagonal matrix with $\Phi_i$ on the diagonal. The \emph{reduced} observable is then given by $\tilde{\mbf{y}}_i \approx \Phi_i^\top \mbf{y}_i$. Recall that $\mbf{A} \in \mbb{R}^{N\times ON}$ is nonsquare, and hence a suitable choice for the test basis for projection is $\mbf{\Psi} = \mbf{A}\mbf{\Phi}$. Note that this choice of the test basis is equivalent to a Galerkin projection ($\mbf{\Psi} = \mbf{\Phi}$) on the normal equations; in other words,  on $\mathbf{A}^\top , \mathbf{A} \mbf{y} = \mathbf{A}^\top \mbf{f}$. Let $\mathbf{B} = \mathbf{A}^\top\mathbf{A} \in \R^{ON \times ON}$; then the projection leads to

\begin{equation}
\mbf{\Phi}^\top \mbf{B} \mbf{\Phi} \tilde{\mbf{y}} = \mbf{\Phi}^\top \mbf{A}^\top \mbf{f}.
\label{e:projection}
\end{equation}

Setting $\tilde{\mbf{f}} = \mbf{\Phi}^\top \mbf{A}^\top\mbf{f} \in \mathbb{R}^k$ and $\tilde{\mbf{B}} = \mbf{\Phi}^\top \mbf{B} \mbf{\Phi} \in \mathbb{R}^{k \times k}$ leads to the reduced-order model

\begin{equation}
\tilde{\mbf{B}} \tilde{\mbf{y}}=\tilde{\mbf{f}}.
\label{e:ROM}
\end{equation}

 The ROM given by Equation (\ref{e:ROM}) is now a $k\times k$ system, where $k<<N$, and is solved along with the constraints presented in Equation (\ref{e:nl_cons_generalform}), posed as a nonlinear program as shown below.

\begin{equation}
\begin{aligned}
& \underset{\tilde{\mbf{y}}}{\text{minimize}}
&& \frac{1}{2}\|\tilde{\mbf{B}} \tilde{\mbf{y}} - \tilde{\mbf{f}} \|_2^2 \\
& \text{s.t.}
&& \mbf{\Phi}^\top h_i(\mbf{\Phi} \tilde{\mbf{y}})=\mbf{0},~i=1,\hdots,O-S
\end{aligned}
\label{e:conmin}
\end{equation}

The optimization problem in Equation (\ref{e:conmin}) needs special treatment in order to handle the nonlinear constraints that still depend on the full state of observables. This is efficiently done by using the DEIM  mentioned earlier; see Appendix~\ref{A:DEIM} for details on implementation for a specific example. The ROM in \eqref{e:conmin} is solved via sequential quadratic programming~\cite{schittkowski1986nlpql} with the objective function and constraint tolerances set to $10^{-6}$ and the number of function evaluations  limited to $4\times 10^6$. The initial guess to the solution of \eqref{e:conmin} is given as the nearest flow snapshot to the test parameter at which prediction is sought.

\subsection{ROM Interpolation}
\label{ss:ROM_Interp}
\begin{figure}[htb!]
\centering
\begin{subfigure}{0.5\textwidth}
\includegraphics[width = 3in,clip = 2cm 2cm 2cm 2cm]{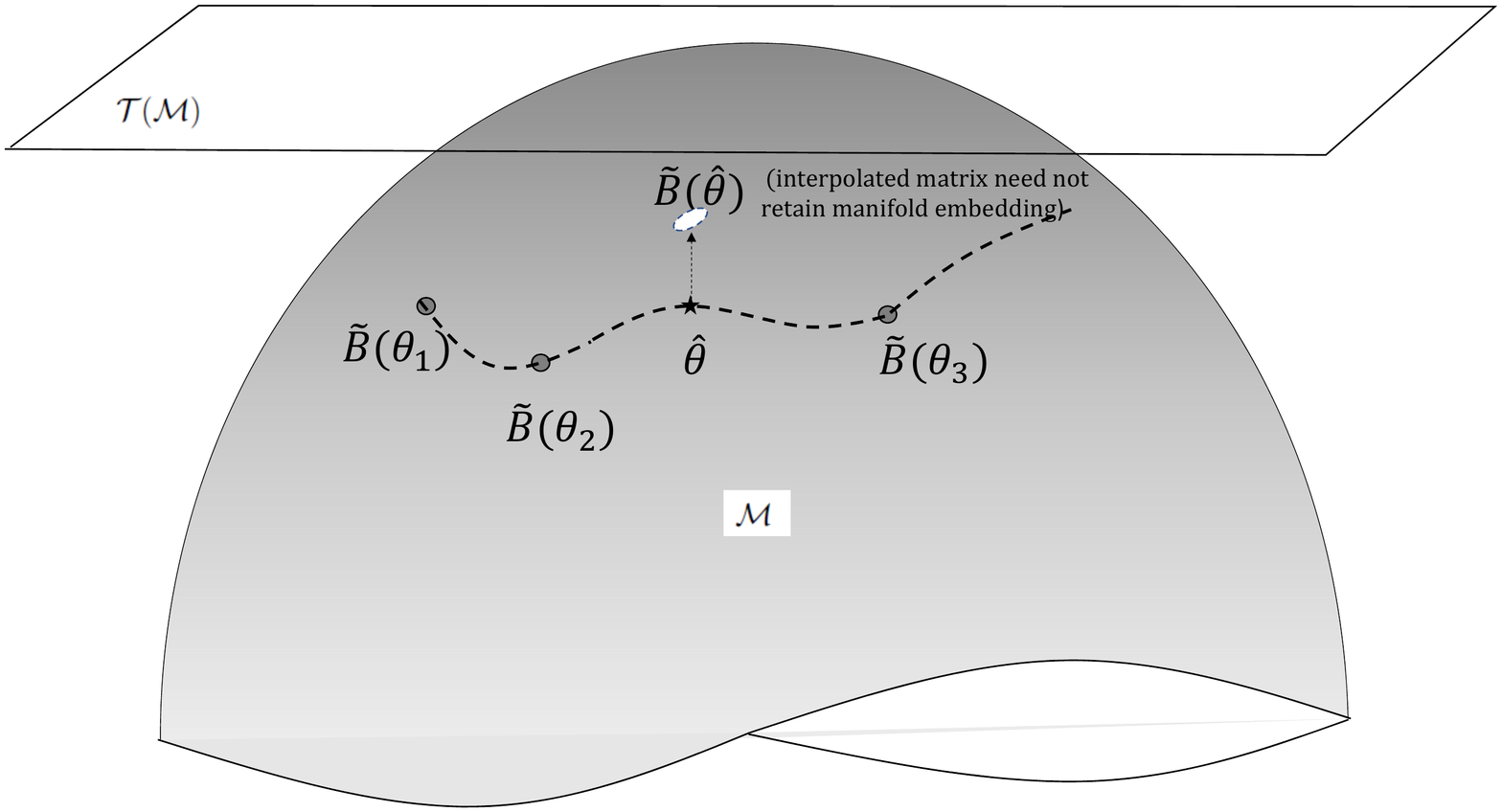}
\end{subfigure}~
\begin{subfigure}{0.5\textwidth}
\includegraphics[width = 3in,clip = 2cm 2cm 2cm 2cm]{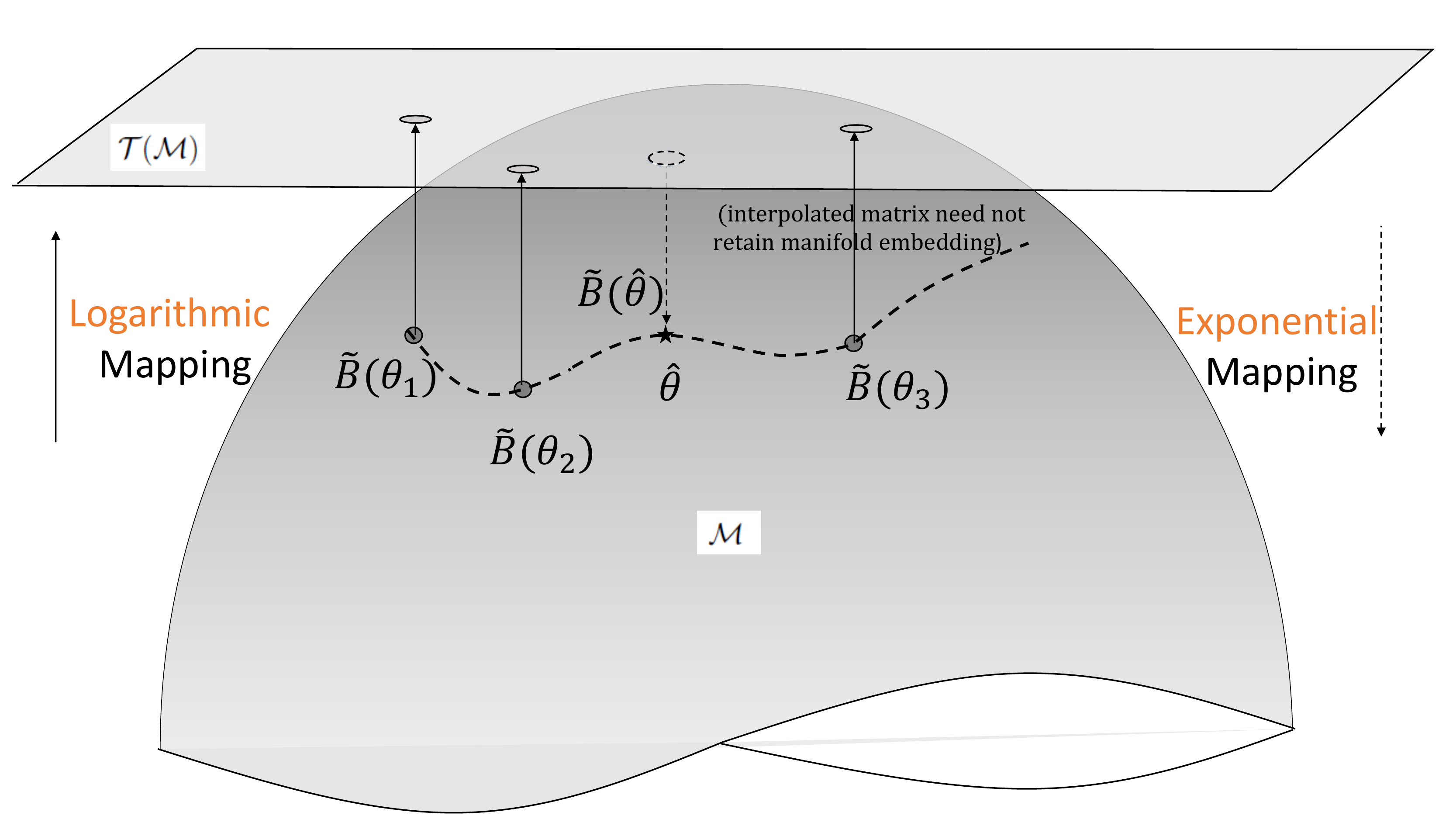}
\end{subfigure}
\caption{Graphical representation of a manifold $\mathcal{M}$ and the embedding of parametric matrices $\tilde{\mbf{B}}(\bs{\theta})$. A direct elementwise interpolation of $\tilde{\mbf{B}}$ at $\hat{\bs{\theta}}$ may not necessarily result in a matrix $\in \mathcal{M}$, and is carried out after mapping to the tangent plane, $\mathcal{T}(\mathcal{M})$~\cite{renganathan2018methodology}}
\label{f:Manif_Intro}
\end{figure}

The ROM in \eqref{e:ROM} corresponds to one parameter snapshot since $\tilde{\mbf{B}}$ and $\tilde{\mbf{f}}$ are parameter dependent. Therefore, the approach generates a database of ROMs for a predetermined set of parameter snapshots, which are later interpolated to predict the state at a new parameter. The interpolation is carried out in a manner that retains the inherent structure and properties of the matrix $\mbf{\tilde{B}}$ AFTER interpolation. The general principle  is to map the matrices to a plane that is locally tangent to the manifold in which they are originally embedded. The anchor point on the manifold, which is the point of tangency, is chosen to be one of the matrices themselves. While this choice is arbitrary, in this work we use the matrix that corresponds to the nearest (in the \emph{standardized} Euclidean sense) parameter snapshot to the test parameter.
The traditional Euclidean space interpolation (where typical vector operations are valid) is then carried out in the tangent plane, after which they are mapped back to the manifold. The mapping to and from the tangent plane is carried out via logarithmic and exponential relationships, as depicted in Figure~\ref{f:Manif_Intro}. Specific details  follow.

The matrix $\tilde{\mbf{B}} = \mbf{\Phi}^\top \mbf{A}^\top \mbf{A} \mbf{\Phi}$ in Equation~\ref{e:conmin} is symmetric positive definite (SPD) for the following reasons. First, $\mbf{A}^\top \mbf{A}$ is a covariance matrix and hence is symmetric positive semi-definite (see~\cite{golub2012matrix}, sec. 5.3 ). Furthermore, multiplication by orthogonal matrix $\mbf{\Phi}$ of rank $k$, where $k < rank(A)$, ensures that $\tilde{\mbf{B}}$ is SPD. SPD matrices of size $k \times k$ form a special group called the $SPD(k)$ ~\cite{Rahman2005, barachant2010riemannian}, and the manifold they are embedded in is denoted as $\mathcal{M}$. Also, for the set of all SPD matrices $\tilde{\mbf{B}}~\in \mathcal{M}$, the tangent plane is the set of all \emph{symmetric} matrices $\tilde{\mbf{B'}}$ ~\cite{barachant2010riemannian}. Any metric ($\mathcal{M}_f$) defined on $SPD(k)$ for any two matrices uses the following functional relationship

\begin{equation}
\mathcal{M}_f(\mbf{B}_1, \mbf{B}_2) = \mbf{B}_1^{1/2} f\left( \mbf{B}_1^{-1/2} \mbf{B}_2 \mbf{B}_1^{-1/2} \right) \mbf{B}_1^{1/2},
\label{e:SPD_f}
\end{equation}
which leads to the following results for the exponential and logarithmic mapping for $SPD(k)$~\cite{Rahman2005} where, $\tilde{\mbf{B}}_0$ is the anchor point and $\tilde{\mbf{B}'}$ is the point whose mapping is desired.

 {Given parameter snapshots $\lbrace \bs{\theta}_i \rbrace, i=1,\hdots,M$ and the corresponding ROMs $\lbrace \tilde{\mbf{B}}_i, \tilde{\mbf{f}}_i\rbrace, i=1,\hdots,M$, the reduced matrix $\tilde{\mbf{B}}(\hat{\bs{\theta}})$ is desired at a new parameter $\hat{\bs{\theta}}$ (unseen by training set). We first pick $\varrho < M$ candidate parameter snapshots and ROMs $\lbrace \bs{\theta}_i, \tilde{\mbf{B}}_i, \tilde{\mbf{f}}_i \rbrace, i=1,\hdots,\varrho$, necessary to perform the interpolation, which are again chosen as the nearest neighbors to $\hat{\bs{\theta}}$. Note that the size of the candidate set $\varrho$ depends on the interpolation being performed; we use second-order multivariate Lagrange polynomials that, for $d$ variables, require $\varrho = \binom{d+2}{d}$ candidate points; see \cite{renganathan2018koopman} for details. We then pick an anchor point $\tilde{\mbf{B}}_0 = \tilde{\mbf{B}}(\hat{\bs{\theta}}_0)$, where $\bs{\theta}_0$ is chosen as the \emph{nearest} neighbor to $\hat{\bs{\theta}}$ within the candidate set.} Then, the exponential mapping of $\tilde{\mbf{B}}'$ from tangent plane to $\mathcal{M}$ at $\tilde{\mbf{B}}_0 \in \mathcal{M}$, to $\mathcal{M}$ is given by 

\begin{equation}
\mbf{Exp}_{\tilde{\mbf{B}}_0} \tilde{\mbf{B}} = \tilde{\mbf{B}}_0^{1/2} \left(\tilde{\mbf{B}}_0^{-1/2} exp(\tilde{\mbf{B}}') \tilde{\mbf{B}}_0^{-1/2} \right) \tilde{\mbf{B}}_0 ^{1/2} 
\label{e:ExpSPD}
\end{equation}
and the logarithmic mapping of $\tilde{\mbf{B}} \in \mathcal{M}$ to tangent plane to $\mathcal{M}$ at $\tilde{\mbf{B}}_0 \in \mathcal{M}$ is 
\begin{equation}
\mbf{Log}_{\tilde{\mbf{B}}_0} \tilde{\mbf{B}'} = \tilde{\mbf{B}}_0^{1/2} log \left(\tilde{\mbf{B}}_0^{-1/2} \tilde{\mbf{B}} \tilde{\mbf{B}}_0^{-1/2} \right) \tilde{\mbf{B}}_0^{1/2} .
\label{e:LogSPD}
\end{equation}

 {The candidate ROMs are first mapped to the tangent space via \eqref{e:LogSPD}, following which they are interpolated element-wise (using Lagrange polynomials) to find $\tilde{\mbf{B}}'(\hat{\bs{\theta}})$. Then $\tilde{\mbf{B}}'(\hat{\bs{\theta}})$ is mapped back to the manifold $\mc{M}$, via \eqref{e:ExpSPD} to give $\tilde{\mbf{B}}(\hat{\bs{\theta})}$ which is guaranteed to be SPD. On the other hand, $\tilde{\mbf{f}}(\hat{\bs{\theta}})$ is computed by directly interpolating the candidate $\tilde{\mbf{f}}_i$'s elementwise.}
In summary, for new parameter instances outside of the training set, the ROM database is first interpolated via the method described in this section following which \eqref{e:conmin} is solved via SQP. We now proceed to introduce the machine learning-enabled approach which is independent of the system matrices $\lbrace \tilde{\mbf{B}}, \tilde{\mbf{f}} \rbrace$ and relies directly on learning the variation of $\tilde{\mbf{u}}$ in the $\bs{\theta}$ space.

\section{Machine-Learning-Enabled Model Order Reduction}
\label{s:DNN}

In the recent past, several studies have examined the integration of machine learning techniques within the projection-based ROM methodology. These have generally focused on improving the capture of transient dynamics within the reduced space spanned by the POD bases \cite{wan2018data,wang2019non,mohan2018deep,maulik2019time} or on determining projections where advective dynamics may be captured more confidently \cite{murata2020nonlinear,lusch2018deep,gin2019deep}.  In this article, we introduce a machine learning framework to predict the coefficients (the reduced state) of the POD modes obtained from our simulation database where all snapshots generated for reduced-basis identification are at \emph{steady state}. We remind the reader that the POD modes were obtained by using the training dataset (corresponding to $M=80$ simulations only). We outline a learning task that seeks to learn a map between inputs $\bs{\theta}$ and our outputs given by the coefficients of these POD modes. Therefore our training data set consists of coefficients obtained by projecting the training snapshots at steady state for a variety of control parameters onto their global POD modes. For capturing approximately 97\% of the total variance in the pressure fields, we retain $k=16$ POD modes for the pressure snapshots. The eigenvalue-variance decay plot is shown on the left of Figure \ref{Fig_ML_1}. We note that for both the 2- and 8-parameter cases, 16 modes roughly captured 97\% of the energy and our output space dimensionality was fixed for both assessments. We also note that the $k$ chosen for the projection-based approach is to capture approximately 99\% of the variance in the state. However, we restrict it to 97\% here in order to keep the output dimensionality tractable and hence avoid overfitting. In other words, $k=16$ is the highest output dimensionality we are able to achieve for the given training dataset for the best possible accuracy.

\begin{figure}
    \centering
    \includegraphics[width=0.8\textwidth]{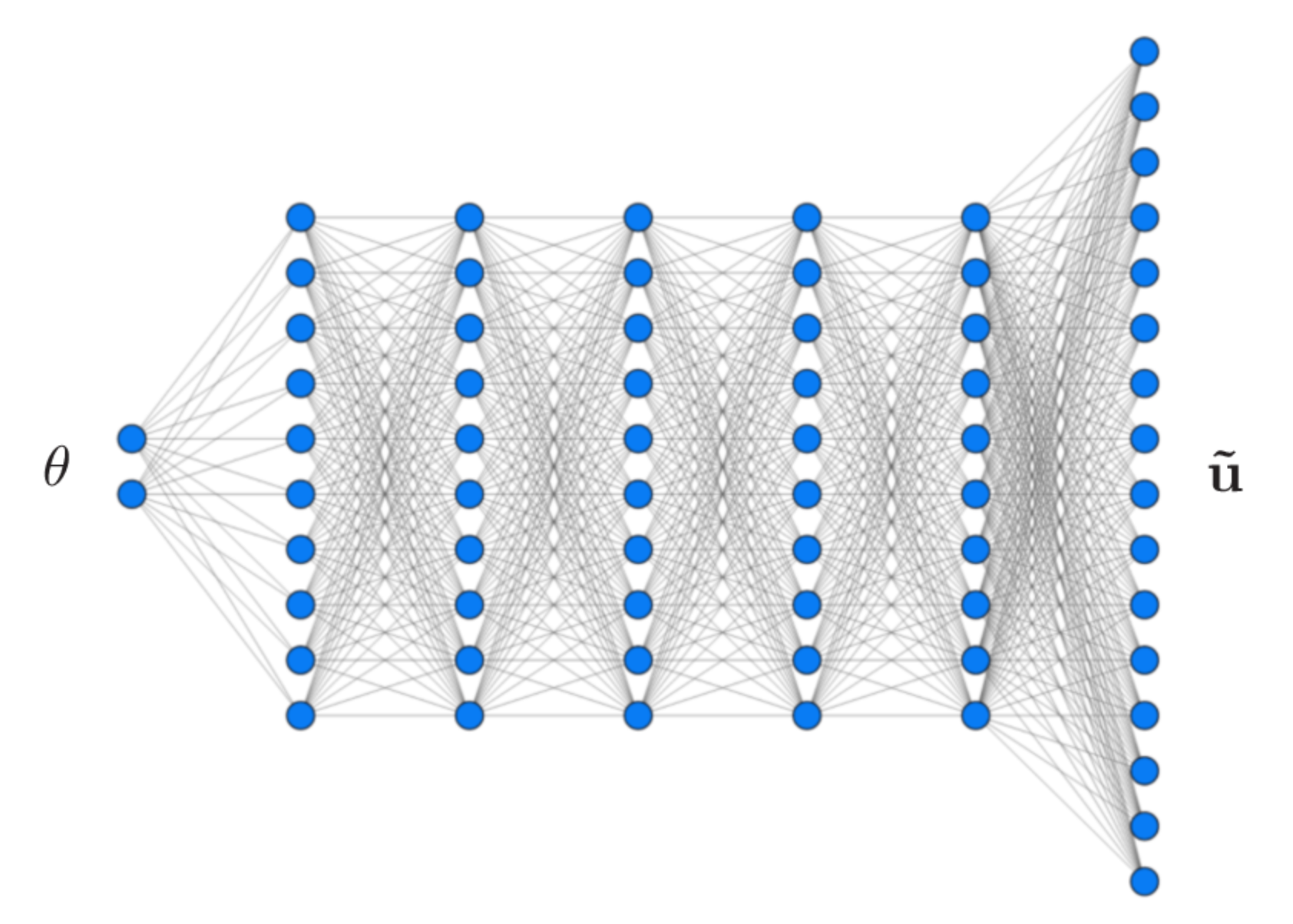}
    \caption{Deep neural network schematic for POD coefficient reconstruction for our ROM. Note that this architecture is solely for the purpose of representation and our proposed framework utilizes 6 hidden layers and 50 neurons in each layer.}
    \label{NN_Schematic}
\end{figure}

To build our reduced-order model, we utilize a multilayered perceptron initially developed by Rumelhart \cite{rumelhart1986learning}. The multilayered perceptron (more commonly known as the artificial neural network) is modeled on the structure of biological neural networks and has recently seen widespread use in multiple fluid mechanics applications. A deep neural network consists of multiple minimal units that perform the following operation,
\begin{align}
    m_{i}^{(l)}=\varphi\left(\sum_{j} H_{ij}^{(l)} m_{j}^{(l-1)}\right),
\end{align}
where $H_{ij}^l$ is a trainable ``weight'' of the $l$th layer and $m_i^{l-1}$ is the input from the previous layer if $l > 1$ or the inputs ($\bs{\theta}$) themselves if $l=1$. The function $\varphi$ is generally chosen to be monotonically increasing and is often called the \textit{activation}. For deep neural networks, a common activation is given by
\begin{align}
    \varphi(x) = \max(x,0).
\end{align}
This activation is also commonly called the rectified linear activation unit (or ReLU)  \cite{nair2010rectified}. Details on the specific choices and their influence on networks can also be found in \cite{haykin1994neural}. A visual schematic of a sample neural network utilized in this study is shown in Figure \ref{NN_Schematic}. We note that the choice of a fully connected neural network stems from the knowledge of global interactions between POD basis coefficients parametrized by the control parameters $\bs{\theta}$. 

The MLP framework described above is used to train nonlinear maps given by 
\begin{equation*}
    \mbb{M}: \bs{\theta} \rightarrow  \tilde{\mbf{u}},
\end{equation*}
where $\bs{\theta} \in \mbb{R}^d$ and $\tilde{\mbf{u}} \in \mbb{R}^k$. We  consider two cases: the first map utilizing a $d=2$ dimensional input given by  $\bs{\theta} = [Ma, \alpha] \in \mathbb{R}^2$, and the second case mapping from a higher-dimensional input $d=8$ (a more practical predictive task) that captures the airfoil shape. To train these maps, we minimize a cost functional 
\begin{align}
    \label{ML_OF}
    \mbf{H}^* = \underset{\mbf{H}}{\text{arg min}} \quad \frac{1}{2} \| \tilde{\mbf{U}} - \tilde{\mbf{U}}^p \|_2 ^2,
\end{align}
where $\mathbf{H}$ is the set of trainable parameters of the nonlinear map (determined by the shape of the neural network) and $\tilde{\mbf{U}} = [\tilde{\mbf{u}}_1,\hdots,\tilde{\mbf{u}}_M] \in \mbb{R}^{k \times M}$. The optimization problem is solved by using the Adam optimizer \cite{kingma2014adam} with a learning rate of 0.001. Our batch size is 40 samples for the first map given by $d=2$ and 64 samples for the second map with $d=8$. The $d=2$ case was characterized by 13,716 trainable parameters, and the $d=8$ case had 14,016 trainable parameters. Both frameworks were allowed to train for a maximum number of 2,000 epochs, with an early stopping criterion utilized to prevent overfitting to the training data. This criterion was set at 100 epochs (i.e., if training exceeded 100 epochs without an improvement in the accuracy, optimization would be terminated, and the network parameters would be reset to the previous best estimates).The Glorot uniform initializer \cite{glorot2010understanding} was used for starting points in the optimization process, and a stochastic optimization was performed by choosing the aforementioned batch size; this  helps in efficiently exploring the loss surface and avoid getting stuck in flat minima. As in most deep learning applications, however, there is no guarantee that a trained map represents a globally optimally solution. The overall algorithm for training is shown in Algorithm \ref{ML_Algo}.

\begin{algorithm}[H]
\SetAlgoLined
\KwResult{Trained map $\mathbb{M}$}
 Given steady-state training pressure snapshots at different $\bs{\theta}$\;
 1. Compute POD basis vectors using Equation \ref{e:svd}\;
 2. Obtain POD coefficients by projecting training pressure snapshots on truncated POD basis vectors\;
 3. Collect parameters $\bs{\theta}$ and coefficients as training data set\;
 4. Train a deep neural network by optimizing for cost function given by Equation \ref{ML_OF} using the ADAM optimizer\;
 5. Assess accuracy of trained $\mathbb{M}$ on testing dataset for unseen $\bs{\theta}$.
 \label{ML_Algo}
 \caption{Machine learning steady-state pressure fields for transonic flow over RAE2822}
\end{algorithm}

\section{Numerical Experiments}
\label{s:gov_eqns}
We now introduce the compressible Euler equations upon which we demonstrate our proposed method. Following that, we present the RAE2822 airfoil which is used as the test case and the associated parametrization used in this work.

\subsection{Compressible Euler Equation}
The Euler equation governing the two-dimensional, compressible inviscid flow past an airfoil is the chosen FOM on which we perform model reduction. The equation in conservation form is 

\begin{equation}
\nabla_x \mbf{F} + \nabla_y \mbf{G} = 0, 
\label{e:Euler_cons}
\end{equation}

where
\begin{equation*}
\begin{aligned}
&& \mbf{F} = 
\begin{bmatrix}
\rho u \\
\rho u^2 + p \\
\rho uv \\
\rho uH
\end{bmatrix},~ \mbf{G} = \begin{bmatrix}
\rho v \\
\rho uv \\
\rho v^2 + p \\
\rho vH
\end{bmatrix} \\
&& H = E + \frac{p}{\rho}\\
&& \rho E = \frac{1}{2} \rho (u^2 + v^2) + \frac{p}{\gamma -1}.
\end{aligned}
\end{equation*}
Here $\rho,u,v,$ and $p$ are the primitive variables, namely, the density, velocity components, and pressure, respectively; $H$ is the enthalpy; $E$ is the internal energy and $\gamma$ is the ratio of specific heats; and  $\nabla_x$ and $\nabla_y$ are the $x$ and $y$ components of the gradient operator $\nabla$, respectively. The farfield ($\infty$) boundary conditions are specified with a flow direction, the Mach number ($Ma$), static pressure ($p$), and static temperature ($T$). The free-stream boundary values on the boundary face of the computational domain are computed based on extrapolation of Riemann invariants under the assumption of irrotational, quasi-1D flow in the boundary-normal direction. The airfoil surface is modeled as an \emph{adiabatic slip wall} where all the primitive and thermodynamic variables are extrapolated from the interior domain via reconstruction gradients. The numerical solution to the nonlinear system is obtained with a coupled implicit finite-volume-based solver with second-order spatial discretization. The gradients are computed with the hybrid Gauss--least-squares method and the Venkatakrishnan limiter \cite{venkatakrishnan1995convergence}. All the FOM snapshots are obtained via the commercial black-box computational fluid dynamics solver Star-CCM+. The nonintrusive approach to model reduction of the Euler equations is detailed in Appendix~\ref{a:Euler_MOR}.

\subsection{The RAE2822 Airfoil}
\label{Problem_Def}
\begin{figure}
  \begin{subfigure}{0.5\textwidth}
  \includegraphics[width=1\linewidth]{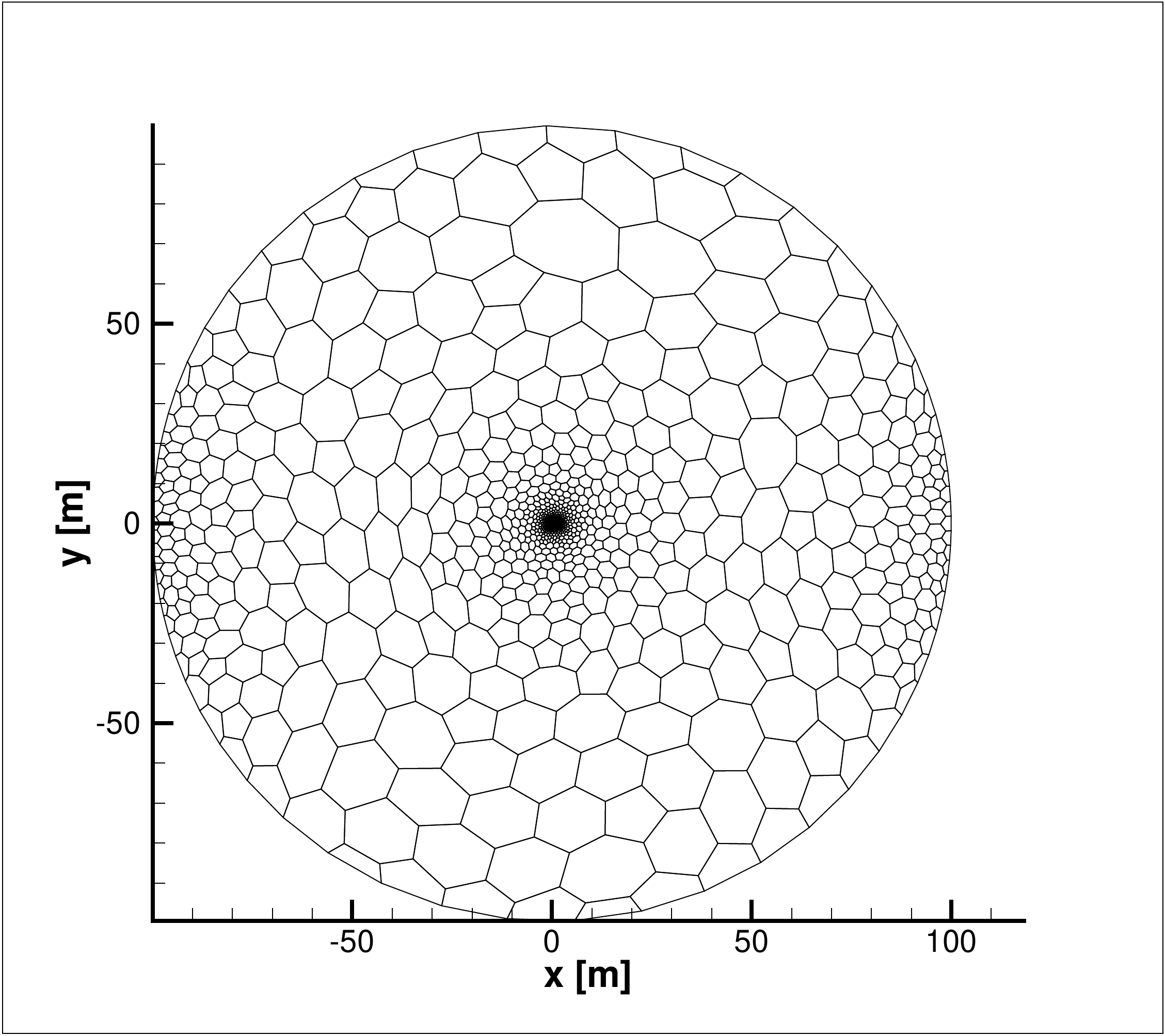}
  \caption{Flow domain with mesh}
  \label{f:RAE2822_Grid}
  \end{subfigure}%
  \begin{subfigure}{0.5\textwidth}
  \includegraphics[width=1\linewidth]{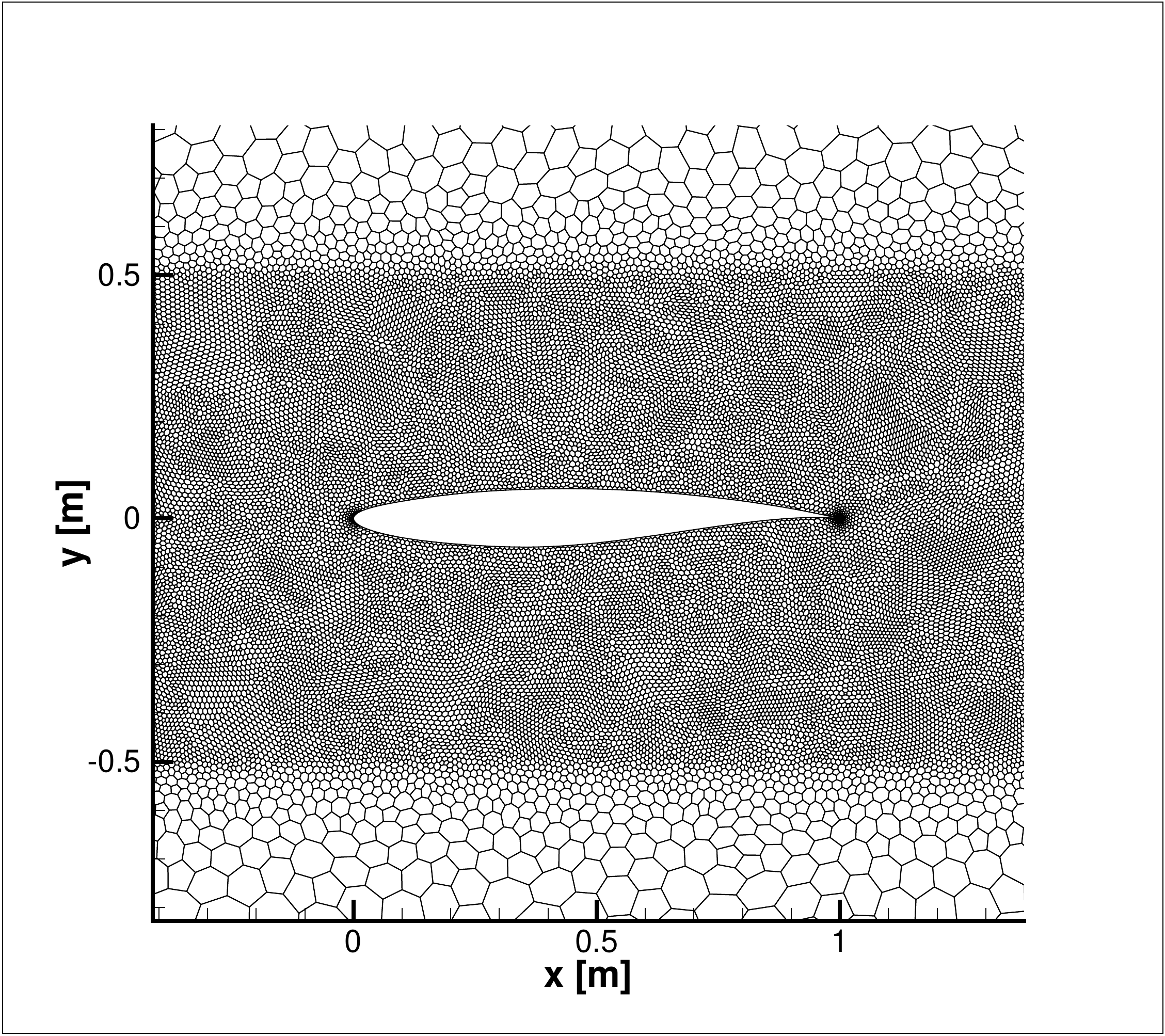}
  \caption{Near field mesh}
  \label{f:RAEMesh}
  \end{subfigure}%
  \caption{Computational grid for the airfoil test case}
  \label{f:airfoil_grid}
\end{figure}

The RAE2822 is a commonly used canonical test case for aerospace engineering problems under transonic flight conditions. A spherical domain of $100$ airfoil chord length radius is used to model the fluid domain, which is meshed with 27,857 polyhedral mesh elements with near-field refinement to capture the shock (see Figure~\ref{f:airfoil_grid}). In this work, the main quantity of interest is the normalized pressure, also known as the \emph{coefficient of pressure} ($C_P$), which is defined as \begin{equation}
C_P = \f{p - p_\infty}{\f{1}{2} \rho_\infty (Ma_\infty\times a_\infty)^2} ,
\label{e:cp}
\end{equation}
which follows from the fact that this is a two-dimensional simulation and the chord length is set to unity. Additionally we also consider output quantities of interest namely the lift and drag coefficients define as
\[C_l = \frac{L}{\frac{1}{2}\rho_\infty (Ma_\infty\times a_\infty)^2} \quad 
C_d = \frac{D}{\frac{1}{2}\rho_\infty (Ma_\infty\times a_\infty)^2},\]
where $L$ and $D$ are the forces in the direction perpendicular and parallel to the freestream $(\infty)$ direction respectively.

We investigate two  parametric cases. First, the free-stream Mach number ($Ma$) and the flow incidence angle or \emph{angle of attack} $\alpha$ are chosen parameters at a fixed airfoil shape,  $\bs{\theta} \subset \bs{\Theta} \in \R^2$, where $0.8 \leq \theta_1 = Ma \leq 0.9, 0 \leq \theta_2 = \alpha \leq 2.0$. Then, the airfoil shape is parametrized by using $\bs{\theta} \subset \bs{\Theta} \in \R^8$ at fixed free-stream boundary conditions (summarized in Table~\ref{t:RAE_freestream}) (explained in the next section). A total of $90$ parameter snapshots are generated via a space-filling design of experiments, namely, the Latin hypercube design~\cite{myers2016response}, of which  80 are used for training. 

\subsection{Airfoil Shape Parametrization}
\label{s:test_cases}
The baseline shapes are parametrized by using \emph{Class Shape Transformation (CST)}~\cite{kulfan2006fundamental, kulfan2008universal}. The CST model of parametrization defines a \emph{class} function $c()$ and a \emph{shape} function $s()$ and the curve being parametrized is specified as their product. The main idea is that whereas the class function  defines a general class of geometry such as airfoils, missiles, or Sears-Haack body, the shape function  defines the unique shape within a particular class of shapes (such as the RAE2822 airfoil). The class function $c(\psi)$ is more generally defined as

\begin{equation}
c_{n_1}^{n_2} (\psi) := \psi ^{n_1}(1 - \psi)^{n_2},
\label{e:class_fn}
\end{equation}
where the variable $0 \leq \psi \leq 1$ represents the nondimensional chordwise distance and $n_1$ and $n_2$ define the specific class (for instance $n_1 = 0.5,~n_2 = 1$). Hence $c_{0.5}^{1.0} (\psi) = \sqrt{\psi} (1 - \psi)$ defines airfoils with rounded leading edge and a sharp trailing edge~\cite{kulfan2006fundamental}. The unique shape of an airfoil is driven by the shape function, specified as follows:

\begin{equation}
s(\psi) = \sum_{i=0}^{n} A_i \psi^i ,
\label{e:shape_fn1}
\end{equation}
where $A_i$ are the coefficients that are also the shape parameters; we refer to $A_i$ as \emph{CST coefficients} in the rest of the paper. The final shape of the airfoil shape ($z_{cs}$) is then given by

\begin{equation}
z_{cs}(\psi) = c(\psi)s(\psi).
\label{e:CST_final}
\end{equation}

The coefficients $A_i,~i=1,\hdots,n_s$ represent the actual parameters of the airfoil shape given $n_s$, the order of the Bernstein polynomials. An $n_s$th-order CST parametrization has $n_s+1$ parameters; if separate parametrizations are sought for the upper and lower surfaces of the airfoil, then the CST parametrization leads to $2(n_s+1)$ parameters to specify the whole shape of the airfoil, where the $n_s$ needs to be determined. Typically, however, $n_s=3-5$ are observed to be adequate to parametrize the airfoil shapes considered in the precursor work~\cite{renganathan2018methodology}. One way to determine $n_s$ and the associated polynomial coefficients is to find the values that minimize certain error (such as $L_2$) between the true shape of the airfoil and the resulting approximation via CST. In this work, the parameters for a given airfoil shape are determined by solving the following minimization problem, after setting $n_1 = 0.5$ and $n_2=1.0$:

\begin{equation}
\lbrace A_i \rbrace_{i=0}^{n_s}, n_s = \underbrace{\text{minimize}}_{n_s, A_i}~ \left ( z_{true}(\psi) - c(\psi)s(\psi) \right ) ^2 .
\label{e:CST_param}
\end{equation}

In practice, \eqref{e:CST_param} can be simplified as 

\begin{equation}
\lbrace A_i \rbrace_{i=0}^{n_s}, n_s = \underbrace{\text{minimize}}_{n_s, A_i}~ \left \| \mb{z}_{true} - \mb{c} \otimes \mb{s} \right \|_2 ^2,
\label{e:CST_param_discrete}
\end{equation}
where $\mb{z}_{true}$, $\mb{c}$ and $\mb{s}$ are the discrete approximations of  $z_{true}(\psi)$, $c(\psi)$ and $s(\psi)$, respectively, at $\tilde{\psi} \in \R^{n_s+1}$, which is the discrete approximation of $\psi$ at $n_s+1$ unique points sampled from $\psi$ spanning $[0,1]$, and $\otimes$ is the elementwise or Hadamard product.  In this way, the smallest possible $n_s$ is determined, thereby avoiding overparametrization. Note that $n_s$ is treated as a continuous variable in \eqref{e:CST_param_discrete} and is then rounded off to the nearest integer. Furthermore, \eqref{e:CST_param_discrete} is solved separately for the upper and lower surfaces of the airfoil.

\begin{figure}
    \centering
    \includegraphics[scale=0.4]{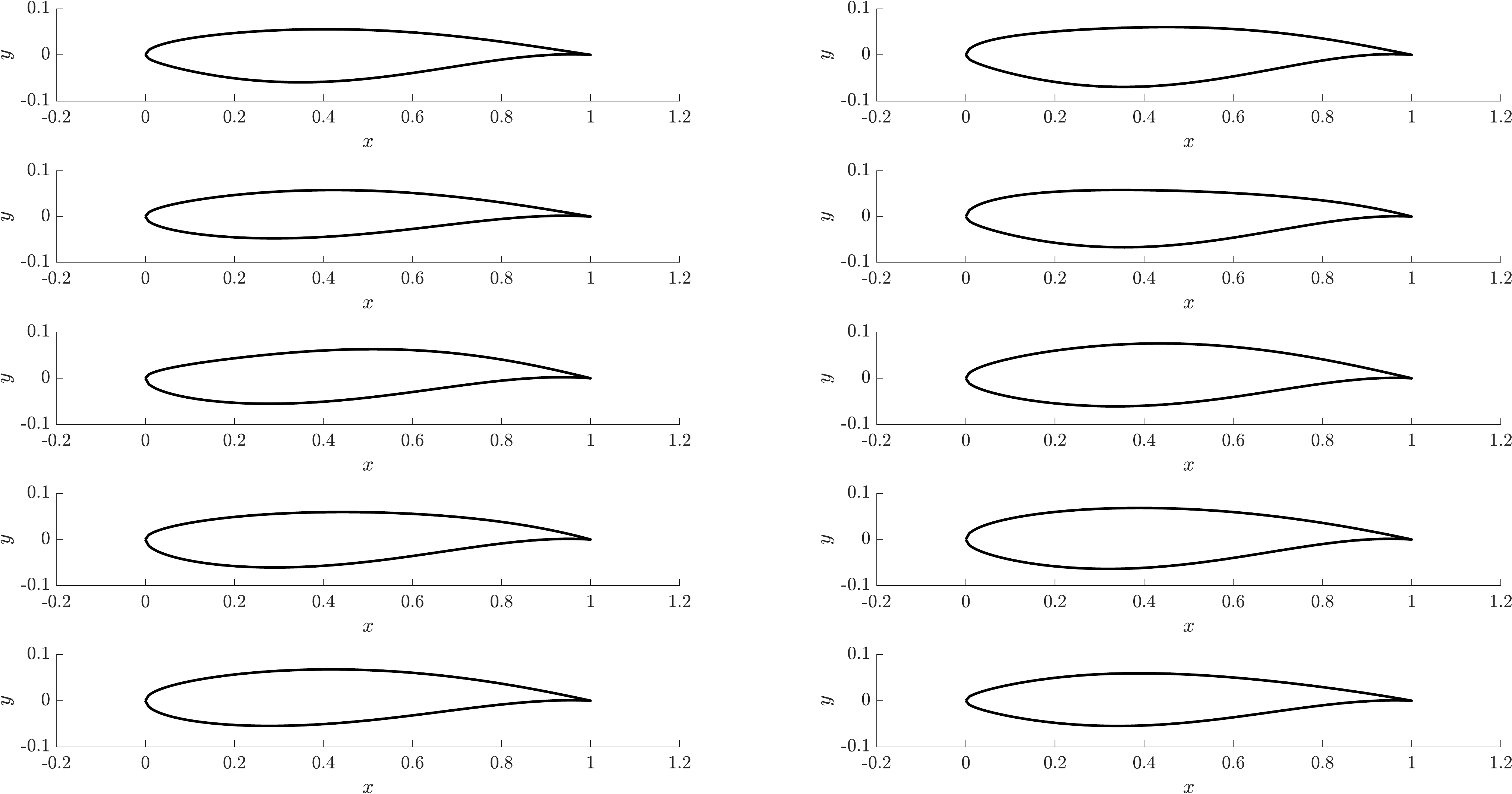}
    \caption{Family of airfoils generated by perturbing ($\pm 30\%$) the baseline CST coefficients of the RAE2822}
\label{f:RAE2822_CST_Family}
\end{figure}

The RAE2822 airfoil is parametrized by using 8 ($n_s = 3$) variables, whose values are denoted $A_{RAE2822}$ in \eqref{e:shape_params}, where the top and bottom rows correspond to the upper and lower surfaces of the airfoil, respectively. Note that \eqref{e:shape_params} represents the \emph{baseline} shape parameters of the RAE2822 airfoil, which are then perturbed to modify the airfoil shape. The parameter vector, $\bs{\theta}$, is expressed as a vector of length 8, for instance by concatenating the top and bottom rows of $A_{RAE2822}$. Bounds on the parameters are carefully chosen to prevent nonphysical (intersecting) airfoil geometries as well as mesh deformation that could potentially lead to FOM solver instabilities due to poor-quality elements. In this regard, a $\pm 30\%$ range is chosen to vary the airfoil shape, a sample of which is shown in Figure~\ref{f:RAE2822_CST_Family}. The airfoil shape changes due to CST parametrization are propagated to the computational mesh via 27 control points subject to free-form deformation (FFD)~\cite{samareh2004aerodynamic}.

\begin{equation}
\begin{aligned}
A_{RAE2822} &= \begin{bmatrix}
 0.1268 & 0.4670 & 0.5834 & 0.2103 \\
-0.1268 &-0.5425 &-0.5096 & 0.0581
\end{bmatrix}
\end{aligned}
\label{e:shape_params}
\end{equation}

\clearpage
\section{Results and Discussion}
\label{Results}

We demonstrate the proposed methodology on the prediction of the transonic flow field past the RAE2822 airfoil. We pick the pressure distribution as the variable for comparison mainly because of  its relevance in aerodynamic design but also because it makes interpretation of compressible flow phenomena such as shocks easier. Furthermore, we  compare the proposed approach in terms of predicting the coefficient of pressure distribution on the surface of the airfoil as well as on the output quantities of interest, namely, the coefficient of drag and lift. We begin this section by providing details about the model diagnostics of the proposed approach.

\subsection{DNN Model Diagnostics}
\label{ss:DNN_MS}

\begin{figure}
\centering
\includegraphics[width=0.49\textwidth]{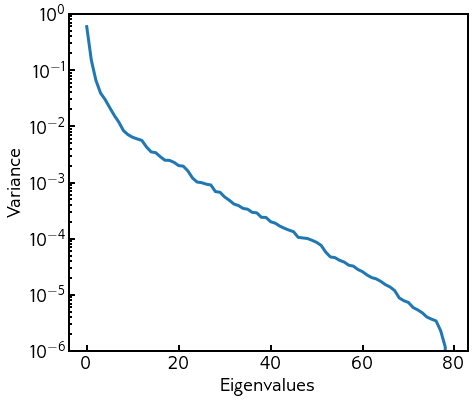}
\includegraphics[width=0.49\textwidth]{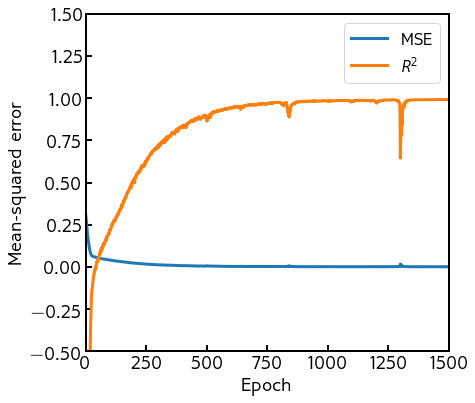}
\caption{Eigenvalue decay for the training dataset (left) showing the need for around 16 modes for adequate variance capture. Also shown is the progress to convergence for the training of our DNN (right), where a high coefficient of determination ($R^2$) is obtained through a training of 1,500 epochs on our training dataset. }
\label{Fig_ML_1}
\end{figure}

In this section we outline results from the training and deployment of the proposed formulation. We  provide quantitative assessments of the training phase on our first map (i.e., parametrizing the POD basis coefficients as a function of $Ma, \alpha$ alone), with similar results obtained for the second map. Results are provided for both prediction tasks. We note that the training of $\mathbb{M}_1$ required approximately 30 seconds on an Intel Core I7 CPU with a naive build of TensorFlow 1.15.

\begin{figure}
\centering
\begin{subfigure}{.5\textwidth}
  \centering
  \includegraphics[width=1\linewidth]{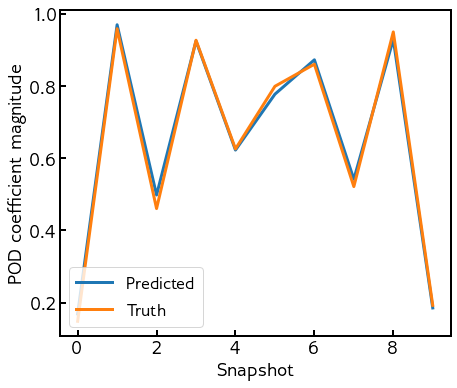}
  \caption{Coefficient 1}
  \label{sf:coeff1}
\end{subfigure}%
\begin{subfigure}{.5\textwidth}
  \centering
  \includegraphics[width=1\linewidth]{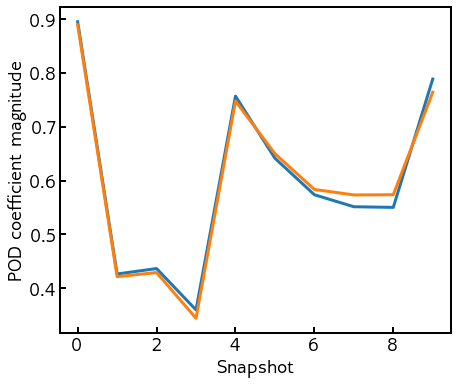}
  \caption{Coefficient 6}
  \label{sf:coeff6}
\end{subfigure}\\
\begin{subfigure}{.5\textwidth}
  \centering
  \includegraphics[width=1\linewidth]{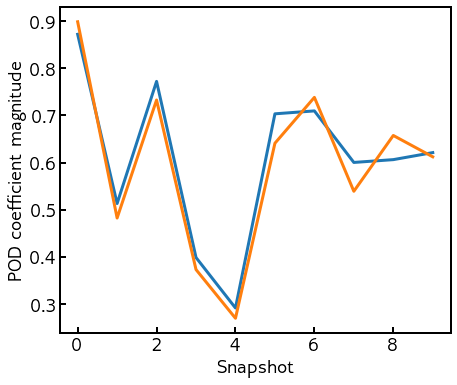}
  \caption{Coefficient 11}
  \label{sf:coeff11}
\end{subfigure}%
\begin{subfigure}{.5\textwidth}
  \centering
  \includegraphics[width=1\linewidth]{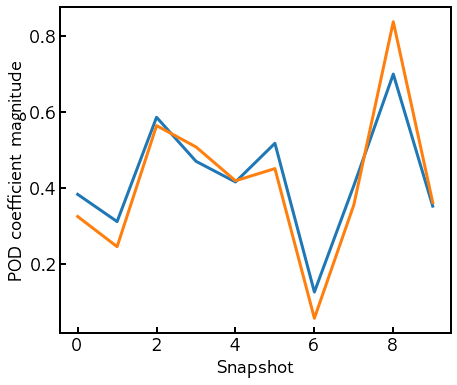}
  \caption{Coefficient 16}
  \label{sf:coeff16}
\end{subfigure}\\
\caption{Prediction ability of the trained framework for the testing dataset. The panels show the ability to predict different POD coefficients of this dataset unseen during training.}
\label{Fig_ML_2}
\end{figure}

\begin{figure}
\begin{subfigure}{.5\textwidth}
  \centering
  \includegraphics[width=1\linewidth]{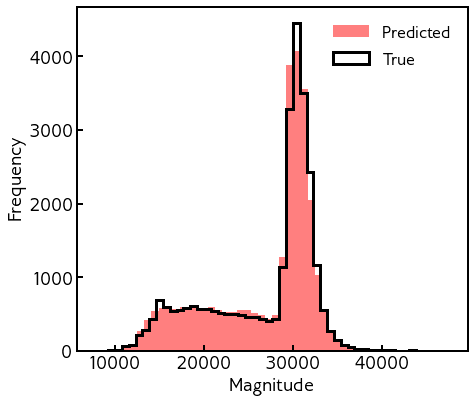}
  \caption{Frequency}
  \label{sf:pdf}
\end{subfigure}%
\begin{subfigure}{.5\textwidth}
  \centering
  \includegraphics[width=1\linewidth]{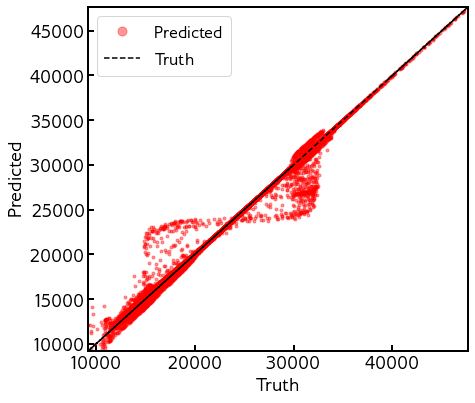}
  \caption{Scatter}
  \label{sf:scatter}
\end{subfigure}
\caption{Prediction ability of the trained framework for one snapshot of the testing dataset. The left panel shows the distribution of the true and predicted values of the absolute pressure. The right panel shows the scatter between true and predicted values.}
\label{Fig_ML_3}
\end{figure}

A first validation of the proposed framework is performed through the use of fivefold cross-validation for the training of the neural network, which provides a trustworthy estimate of the test accuracy of the proposed model. Our average coefficient of determination for this cross-validation was around 0.94. This indicated that the proposed framework was performing in the expected manner and not due to a fortuitous selection of training snapshots.  We further validate the training of our framework through quantitative assessments on the testing dataset, as shown through the comparison of POD coefficient values. A representative progress to convergence for the framework on the training data is shown on the right of Figure \ref{Fig_ML_1}. Predictions for four of these coefficients are shown in Figure \ref{Fig_ML_2}. The proposed framework is seen to perform  accurately. We also perform statistical assessments of the predicted snapshots obtained by reconstructing the field using the predicted POD coefficients. These are shown in Figure \ref{Fig_ML_3} and demonstrate that the distribution of the true field is recovered accurately. A scatter plot is also presented showing that the vast majority of predicted values of the absolute pressure lie close to their true values. We note that all these assessments are performed for snapshots that the learning framework has not seen previously. 

\clearpage
\subsection{Flow Field Prediction}
\label{ss:ff_prediction}
We first show the results for freestream boundary parameters ($\bs{\theta} = [Ma, \alpha]$) with the airfoil shape fixed at the baseline CST coefficients given in \eqref{e:CST_final}. Note that parameter variation in this regime induces distinct shock patterns that serve as a good experiment to evaluate the proposed approach. We then show the results for varying airfoil shape parameters (via the CST coefficients) at fixed freestream boundary conditions. This work finds applications in many-query problems such as aerodynamic shape optimization and uncertainty propagation.

\subsubsection{Freestream Boundary Parameters}
\begin{figure}
\centering
  \begin{subfigure}{.45\textwidth}
  \centering
  \includegraphics[width=1\linewidth]{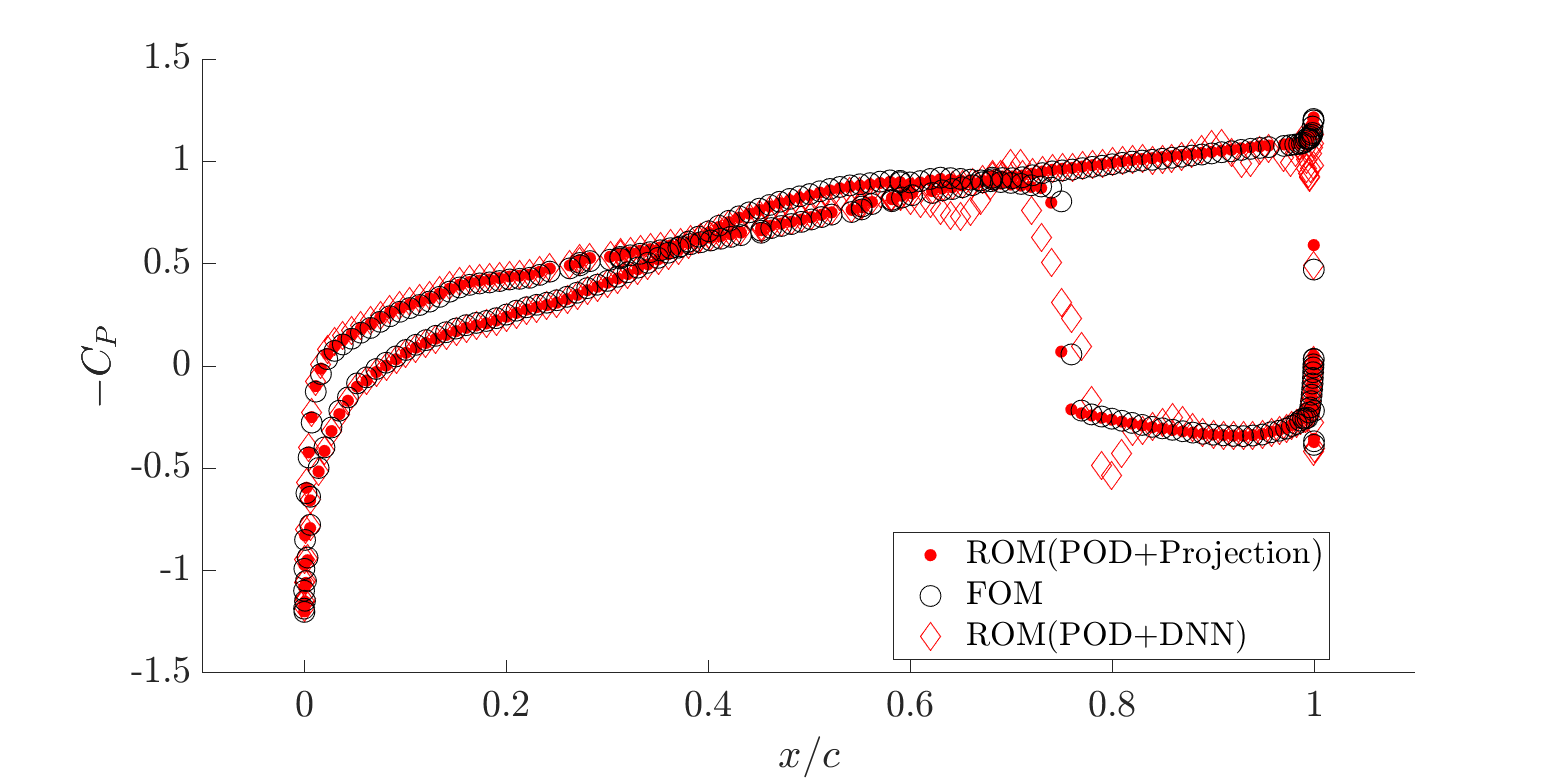}
  \caption{$Ma=0.887$, $\alpha=0.944$}
  \label{sf:DNN81}
  \end{subfigure}%
  \begin{subfigure}{.45\textwidth}
  \centering
  \includegraphics[width=1\linewidth]{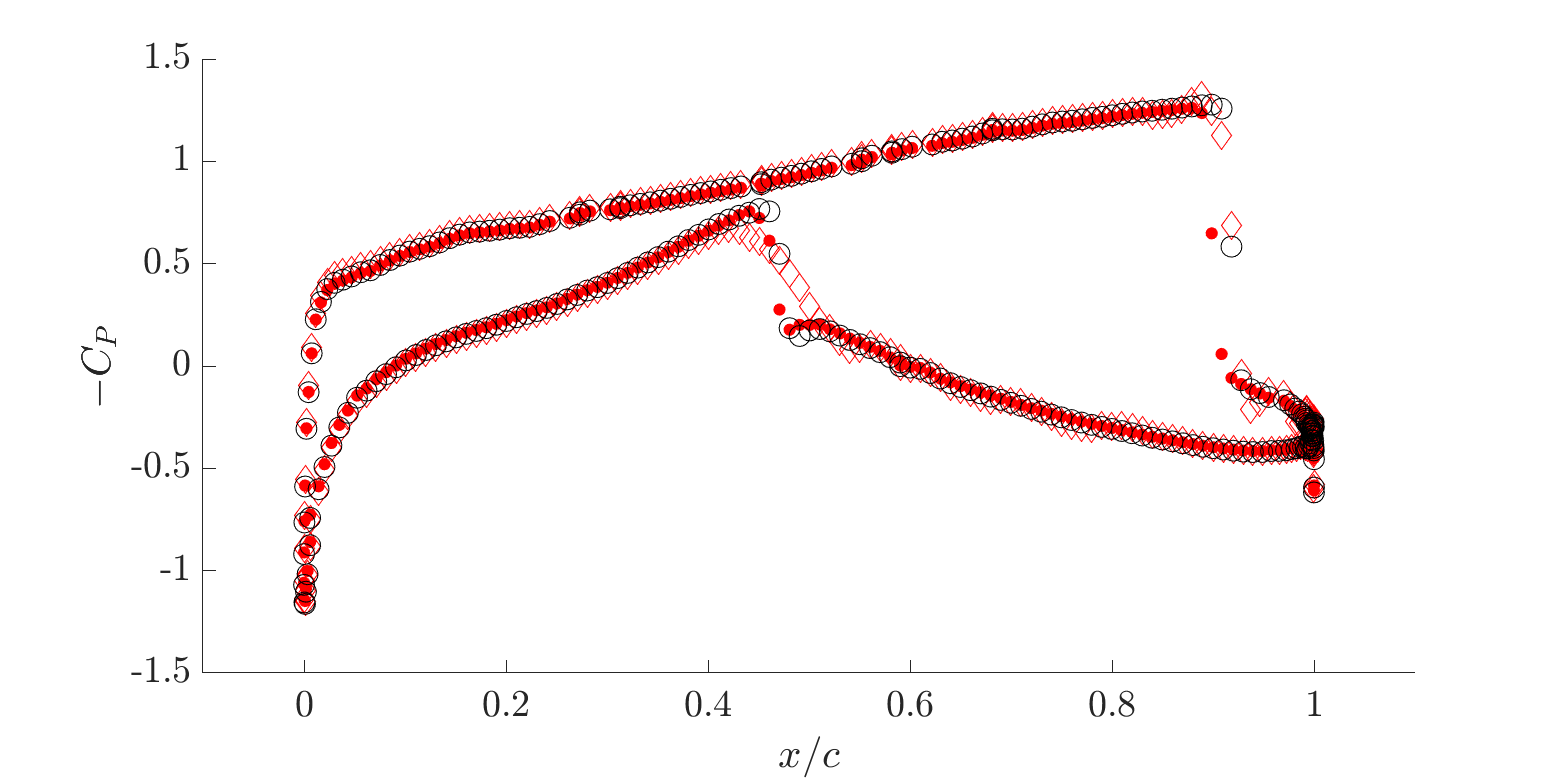}
  \caption{$Ma=0.816$, $\alpha=1.011$}
  \label{sf:DNN82}
  \end{subfigure}\\
  \begin{subfigure}{.45\textwidth}
  \centering
  \includegraphics[width=1\linewidth]{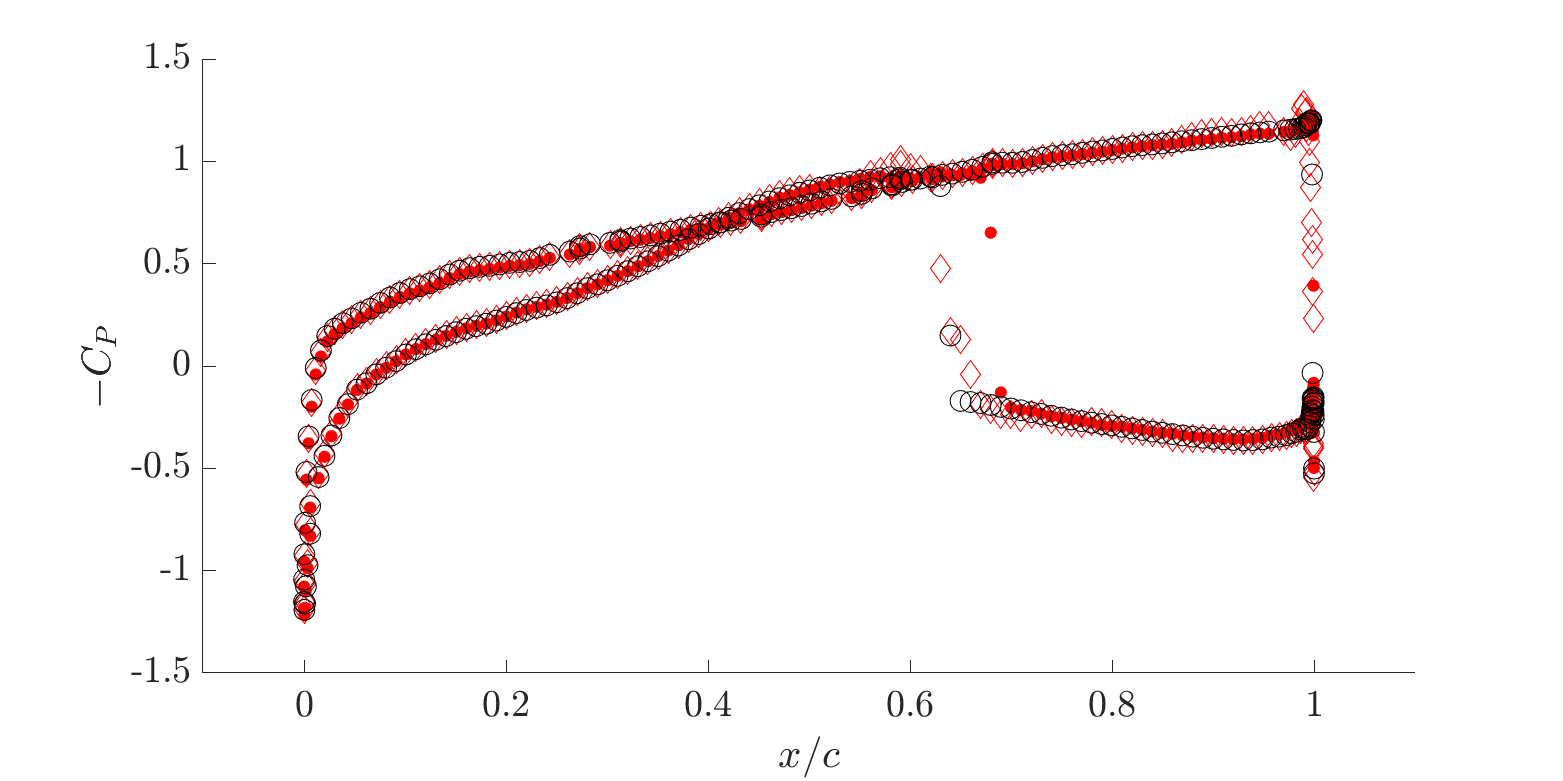}
  \caption{$Ma=0.866$, $\alpha=1.146$}
  \label{sf:DNN83}
  \end{subfigure}%
  \begin{subfigure}{.45\textwidth}
  \centering
  \includegraphics[width=1\linewidth]{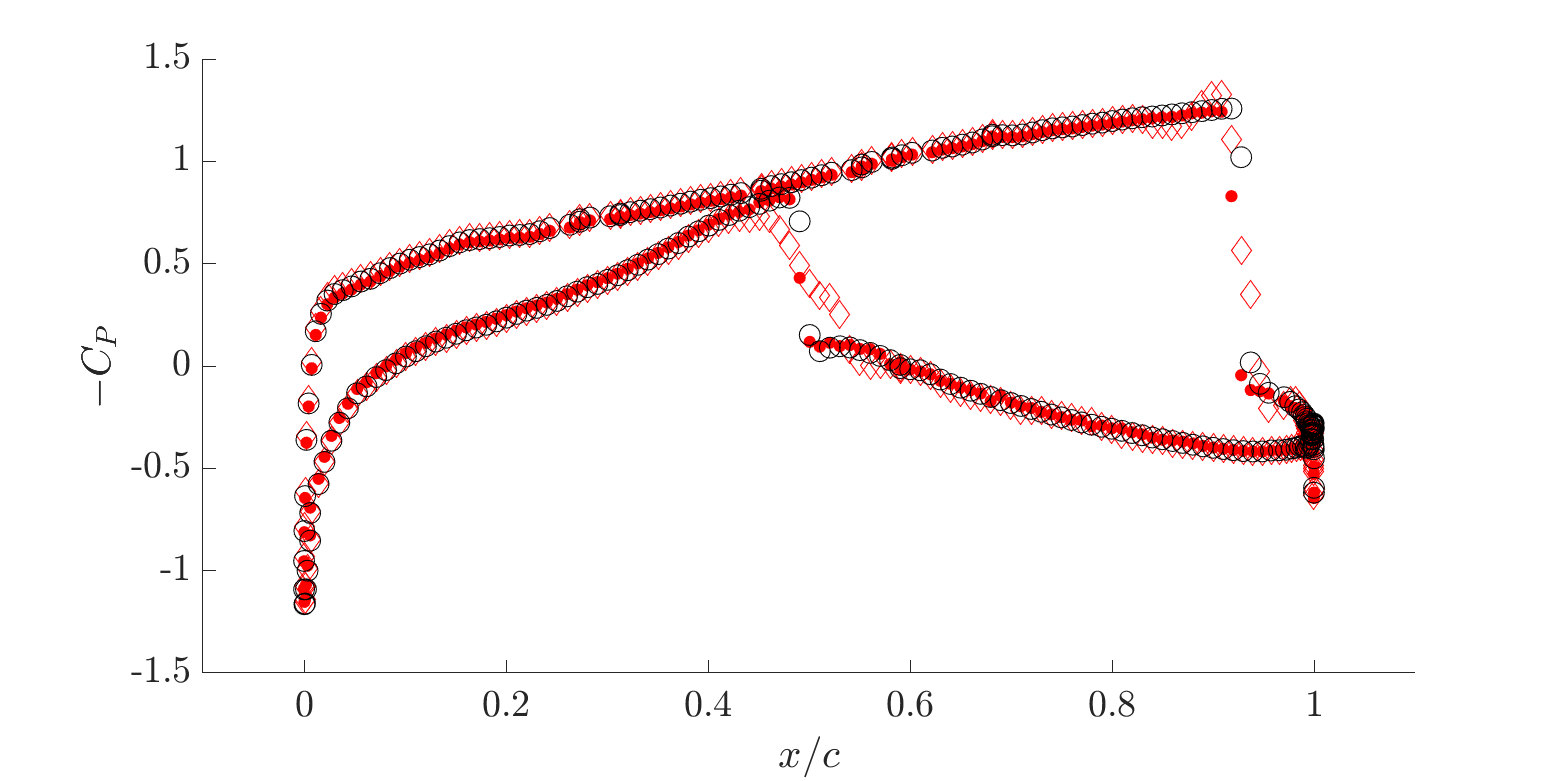}
  \caption{$Ma=0.823$, $\alpha=0.899$}
  \label{sf:DNN84}
  \end{subfigure}\\     
  \begin{subfigure}{.45\textwidth}
  \centering
  \includegraphics[width=1\linewidth]{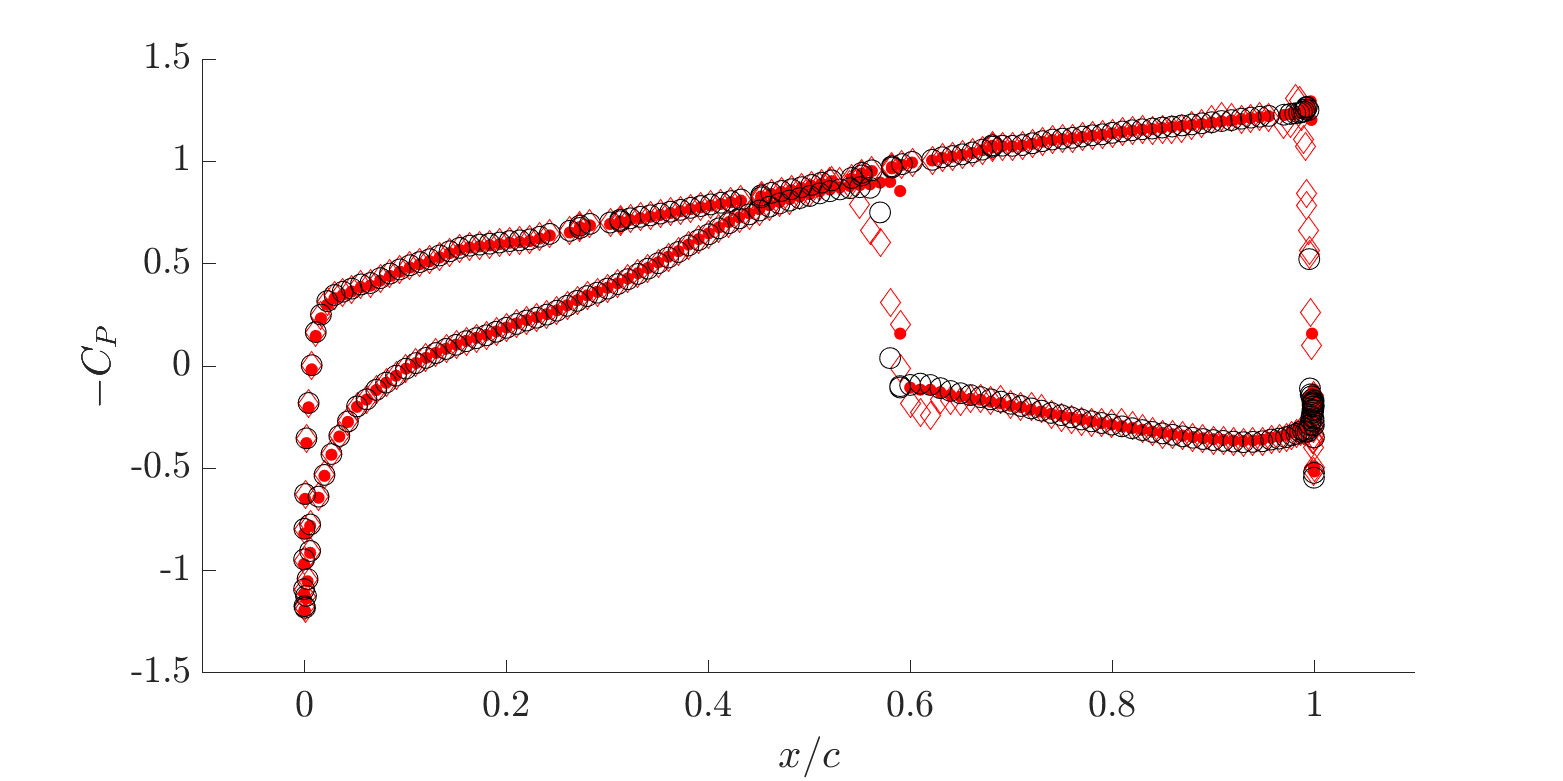}
  \caption{$Ma=0.853$, $\alpha=1.798$}
  \label{sf:DNN85}
  \end{subfigure}%
  \begin{subfigure}{.45\textwidth}
  \centering
  \includegraphics[width=1\linewidth]{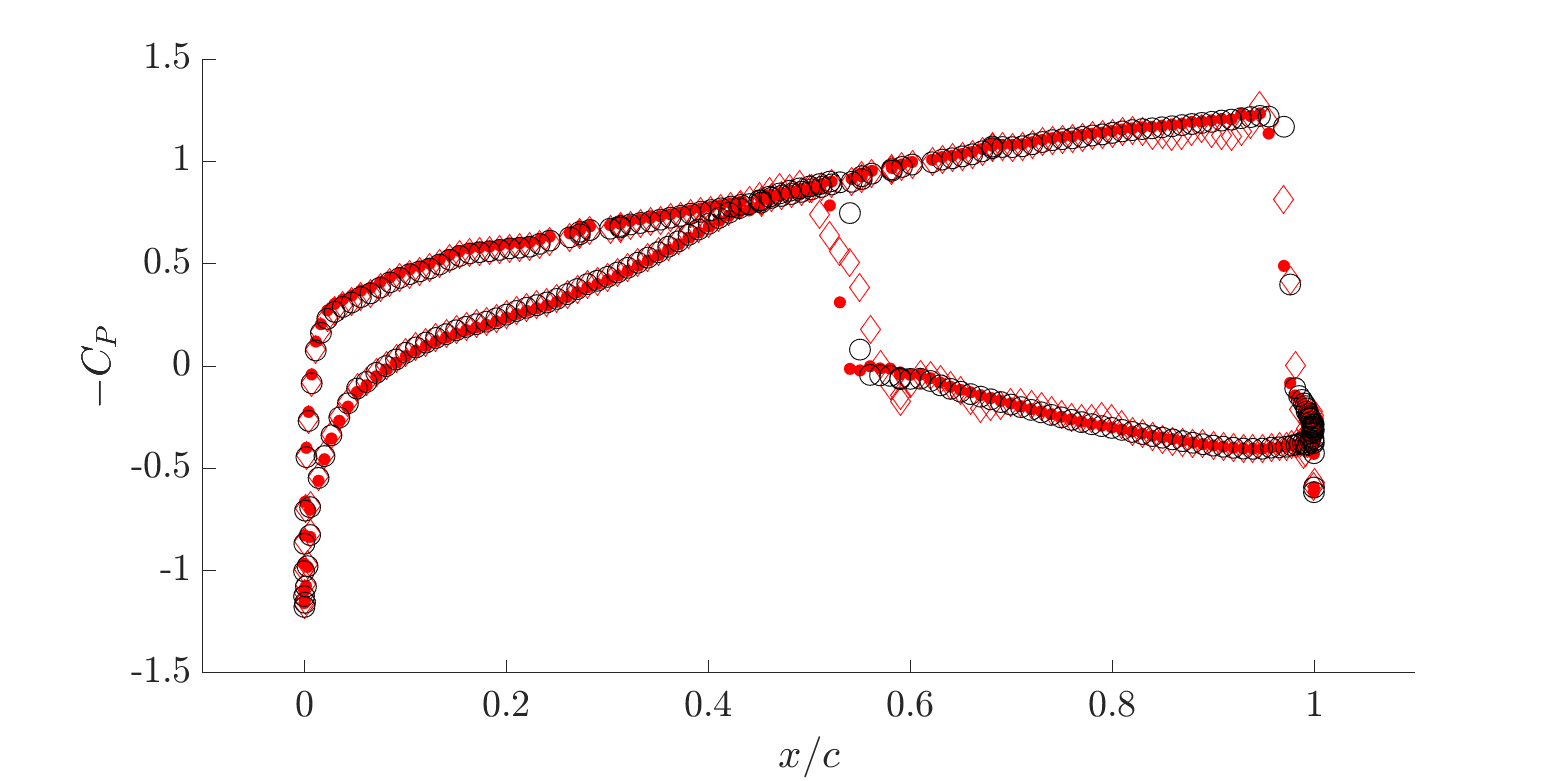}
  \caption{$Ma=0.839$, $\alpha=0.854$}
  \label{sf:DNN86}
  \end{subfigure}\\
  \begin{subfigure}{.45\textwidth}
  \centering
  \includegraphics[width=1\linewidth]{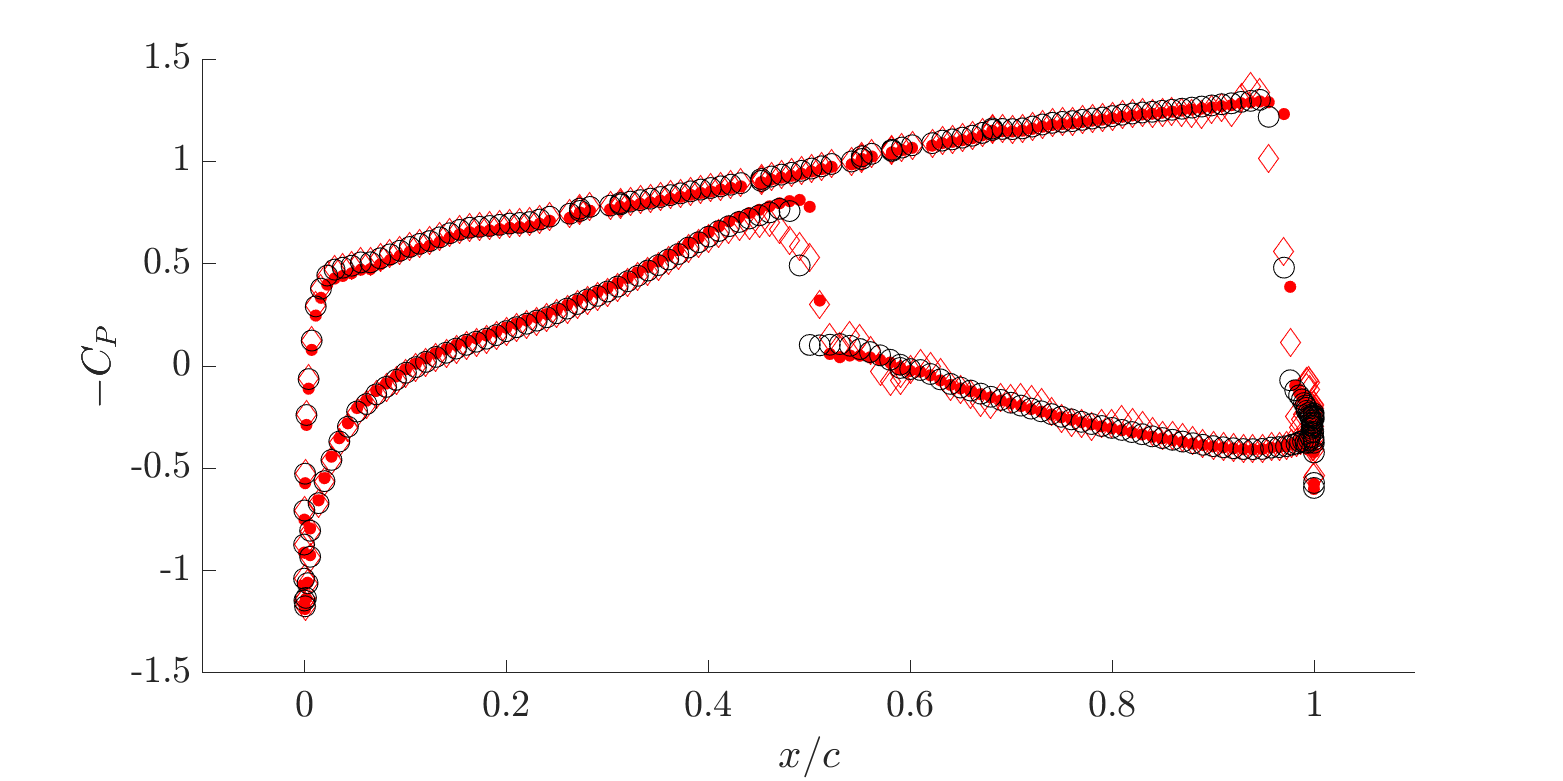}
  \caption{$Ma=0.829$, $\alpha=1.708$}
  \label{sf:DNN87}
  \end{subfigure}%
  \begin{subfigure}{.45\textwidth}
  \centering
  \includegraphics[width=1\linewidth]{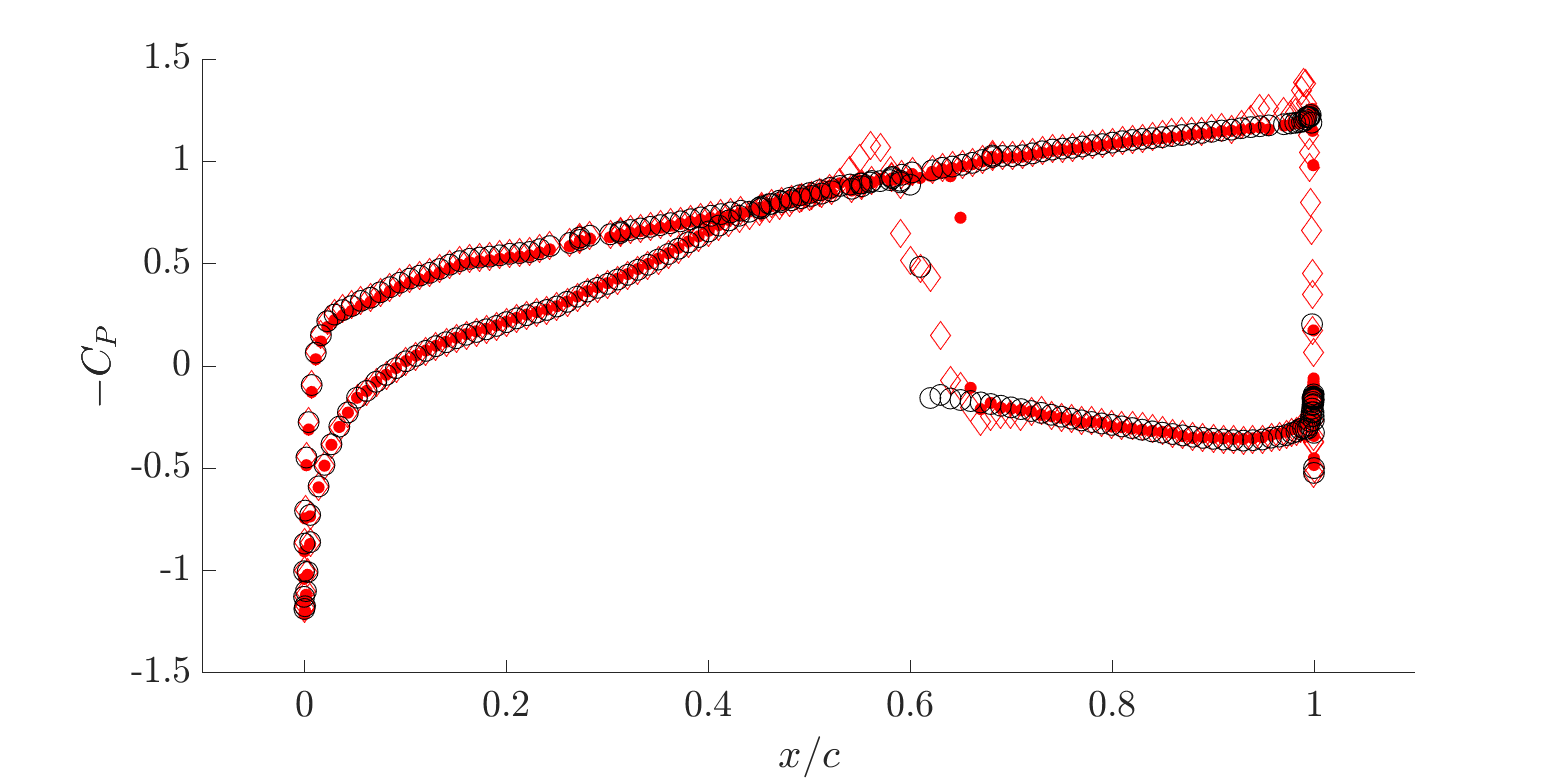}
  \caption{$Ma=0.862$, $\alpha=1.483$}
  \label{sf:DNN88}
  \end{subfigure}\\
  \begin{subfigure}{.45\textwidth}
  \centering
  \includegraphics[width=1\linewidth]{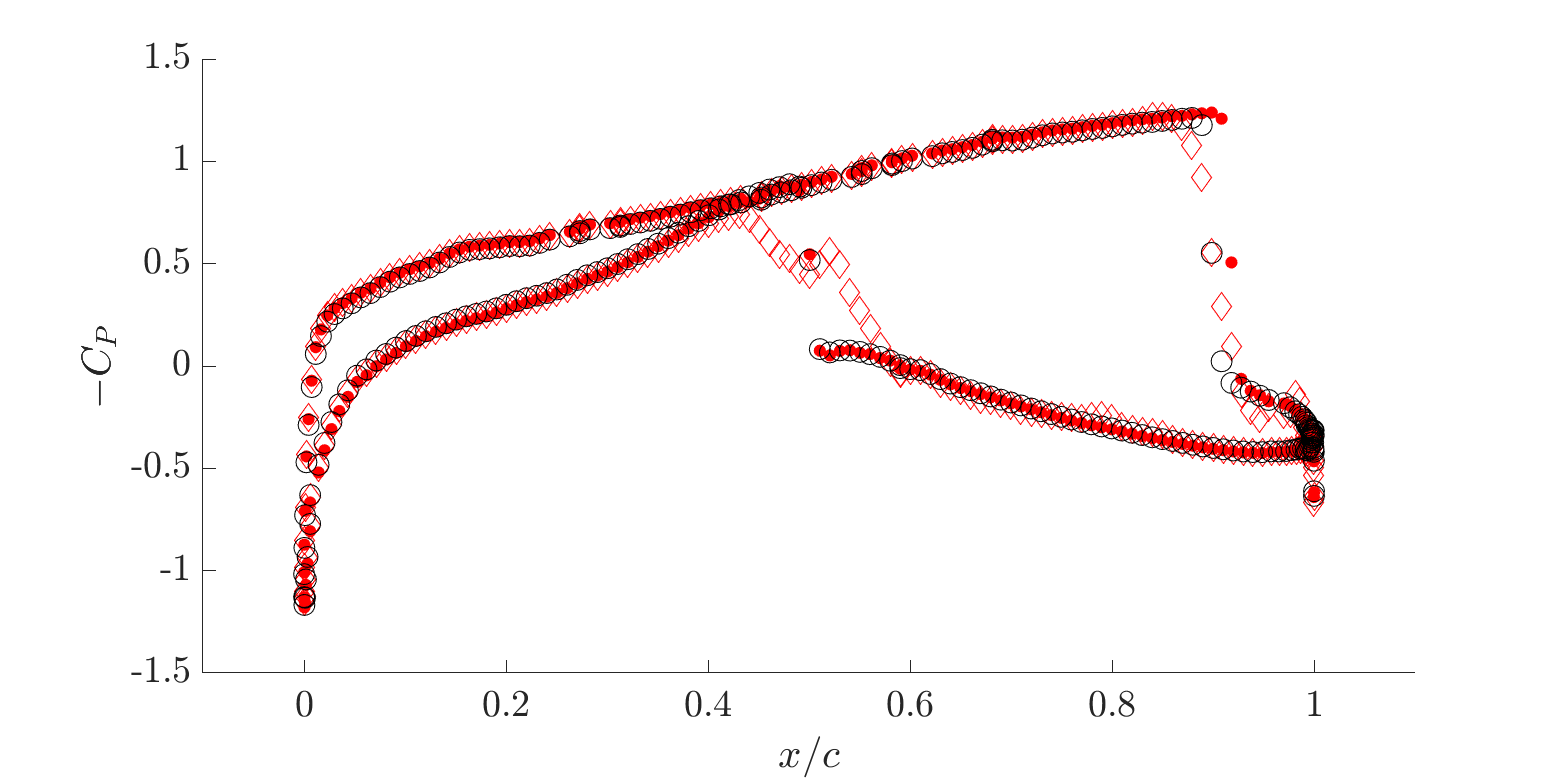}
  \caption{$Ma=0.817$, $\alpha=0.270$}
  \label{sf:DNN89}
  \end{subfigure}%
  \begin{subfigure}{.45\textwidth}
  \centering
  \includegraphics[width=1\linewidth]{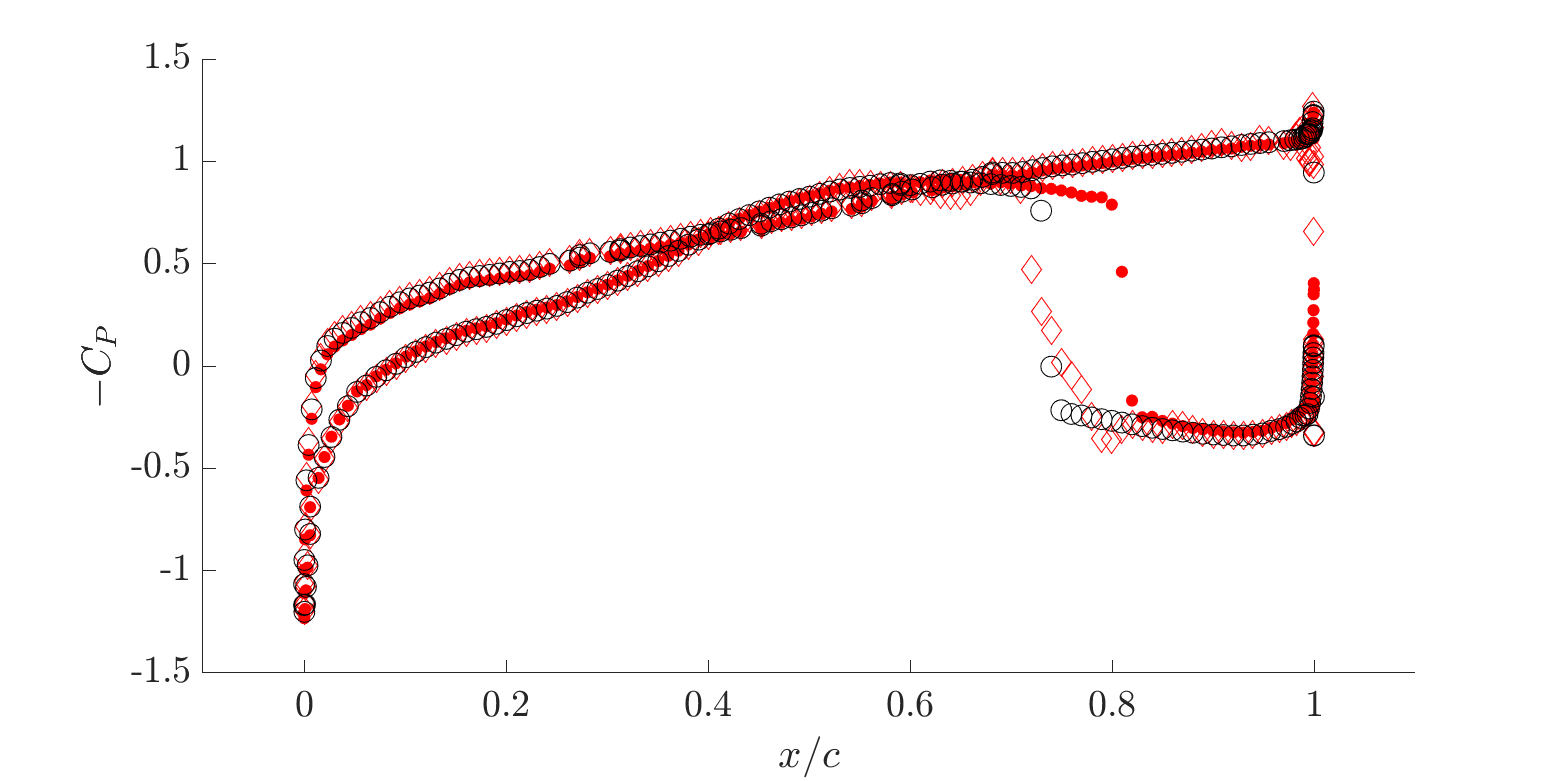}
  \caption{$Ma=0.884$, $\alpha=1.348$}
  \label{sf:DNN90}
  \end{subfigure}
\caption{Airfoil $C_P$ prediction of proposed method with free-stream boundary parameters}  
\label{f:dnn_vs_proj}
\end{figure}

The pressure contours are shown in Figure~\ref{f:contours_fs}, and the airfoil $C_p$ distributions are shown in Figure~\ref{f:dnn_vs_proj}. The plots also compare the predictions of the proposed DNN-based approach with that of \cite{renganathan2018koopman}. The proposed approach clearly adapts well to parametric variation, as demonstrated by how well it captures the location and strength of the shock. Particularly at the flight conditions in Figures~\ref{sf:DNN88} and \ref{sf:DNN90}, it marginally outperforms the projection-based approach. One can see from the contour plots that the DNN predictions are noisy compared with that of the projection-based approach. This performance is possibly due to the lack of sufficient training data; most DNN formulations are known to require a much greater number of training points than the 80 used in this work. Here, we are interested specifically in comparing the DNN-based ROM to the projection-based ROM with identical training data. Nevertheless, the DNN is demonstrably able to generalize the parameter dependence of the pressure distribution in a regime where the flow field shows strong bifurcations. The proposed approach incurs only a small fraction of the offline cost involved in the projection-based approach. This shows great promise with the scalability of the proposed approach when the inputs are high dimensional.  
\begin{figure}[htb!]
\centering
  \begin{subfigure}{.33\textwidth}
  \centering
  \includegraphics[width=1\linewidth]{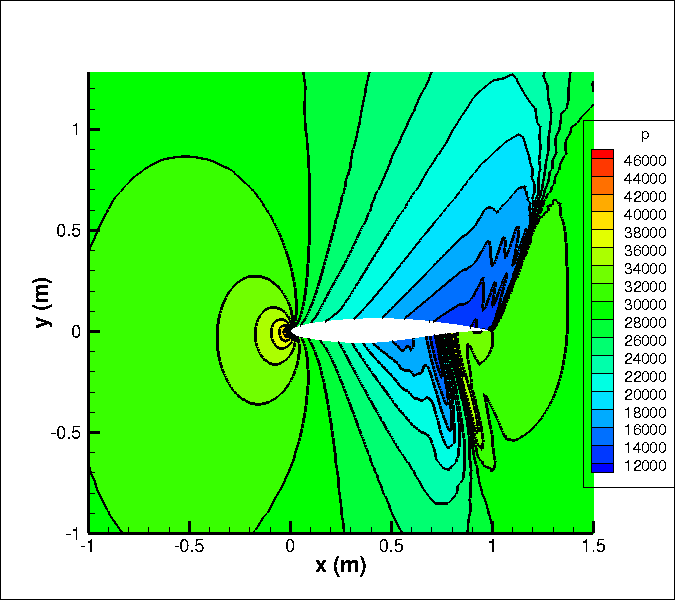}
\end{subfigure}
  \begin{subfigure}{.33\textwidth}
  \centering
  \includegraphics[width=1\linewidth]{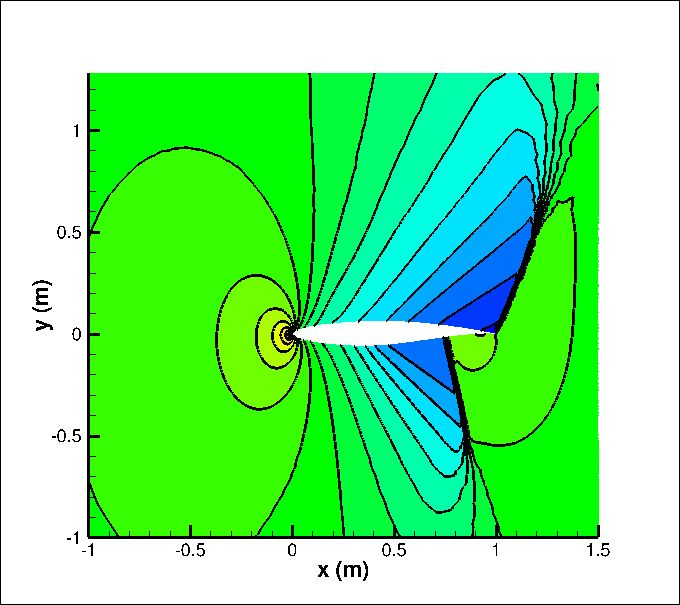}
  \end{subfigure}    \begin{subfigure}{.33\textwidth}
  \centering
  \includegraphics[width=1\linewidth]{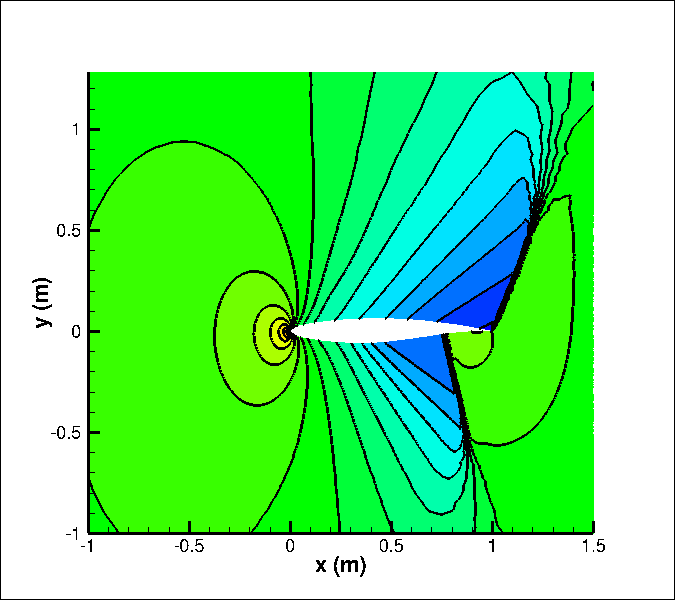}
  \end{subfigure}\\
  \begin{subfigure}{.33\textwidth}
  \centering
  \includegraphics[width=1\linewidth]{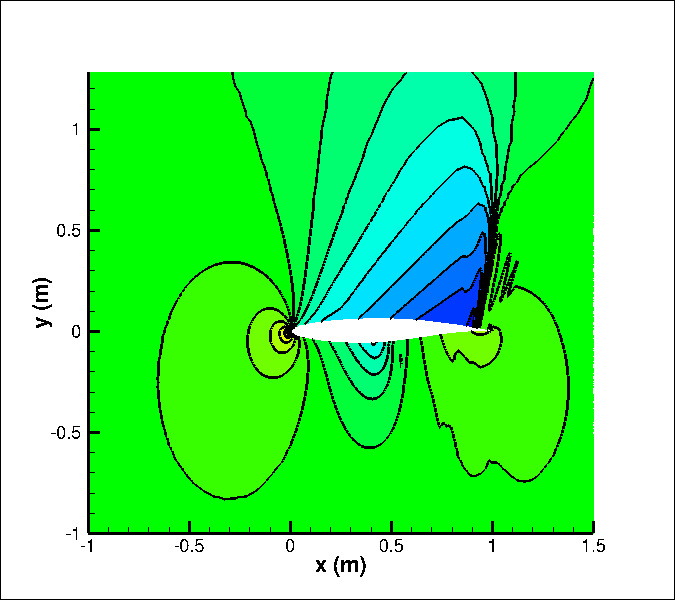}
\end{subfigure}
  \begin{subfigure}{.33\textwidth}
  \centering
  \includegraphics[width=1\linewidth]{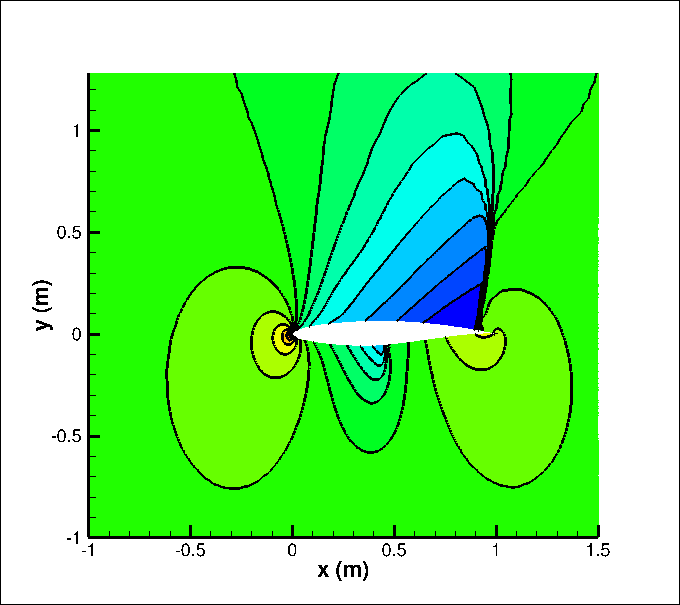}
  \end{subfigure}    \begin{subfigure}{.33\textwidth}
  \centering
  \includegraphics[width=1\linewidth]{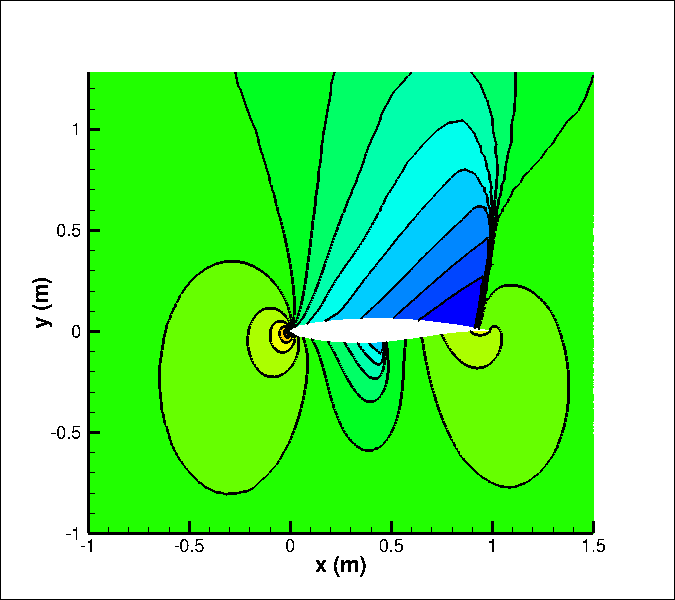}
  \end{subfigure}\\
\end{figure}

\begin{figure}[htb!]
\ContinuedFloat
\centering
    \begin{subfigure}{.33\textwidth}
  \centering
  \includegraphics[width=1\linewidth]{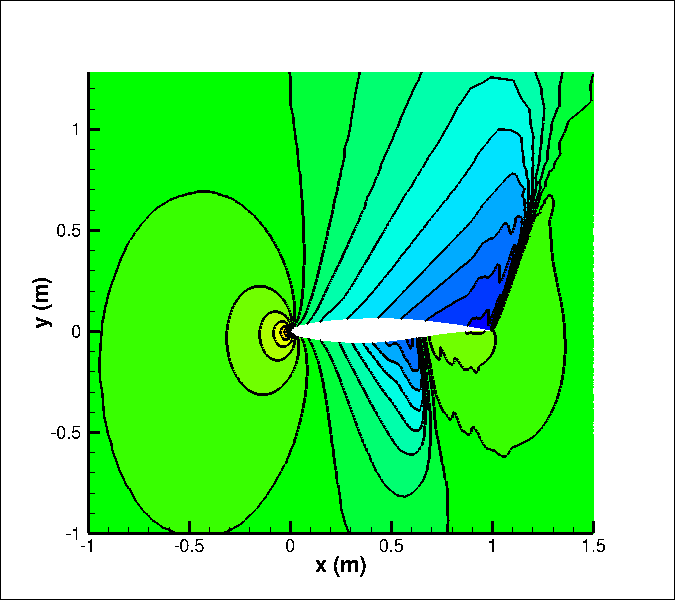}
\end{subfigure}
  \begin{subfigure}{.33\textwidth}
  \centering
  \includegraphics[width=1\linewidth]{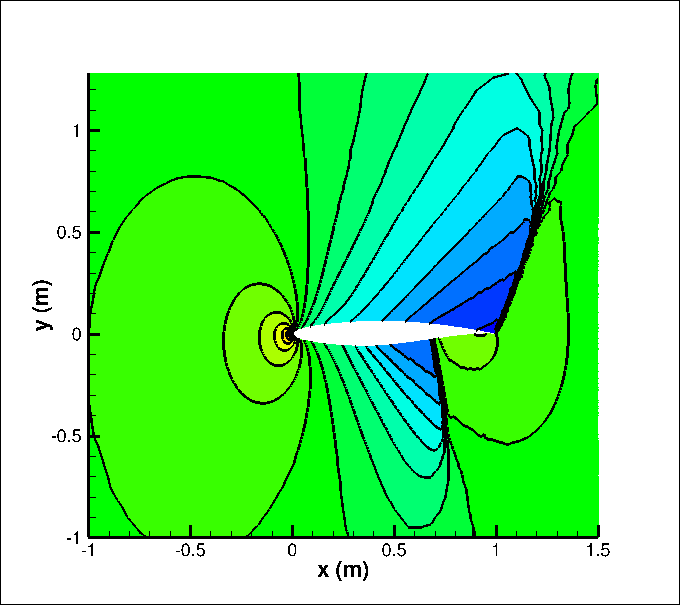}
  \end{subfigure}    \begin{subfigure}{.33\textwidth}
  \centering
  \includegraphics[width=1\linewidth]{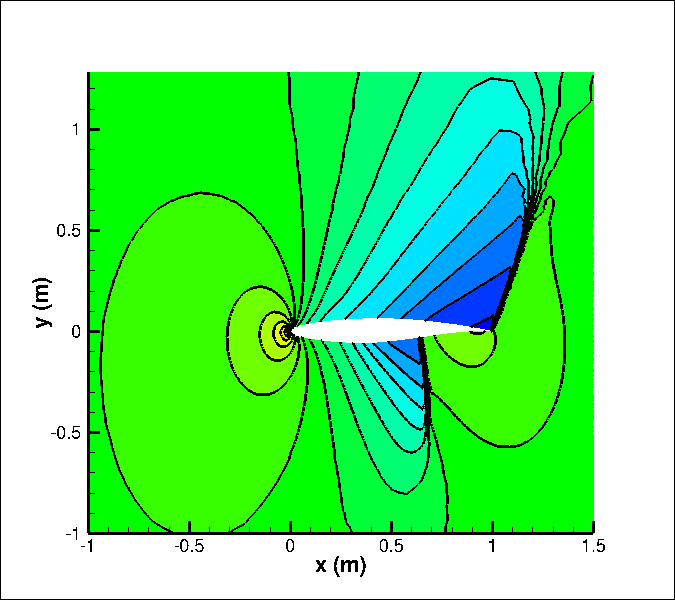}
  \end{subfigure}\\
    \begin{subfigure}{.33\textwidth}
  \centering
  \includegraphics[width=1\linewidth]{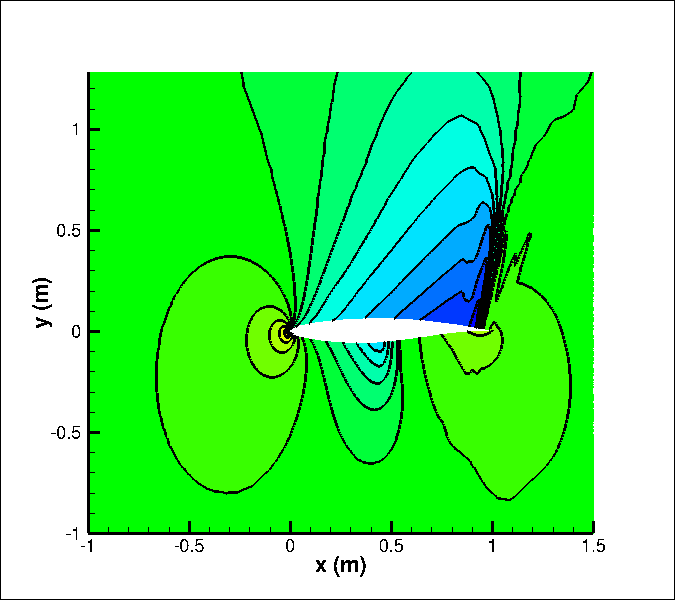}
\end{subfigure}
  \begin{subfigure}{.33\textwidth}
  \centering
  \includegraphics[width=1\linewidth]{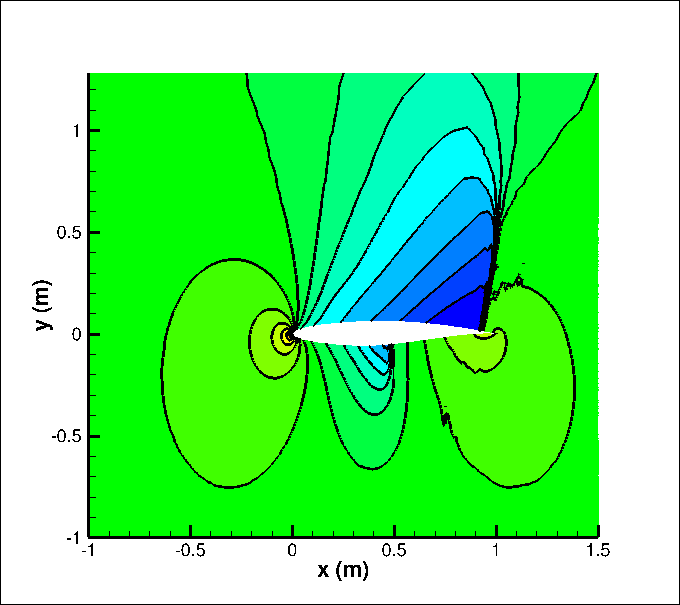}
  \end{subfigure}    \begin{subfigure}{.33\textwidth}
  \centering
  \includegraphics[width=1\linewidth]{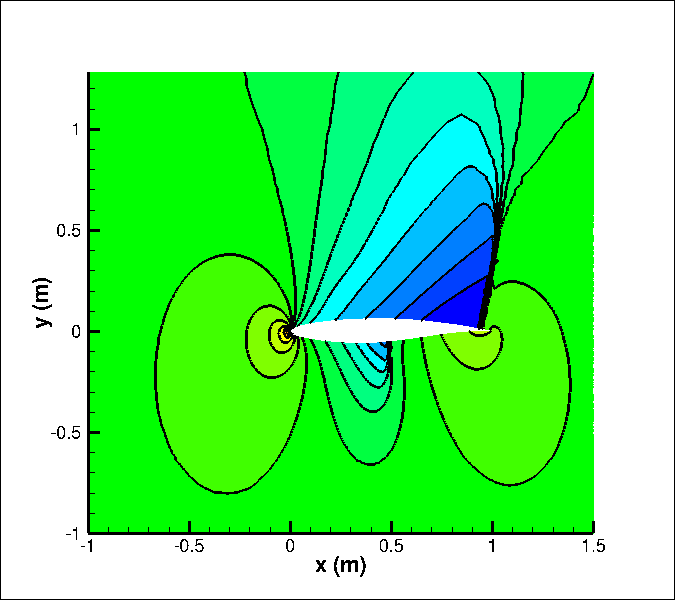}
  \end{subfigure}\\
    \begin{subfigure}{.33\textwidth}
  \centering
  \includegraphics[width=1\linewidth]{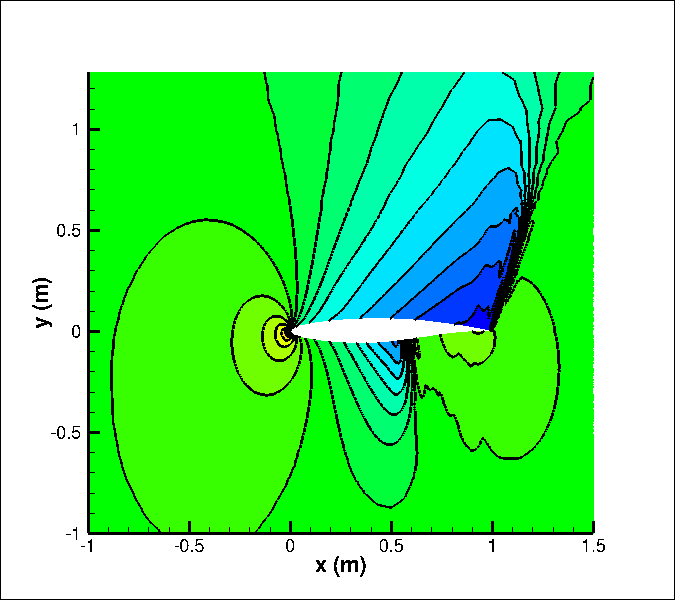}
\end{subfigure}
  \begin{subfigure}{.33\textwidth}
  \centering
  \includegraphics[width=1\linewidth]{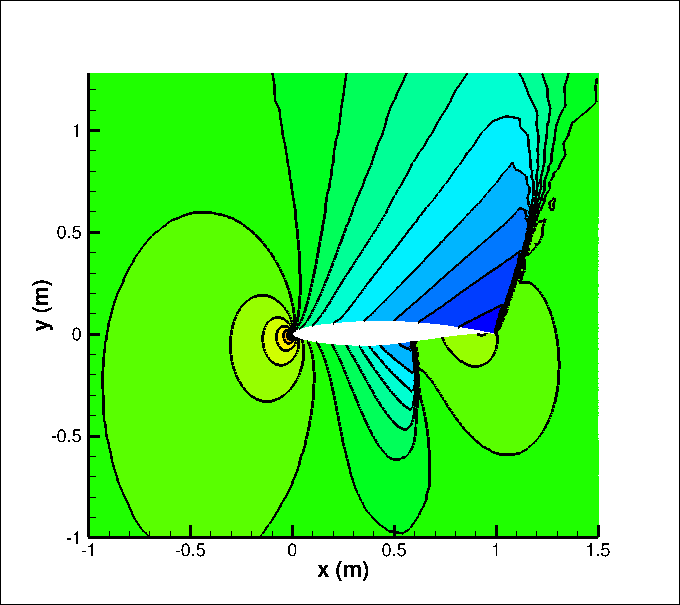}
  \end{subfigure}    \begin{subfigure}{.33\textwidth}
  \centering
  \includegraphics[width=1\linewidth]{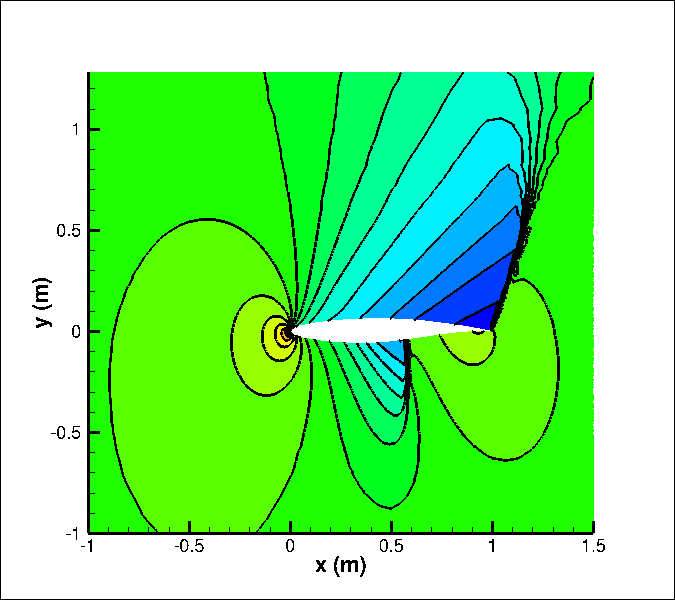}
  \end{subfigure}\\
  \begin{subfigure}{.33\textwidth}
  \centering
  \includegraphics[width=1\linewidth]{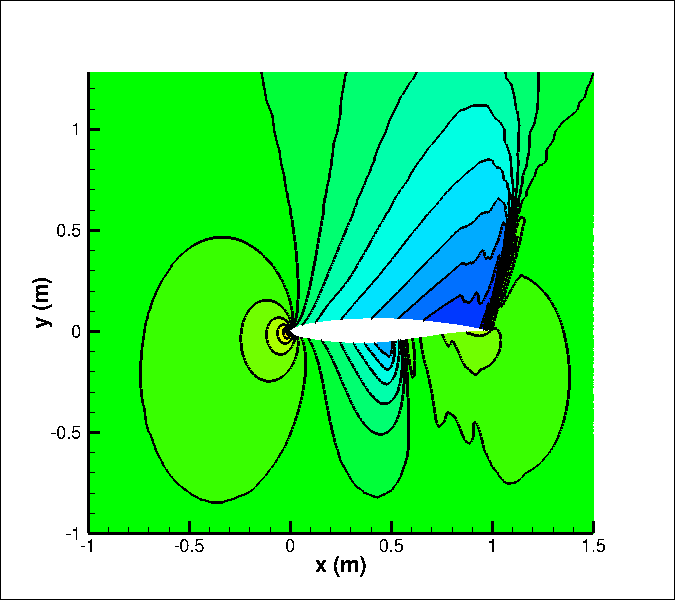}
\end{subfigure}
  \begin{subfigure}{.33\textwidth}
  \centering
  \includegraphics[width=1\linewidth]{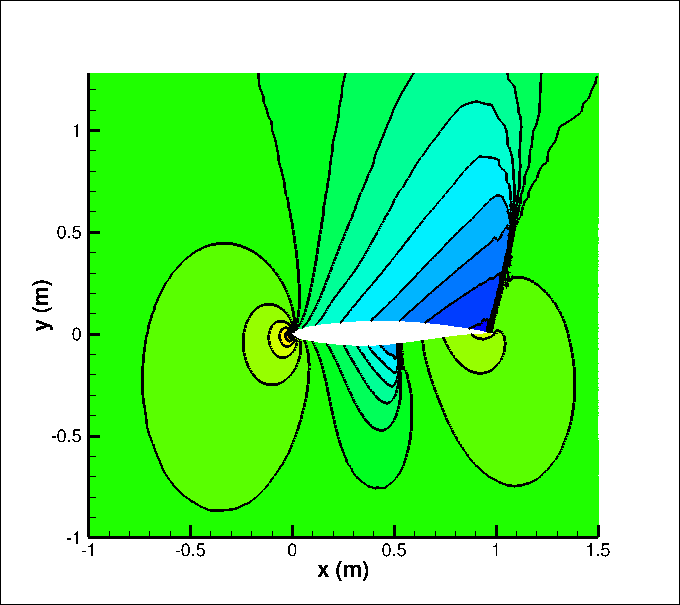}
  \end{subfigure}    \begin{subfigure}{.33\textwidth}
  \centering
  \includegraphics[width=1\linewidth]{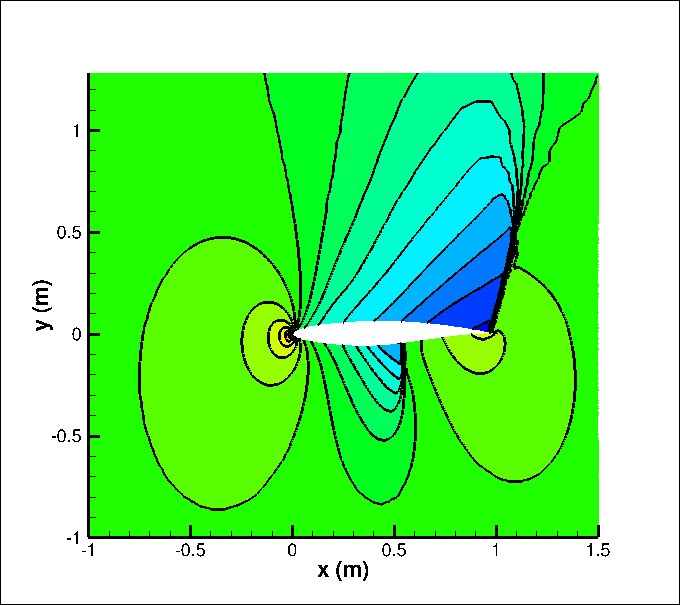}
  \end{subfigure}\\
\end{figure}

\begin{figure}
\ContinuedFloat
\centering
  \begin{subfigure}{.33\textwidth}
  \centering
  \includegraphics[width=1\linewidth]{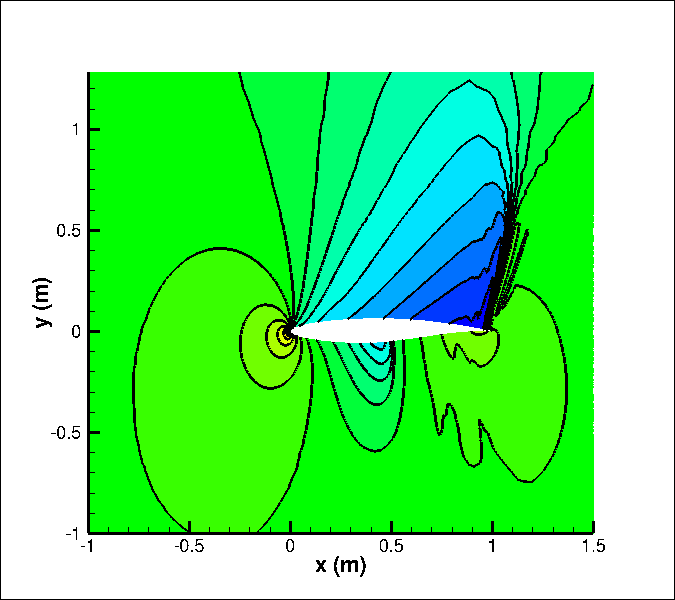}
\end{subfigure}
  \begin{subfigure}{.33\textwidth}
  \centering
  \includegraphics[width=1\linewidth]{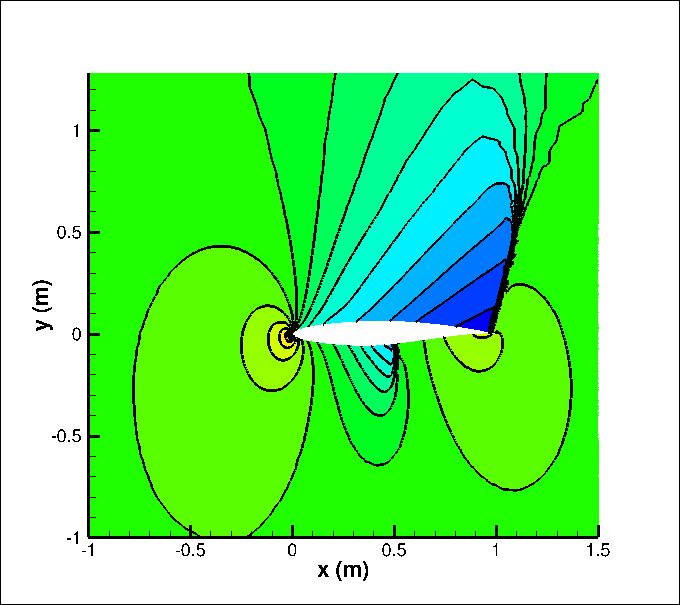}
  \end{subfigure}    \begin{subfigure}{.33\textwidth}
  \centering
  \includegraphics[width=1\linewidth]{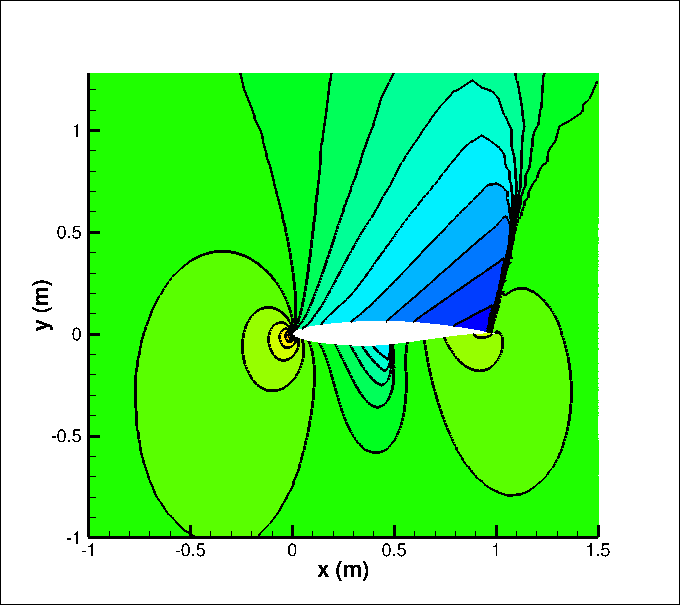}
  \end{subfigure}\\
    \begin{subfigure}{.33\textwidth}
  \centering
  \includegraphics[width=1\linewidth]{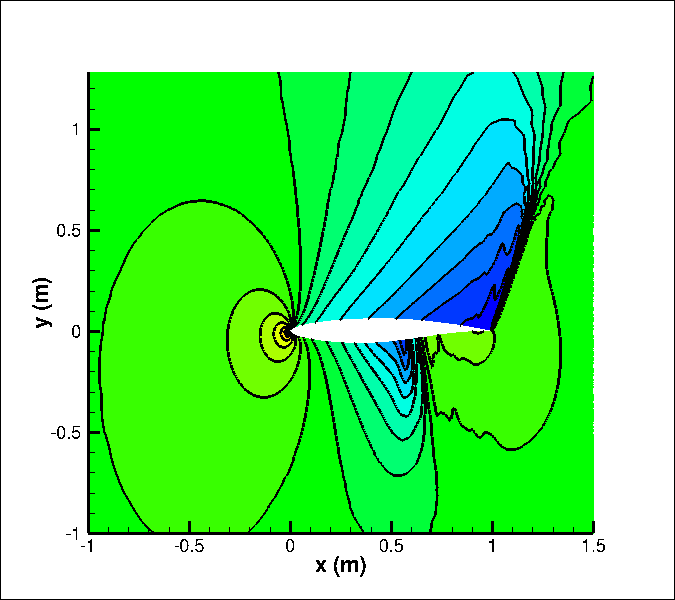}
\end{subfigure}
  \begin{subfigure}{.33\textwidth}
  \centering
  \includegraphics[width=1\linewidth]{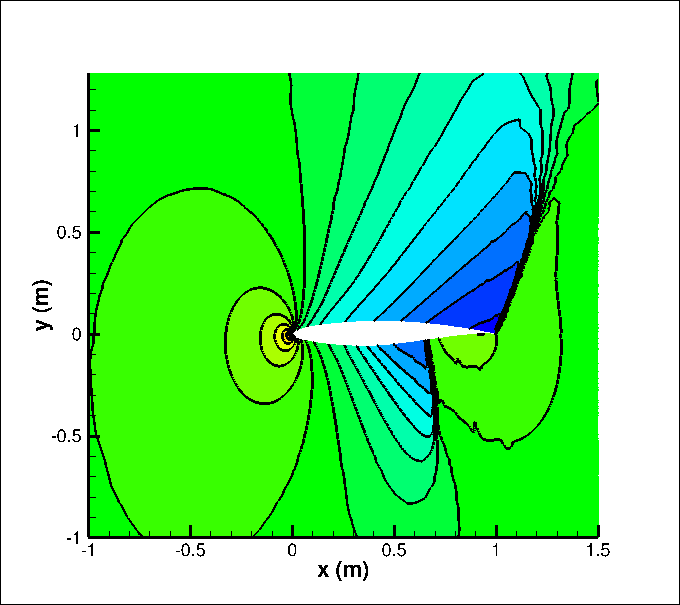}
  \end{subfigure}    \begin{subfigure}{.33\textwidth}
  \centering
  \includegraphics[width=1\linewidth]{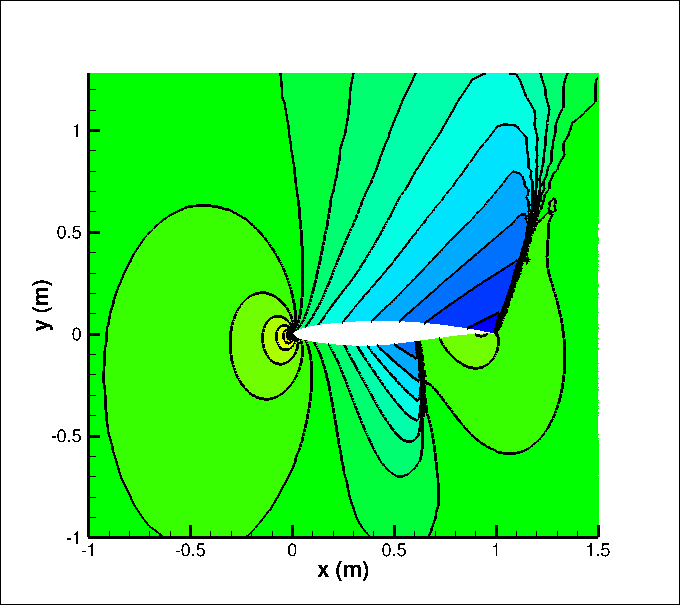}
  \end{subfigure}\\
    \begin{subfigure}{.33\textwidth}
  \centering
  \includegraphics[width=1\linewidth]{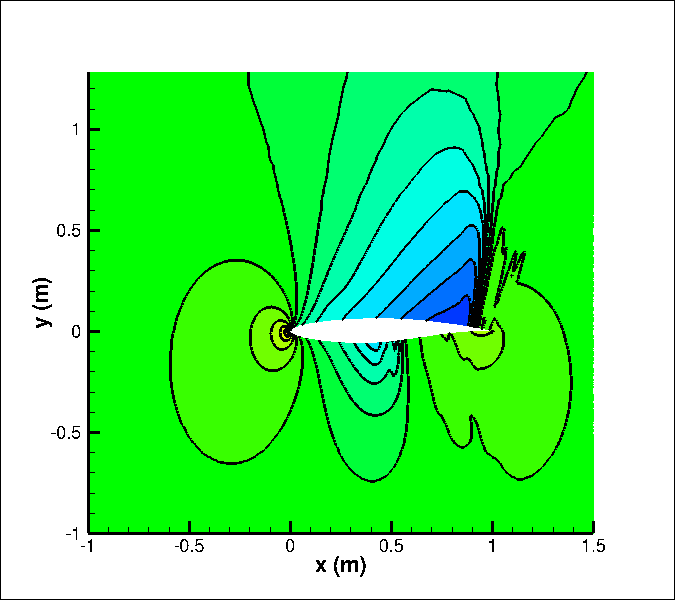}
\end{subfigure}
  \begin{subfigure}{.33\textwidth}
  \centering
  \includegraphics[width=1\linewidth]{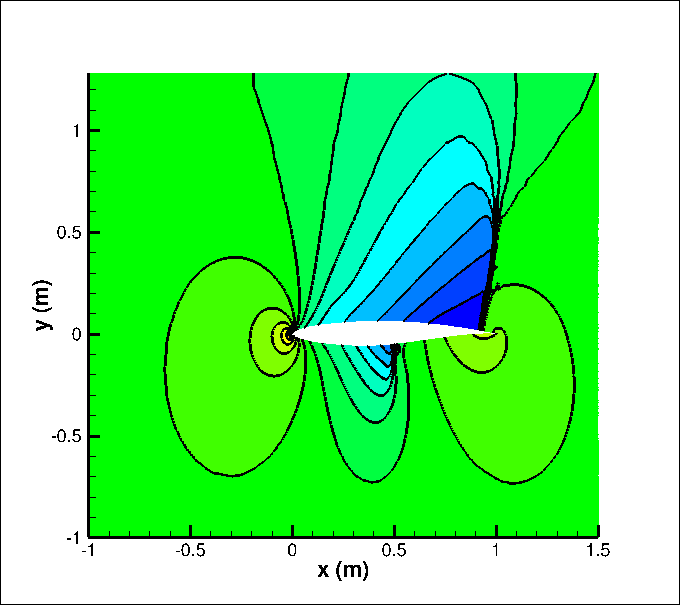}
  \end{subfigure}    \begin{subfigure}{.33\textwidth}
  \centering
  \includegraphics[width=1\linewidth]{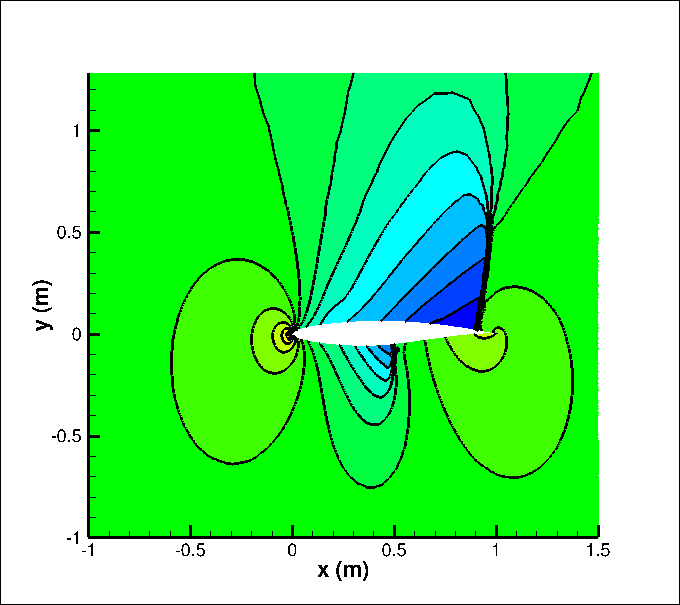}
  \end{subfigure}\\
    \begin{subfigure}{.33\textwidth}
  \centering
  \includegraphics[width=1\linewidth]{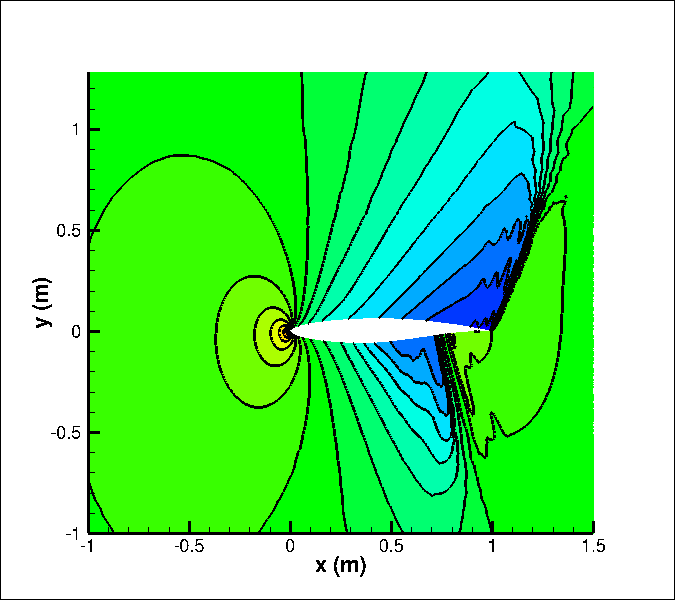}
  \caption*{ROM (POD+DNN)}
\end{subfigure}
  \begin{subfigure}{.33\textwidth}
  \centering
  \includegraphics[width=1\linewidth]{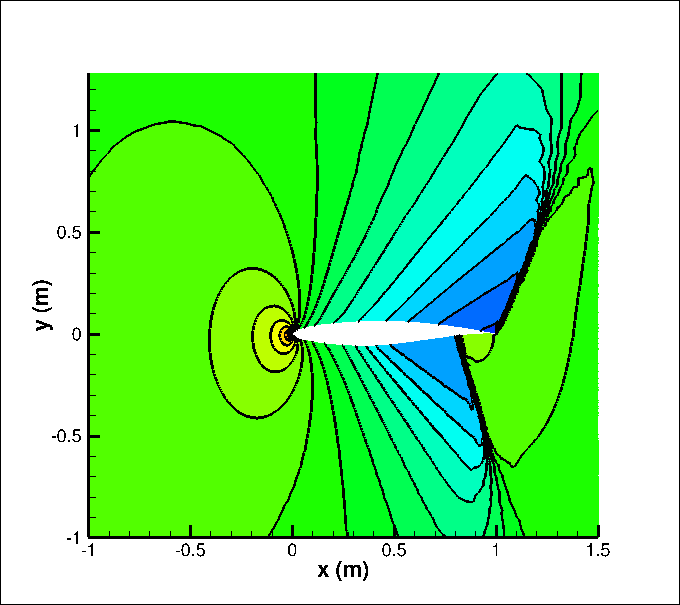}
  \caption*{ROM (POD+Projection)}
  \end{subfigure}    \begin{subfigure}{.33\textwidth}
  \centering
  \includegraphics[width=1\linewidth]{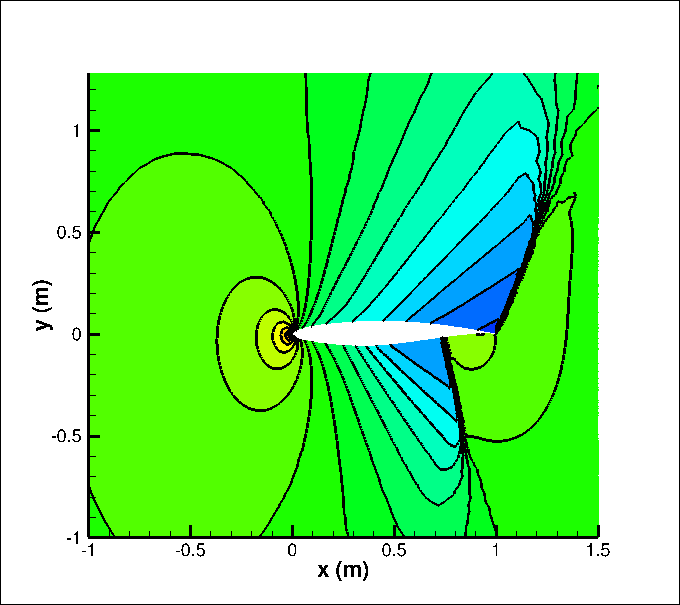}
  \caption*{FOM}
  \end{subfigure}\\
  \caption{Comparison of the proposed approach with free-stream boundary ($\bs{\theta} \in \R^2$) parameters. Contours represent absolute pressure (in pascals) distribution.}
  \label{f:contours_fs}
\end{figure}

\clearpage
\subsubsection{Geometry Boundary Parameters}
\begin{table}
\caption{Ambient flight conditions}
\centering
\begin{tabular}{cc}
\hline
\hline
$P_\infty$    & 28,745 $Pa$\\
$\rho_\infty$ & 0.44 $kg/m^3$\\
$a_\infty$    & 301.86 $m/s$\\
$T_\infty$    & 233.15 $K$ \\
\hline
\end{tabular}
\label{t:RAE_freestream}
\end{table}
We now outline results from the extension of the proposed framework to a higher-dimensional input space that includes parameters that affect the shape of the airfoil, namely, the CST coefficients. These are shown in Figures \ref{f:contours_shape} and \ref{8D_2}. The accurate predictions of the shock location as well as its strength clearly show that the proposed DNN-based approach generalizes the parameter-state relationship in a high-dimensional setting very well. We note that the performance of the proposed approach is improved compared with the freestream boundary parameter case in the preceding section. One of the main reasons  is that for the results in the preceding section the pressure distributions showed strong variations for the chosen parameter range compared with the ranges used for shape parameters. However, the same number of snapshots ($M=80$) is used in both cases. This shows the ability of the proposed approach to learn the parameter-state relationship in a high-dimensional setting that finds a number of applications in aerodynamic design. 


\begin{figure}[htb!]
\centering
  \begin{subfigure}{.33\textwidth}
  \centering
  \includegraphics[width=1\linewidth]{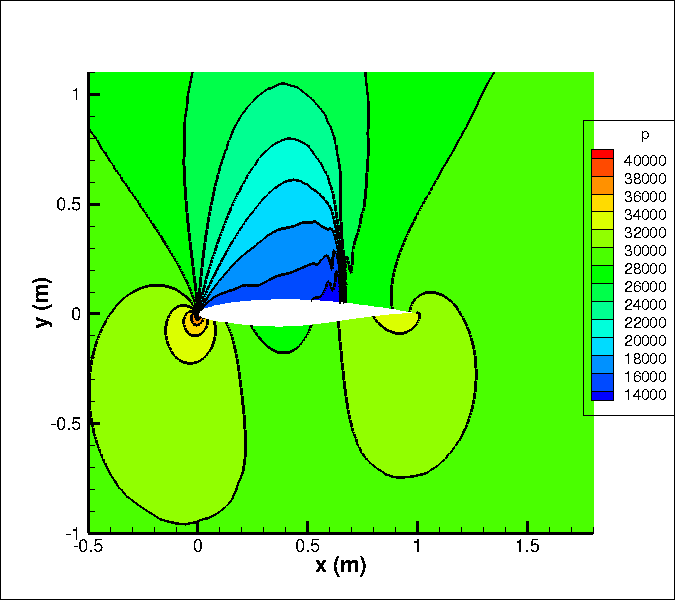}
\end{subfigure}
  \begin{subfigure}{.33\textwidth}
  \centering
  \includegraphics[width=1\linewidth]{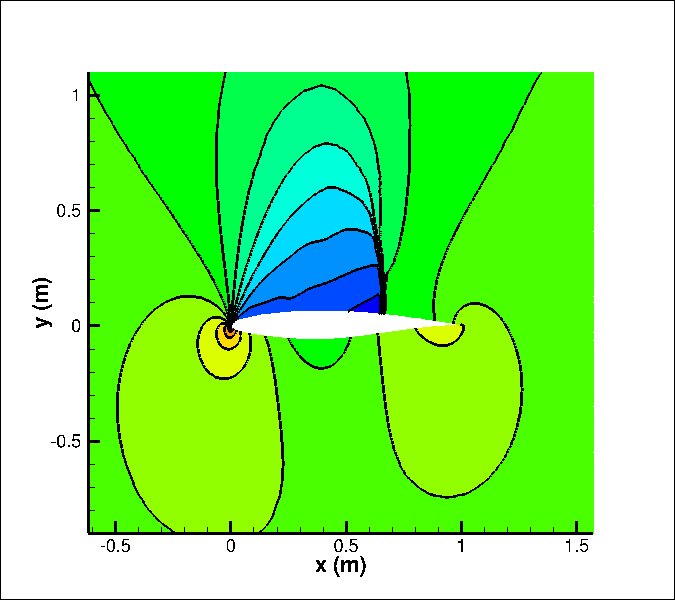}
  \end{subfigure}    \begin{subfigure}{.33\textwidth}
  \centering
  \includegraphics[width=1\linewidth]{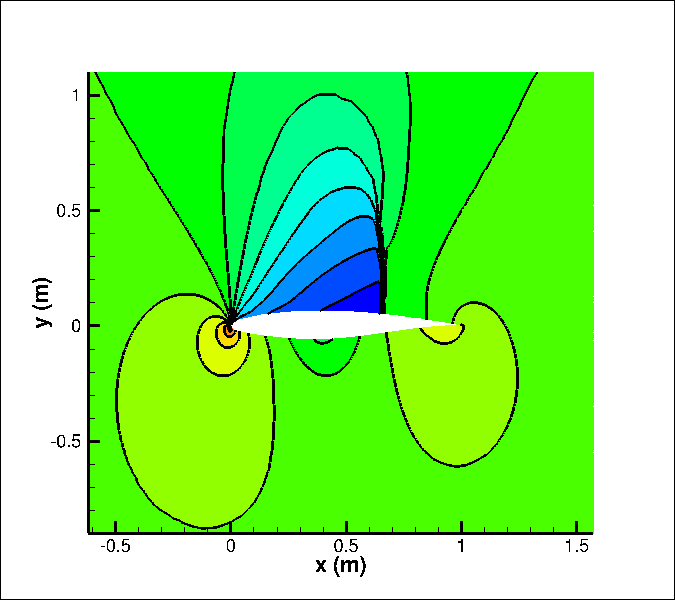}
  \end{subfigure}\\
  \begin{subfigure}{.33\textwidth}
  \centering
  \includegraphics[width=1\linewidth]{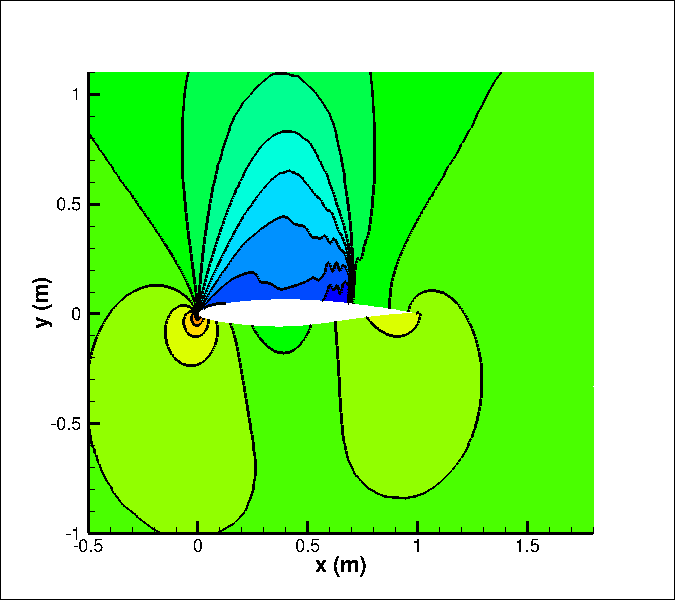}
\end{subfigure}
  \begin{subfigure}{.33\textwidth}
  \centering
  \includegraphics[width=1\linewidth]{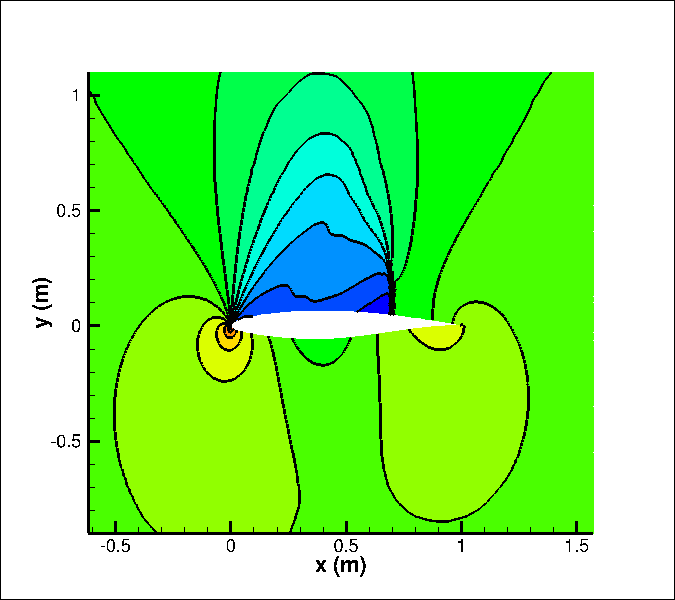}
  \end{subfigure}    \begin{subfigure}{.33\textwidth}
  \centering
  \includegraphics[width=1\linewidth]{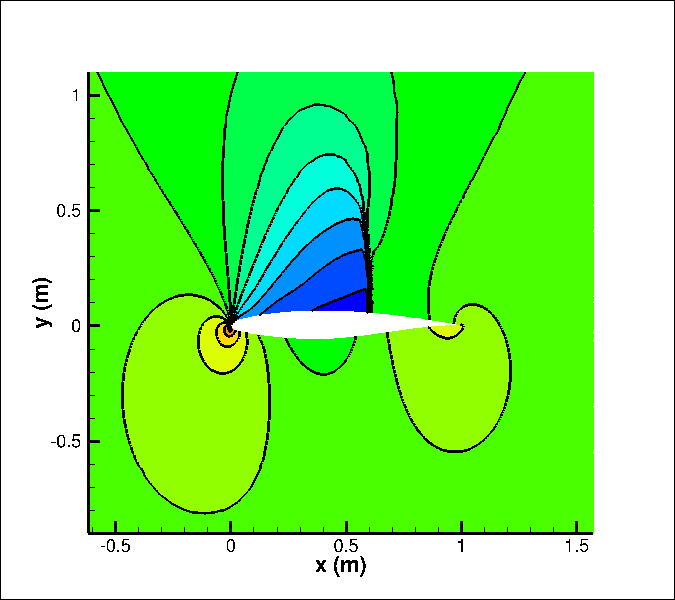}
  \end{subfigure}
  \begin{subfigure}{.33\textwidth}
  \centering
  \includegraphics[width=1\linewidth]{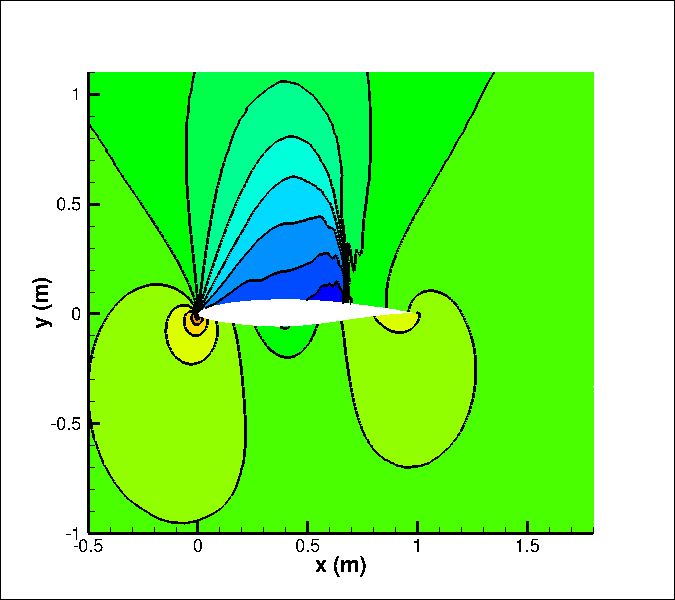}
\end{subfigure}
  \begin{subfigure}{.33\textwidth}
  \centering
  \includegraphics[width=1\linewidth]{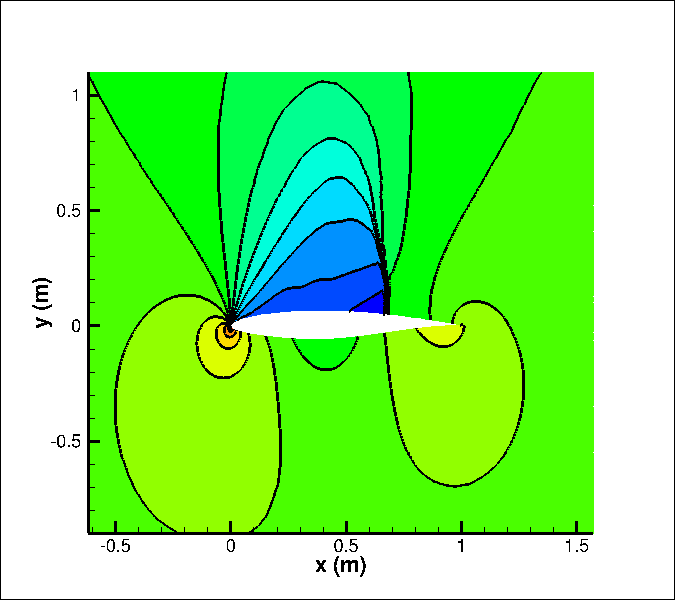}
  \end{subfigure}    \begin{subfigure}{.33\textwidth}
  \centering
  \includegraphics[width=1\linewidth]{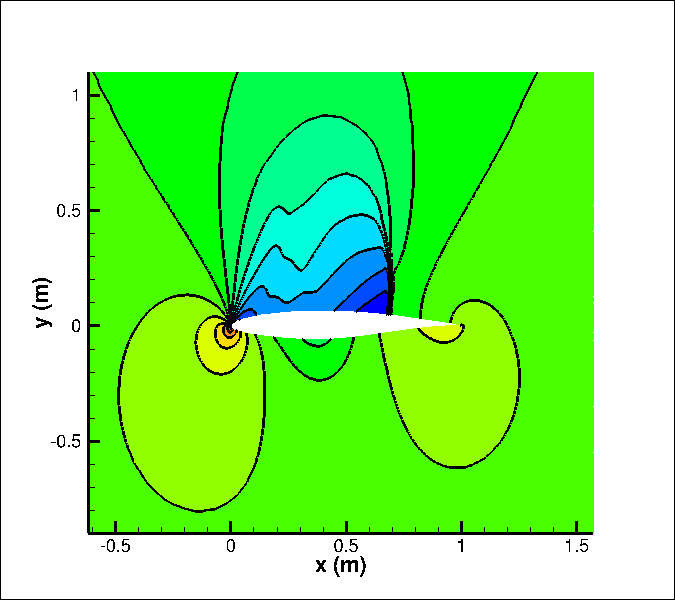}
  \end{subfigure}  
 \end{figure}
 
\begin{figure}[htb!]
\ContinuedFloat
\centering
  \begin{subfigure}{.33\textwidth}
  \centering
  \includegraphics[width=1\linewidth]{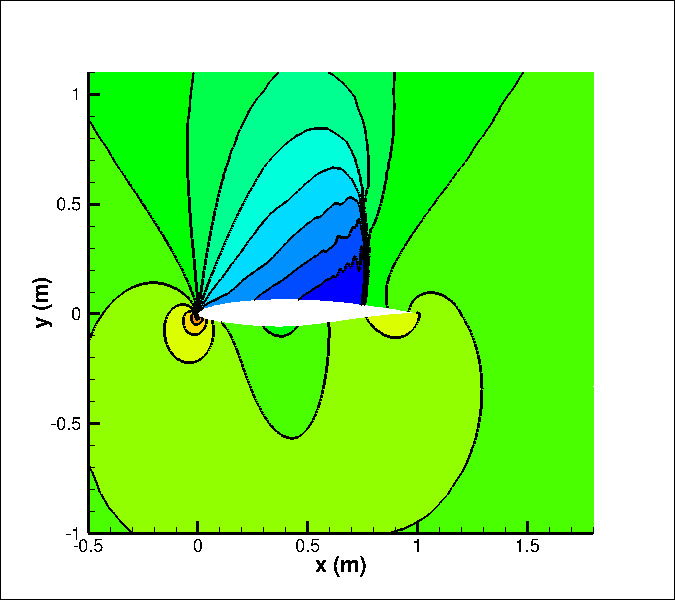}
\end{subfigure}
  \begin{subfigure}{.33\textwidth}
  \centering
  \includegraphics[width=1\linewidth]{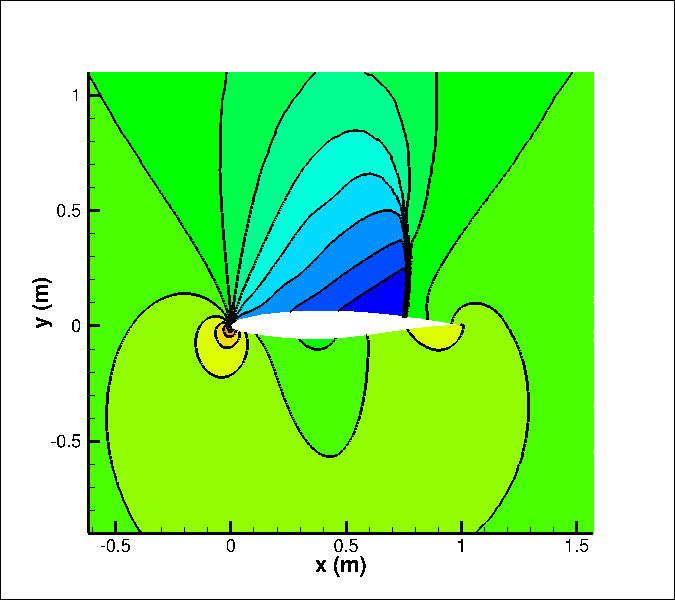}
  \end{subfigure}    \begin{subfigure}{.33\textwidth}
  \centering
  \includegraphics[width=1\linewidth]{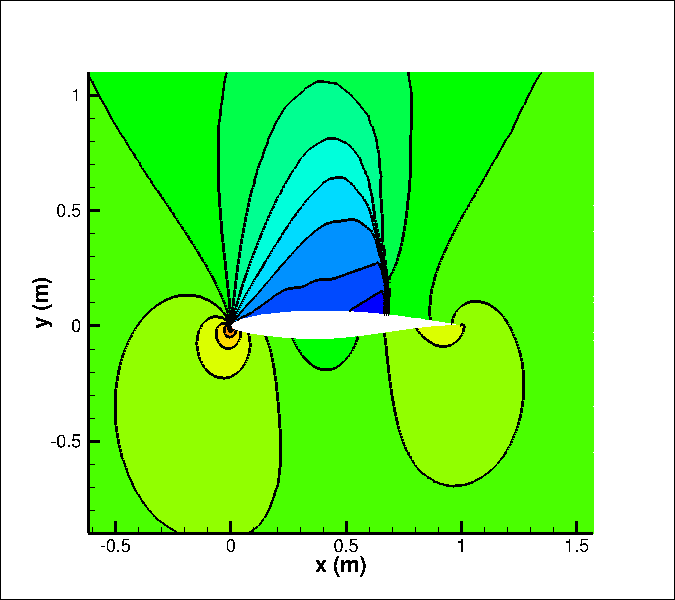}
  \end{subfigure}\\
  \begin{subfigure}{.33\textwidth}
  \centering
  \includegraphics[width=1\linewidth]{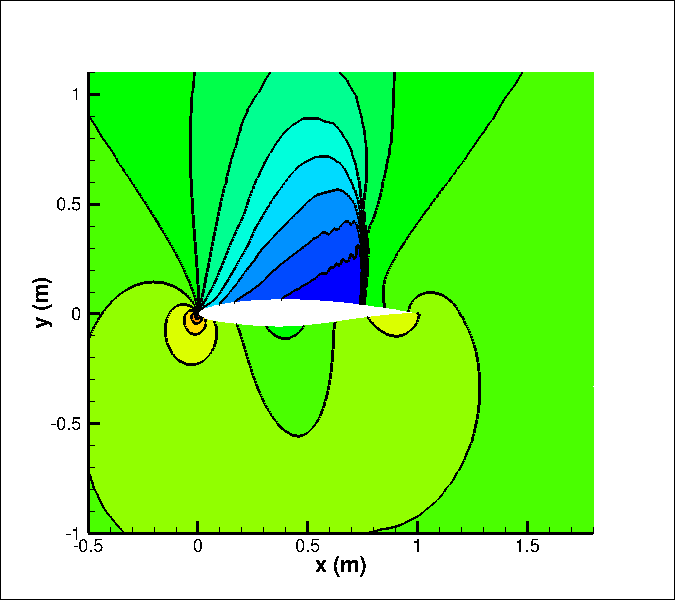}
\end{subfigure}
  \begin{subfigure}{.33\textwidth}
  \centering
  \includegraphics[width=1\linewidth]{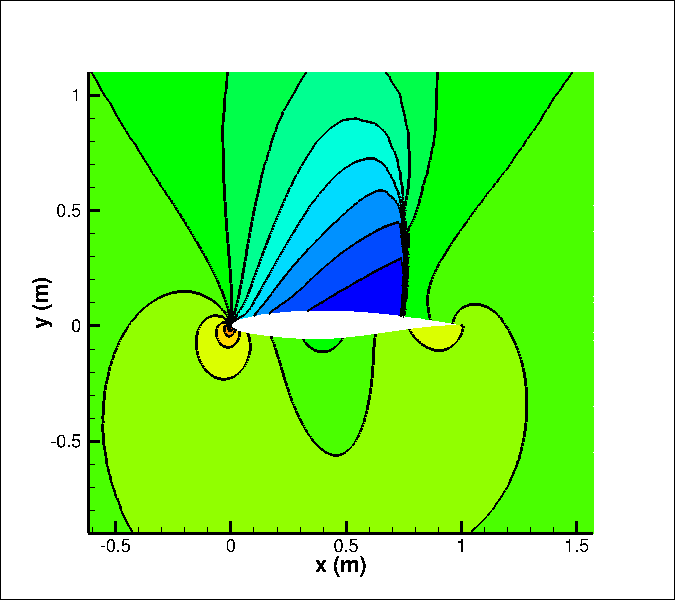}
  \end{subfigure}    \begin{subfigure}{.33\textwidth}
  \centering
  \includegraphics[width=1\linewidth]{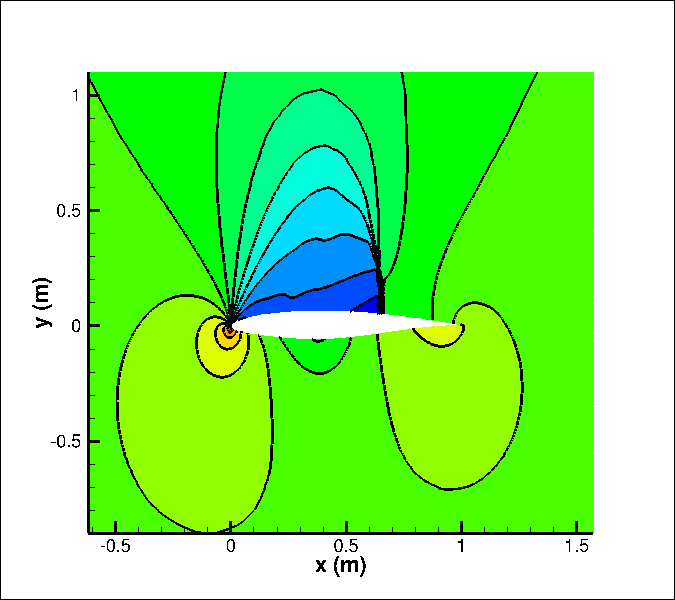}
  \end{subfigure}\\
  \begin{subfigure}{.33\textwidth}
  \centering
  \includegraphics[width=1\linewidth]{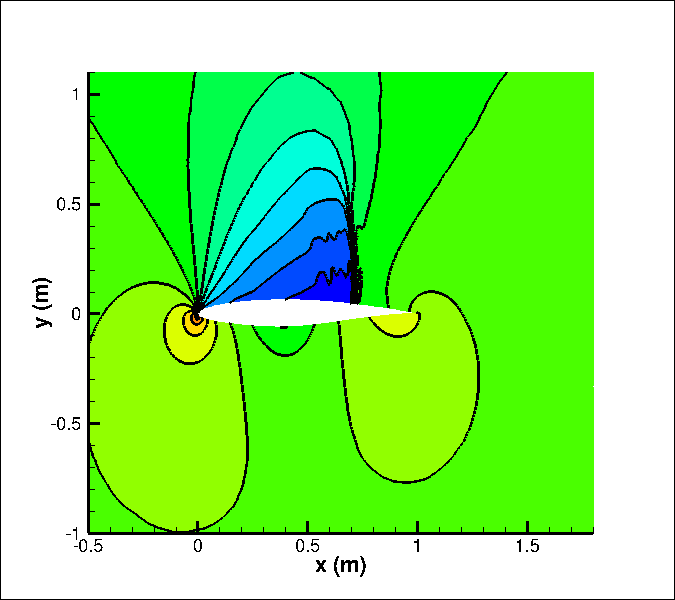}
\end{subfigure}
  \begin{subfigure}{.33\textwidth}
  \centering
  \includegraphics[width=1\linewidth]{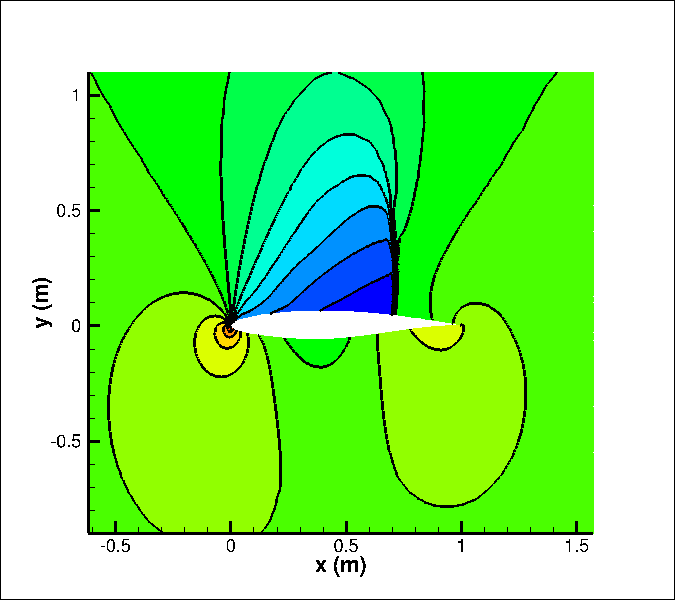}
  \end{subfigure}    \begin{subfigure}{.33\textwidth}
  \centering
  \includegraphics[width=1\linewidth]{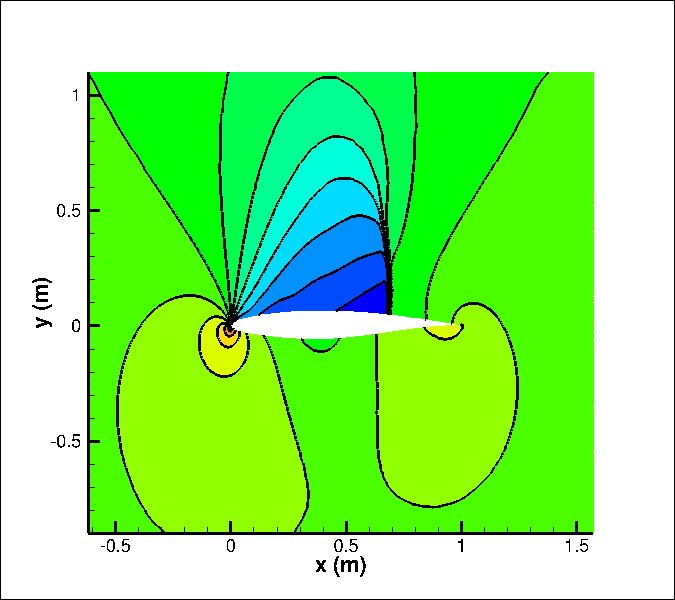}
  \end{subfigure}\\  
   \begin{subfigure}{.33\textwidth}
  \centering
  \includegraphics[width=1\linewidth]{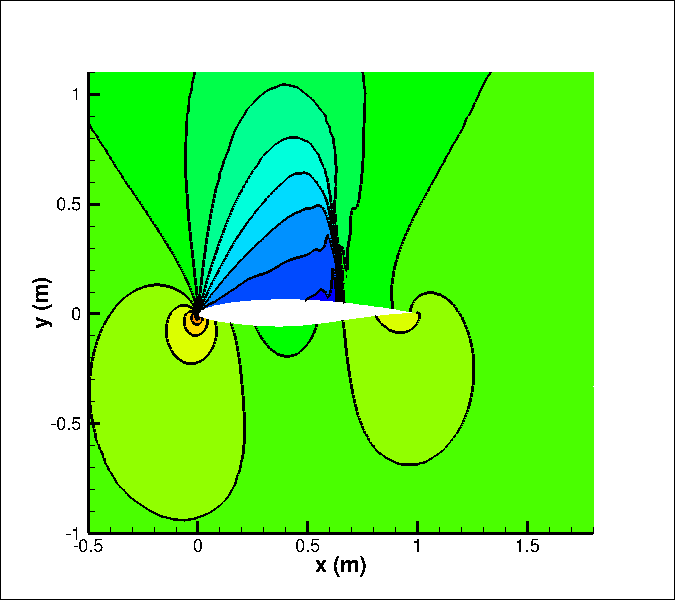}
  \caption*{ROM (POD+DNN)}
\end{subfigure}
  \begin{subfigure}{.33\textwidth}
  \centering
  \includegraphics[width=1\linewidth]{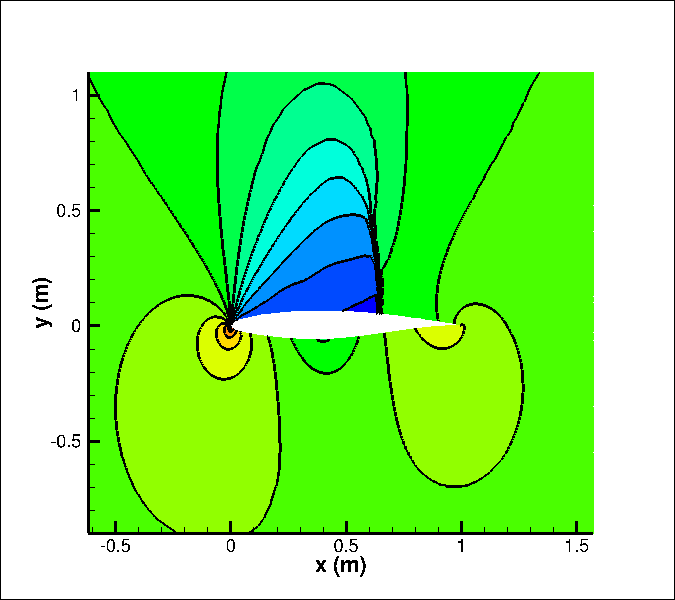}
  \caption*{ROM (POD+Projection)}
  \end{subfigure}    \begin{subfigure}{.33\textwidth}
  \centering
  \includegraphics[width=1\linewidth]{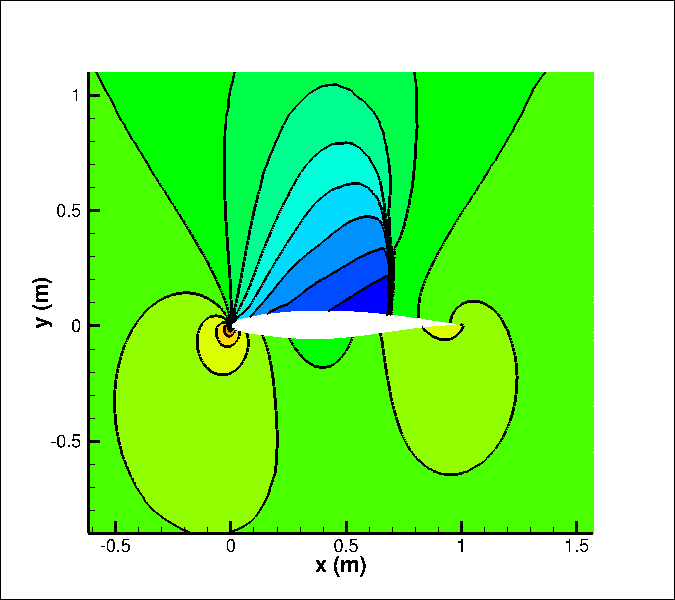}
  \caption*{FOM}
  \end{subfigure}\\
  \caption{Comparison of the proposed approach with high-dimensional ($\bs{\theta} \in \R^8$) shape parameters. Contours represent absolute pressure (in pascals) distribution.}
  \label{f:contours_shape}
 \end{figure}
 
 \begin{figure}[htb!]
\centering
  \begin{subfigure}{.95\textwidth}
  \centering
  \includegraphics[width=1\linewidth]{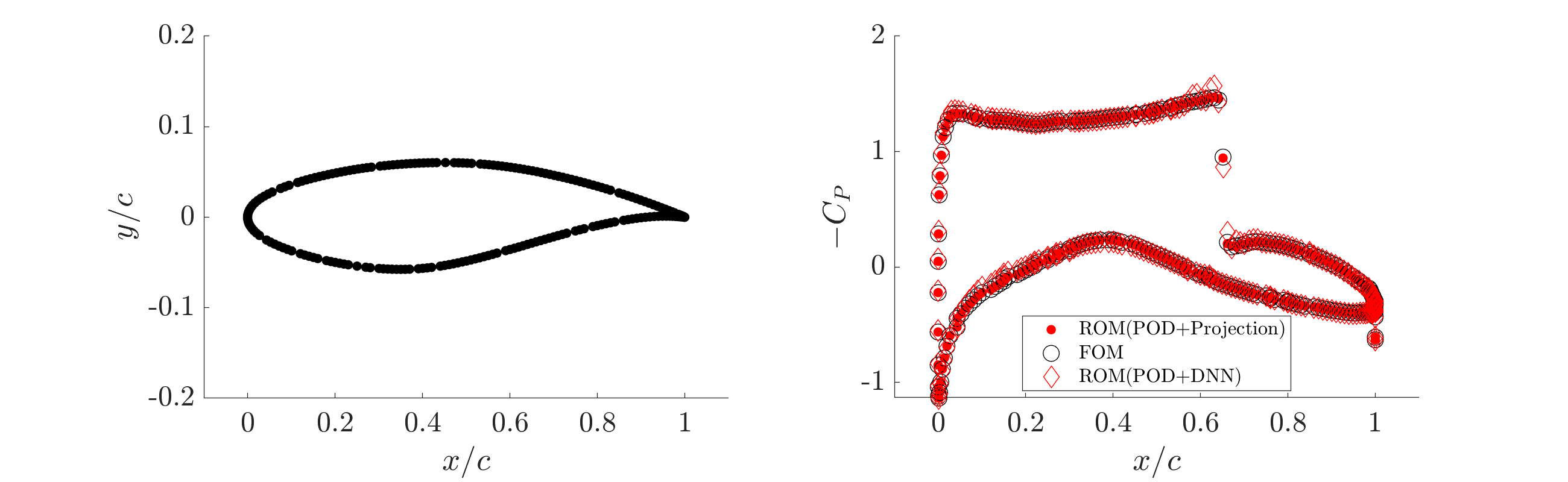}
  \caption{}
  \label{sf:DNN1}
\end{subfigure}\\
  \begin{subfigure}{.95\textwidth}
  \centering
  \includegraphics[width=1\linewidth]{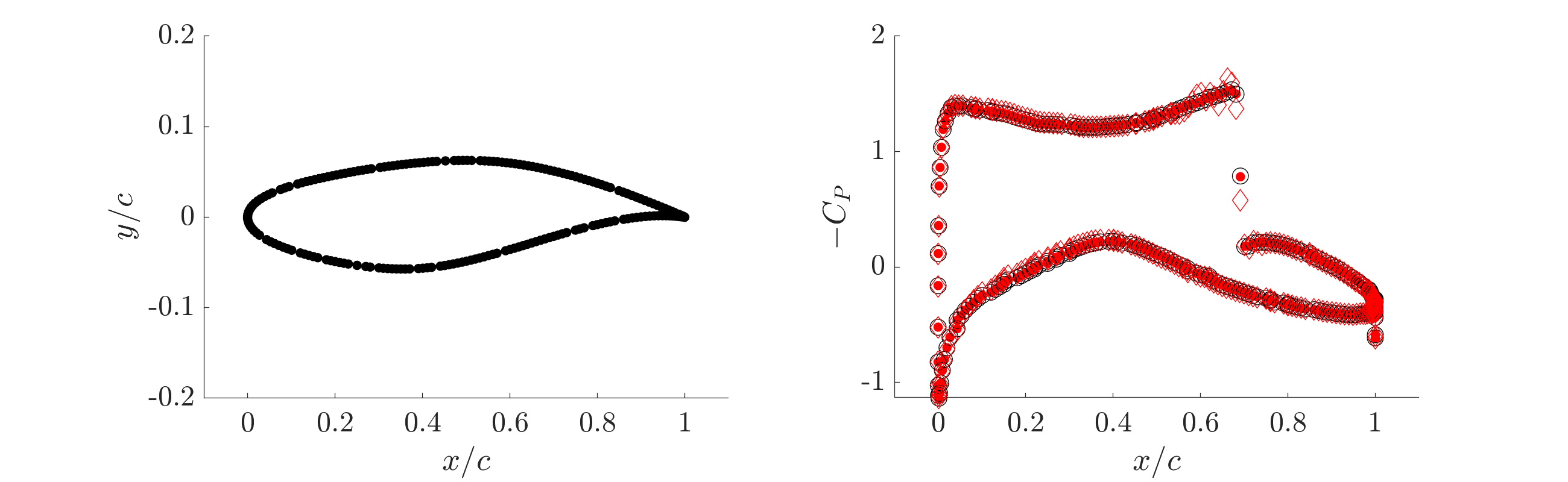}
  \caption{}
  \label{sf:DNN2}
  \end{subfigure}\\
    \begin{subfigure}{.95\textwidth}
  \centering
  \includegraphics[width=1\linewidth]{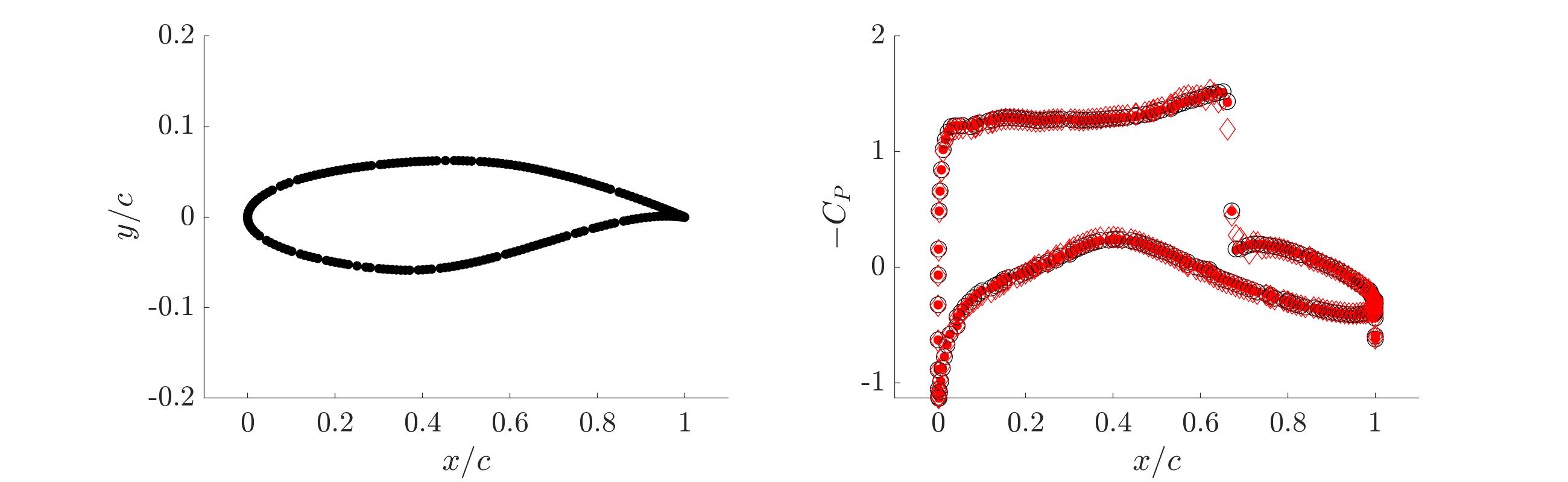}
  \caption{}
  \label{sf:DNN3}
  \end{subfigure}
  \label{8D_1}
 \end{figure}
 
\begin{figure}[htb!]
\ContinuedFloat
\centering
\begin{subfigure}{.9\textwidth}
  \centering
  \includegraphics[width=1\linewidth]{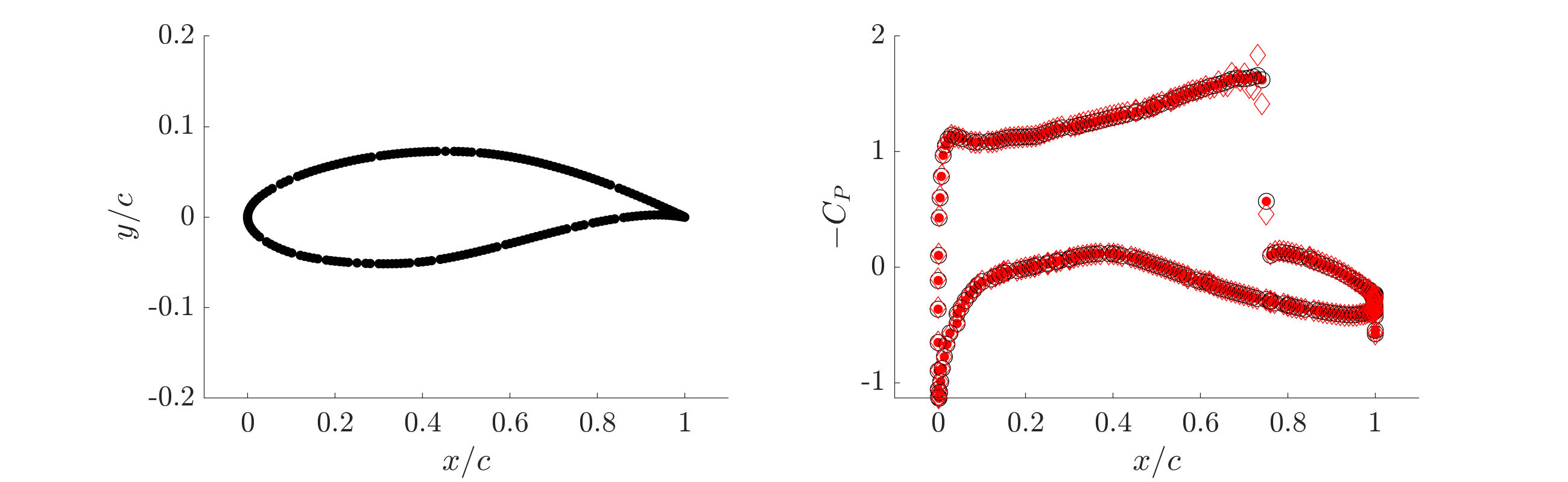}
  \caption{}
  \label{sf:DNN4}
  \end{subfigure}\\
    \begin{subfigure}{.9\textwidth}
  \centering
  \includegraphics[width=1\linewidth]{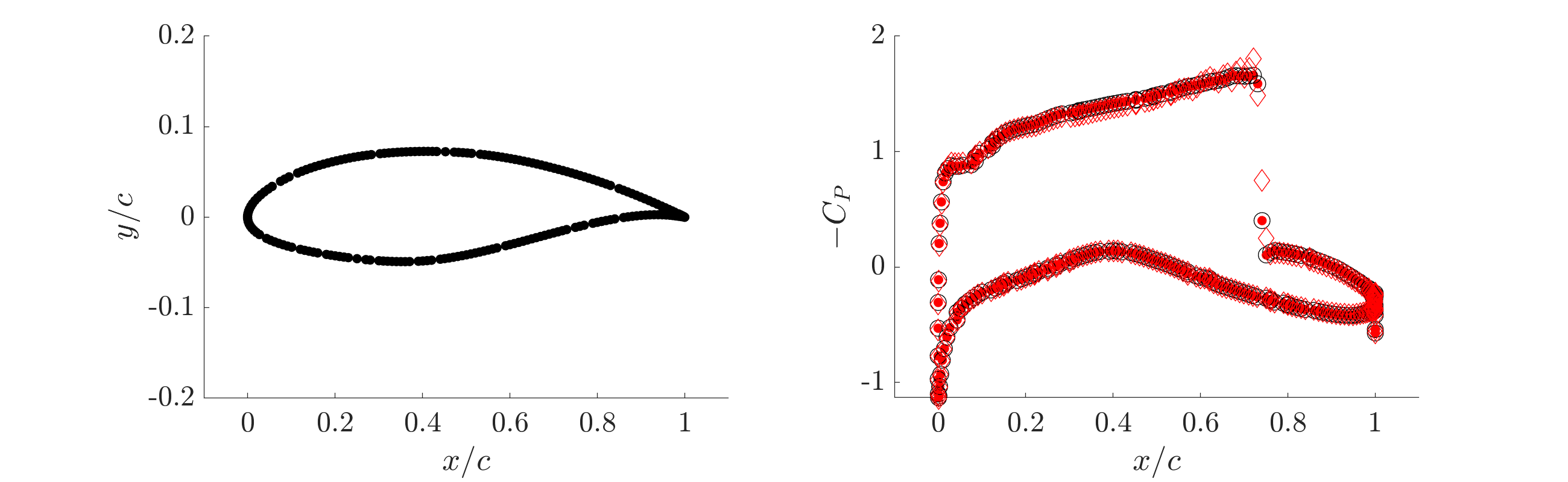}
  \caption{}
  \label{sf:DNN5}
  \end{subfigure}\\
    \begin{subfigure}{.9\textwidth}
  \centering
  \includegraphics[width=1\linewidth]{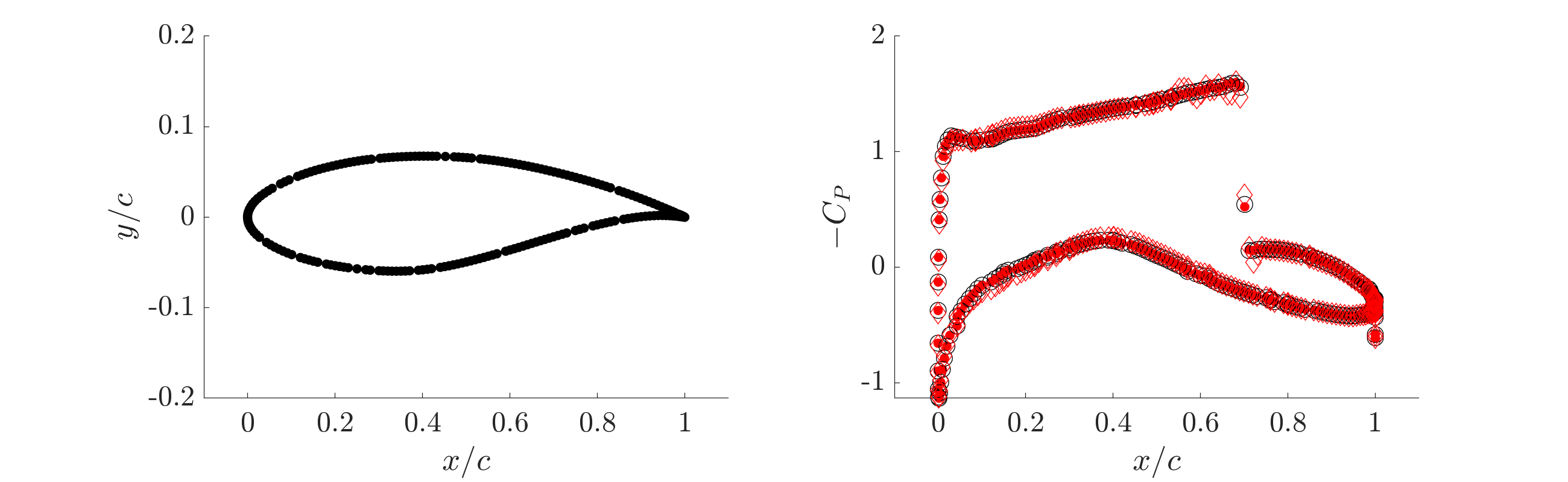}
  \caption{}
  \label{sf:DNN6}
  \end{subfigure}\\
    \begin{subfigure}{.9\textwidth}
  \centering
  \includegraphics[width=1\linewidth]{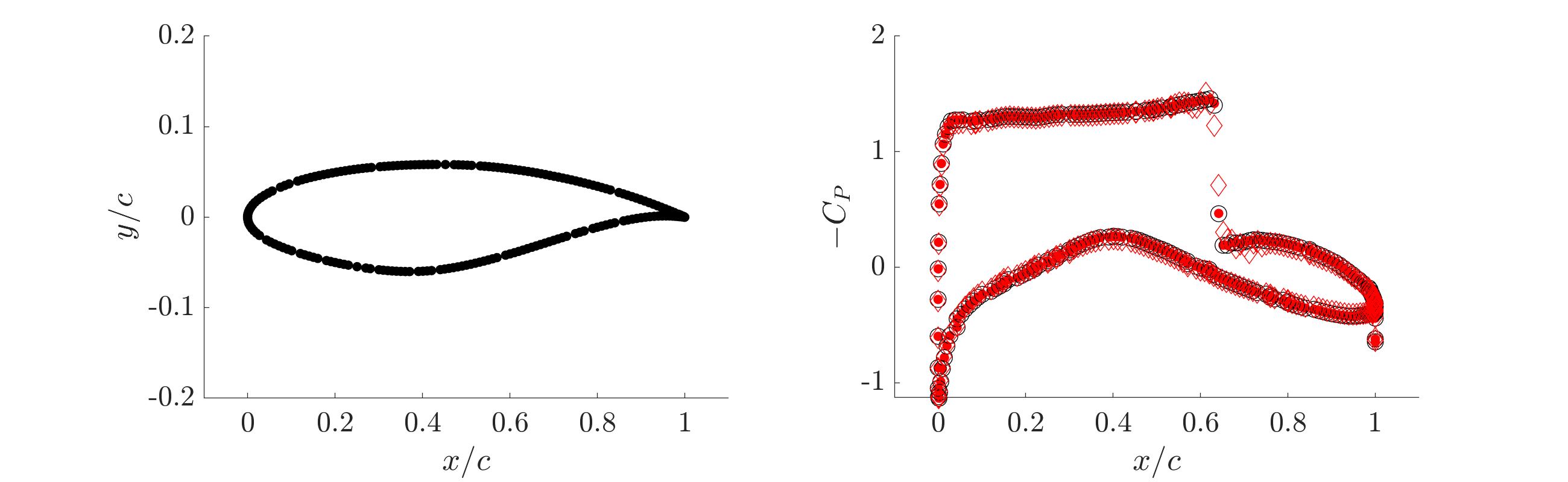}
  \caption{}
  \label{sf:DNN7}
  \end{subfigure}\\
\caption{Airfoil $C_P$ prediction of proposed method with geometry boundary parameters. Each subfigure corresponds to a unique airfoil shape (shown to the left) unseen by the training phase}  
\label{8D_2}
\end{figure}

\clearpage
\subsection{Output Quantities of Interest}
\label{ss:output_qoi}
Aerodynamic design typically requires the emulation of scalar output quantities of interest such as the force and moment coefficients. In this regard, the lift and drag coefficients are compared with the FOM predictions. Note that even though we consider an inviscid flow, the wave drag component is dominant in the transonic regime, which is essentially why we look at the drag coefficient. The results are summarized in Table~
\ref{t:QOIs}. The predictions via the proposed approach are  close to the FOM predictions, which is not surprising since the contour and $C_P$ plots established this already. Overall, although the projection-based approach shows better agreement with the FOM results, the difference is marginal. Furthermore, we repeat again that the offline costs involved in the proposed approach are significantly lower than the projection-based approach.

\begin{table}[htb!]
    \centering
    \caption{Comparison on output quantities of interest}
    \begin{tabular}{ccc|ccc}
    \hline
    \multicolumn{3}{c}{$C_l$} & \multicolumn{3}{c}{$C_d$} \\
    \hline
    DNN-ROM & Proj.-ROM & FOM & DNN-ROM & Proj.-ROM & FOM   \\
    \hline
1.0200  &	1.0169 &	1.0169 &	0.02014 & 0.02144 &	0.02145 \\
1.0722  &	1.0766 &	1.0767 &	0.01841 &	0.02038 &	0.02039 \\
1.0148  &	1.0159 &	1.0158 &	0.02418 &	0.02401 &	0.02399 \\
1.1462  &	1.1432 &	1.1432 &	0.04447 &	0.04470 &	0.04470 \\
1.1540  &	1.1577 &	1.1577 &	0.05319 &	0.05317 &	0.05317 \\
1.0619  &	1.0643 &	1.0642 &	0.04008 &	0.03853 &	0.03851 \\
0.9973  &	1.0074 &	1.0075 &	0.02231 &	0.02376 &	0.02376 \\
\hline
    \end{tabular}
    \label{t:QOIs}
\end{table}

\subsection{Computational Costs}
\label{ss:comp_costs}
We now compare the \emph{offline} computational costs for the proposed approach with the projection-based approach in terms of the algorithmic complexity. As previously mentioned,  projection-based approaches incur heavy offline costs for the projection step, which scales as $\mc{O}(N^2)$. Furthermore, nonintrusive approaches incur additional costs, such as the finite-volume discretization step in \cite{renganathan2018koopman}. We compare only the dominant components of the offline model construction.

The computational costs are summarized in Table~\ref{t:offline_cost}. The absence of the projection step and the operator inference step in nonintrusive approaches clearly makes the proposed approach significantly cheaper to construct compared with traditional ROMs. This also makes the implementation of the proposed approach relatively easier. Note that DNN training does require an efficient optimization method to tune the DNN hyperparameters, although off-the-shelf optimization libraries can be leveraged for this purpose, such as the Adam optimizer used in this work. The online costs (for the evaluation of ROMs) is comparatively negligible and hence is not compared. Overall, a predictive accuracy comparable to the structure-preserving projection-based ROMs and the cheap offline costs make the proposed approach a very good candidate for application in an engineering design setting where many queries to the ROM are necessary.

\begin{table}[htb!]
\centering
\caption{Summary of offline computational cost for ROM}
\begin{tabular}{ccc}
Operation & (Nonintrusive) POD+Proj. & POD+DNN \\
\hline
\hline
\hline
Snapshot Scaling & $\mathcal{O}(N)$ & $\mathcal{O}(N)$\\
POD & $\mathcal{O}(NM^2)$ & $\mathcal{O}(NM^2)$\\
Finite Vol. Discret. & $\mathcal{O}(MN)$ & - \\
Projection & $\mathcal{O}(N^2 M)$ & - \\
\hline
\end{tabular}
\label{t:offline_cost}
\end{table}

\clearpage
\section{Conclusion}
\label{Conclusion}

Projection-based POD-ROMs are known to incur heavy offline costs in constructing the reduced operators, adapting to parametric changes, and capturing nonlinear flows with parameter-dependent discontinuities and/or sharp gradients. These costs have been a particularly limiting factor in their successful application in transonic wing/airfoil design in aerospace engineering. This work proposes a machine-learning-based approach based on deep neural networks to circumvent the expensive projection step and  learn the nonlinear dependence of reduced state on high-dimensional parameters with modest training dataset sizes. Upon application to the flow past the RAE2822 airfoil under inviscid transonic flight conditions,  the proposed approach performs on a par with the projection-based approach while marginally outperforming it in a select few cases. In addition,  the DNN method has low online computational expense and circumvents stability issues of the projection-based approach, thereby increasing its suitability for optimization workflows. 

The proposed approach is tested with two parameters in the freestream boundary and eight parameters in the geometry boundary. The number of high-fidelity snapshots used for both cases are fixed at 80 and are identical to that used by the projection-based approach to allow for a fair comparison. The ability of the proposed approach to learn the parameter dependence with as few as 80 snapshots show great promise in its application towards many-query problems in aircraft design where each full-order snapshot is computationally very expensive to generate. Although online computational costs are not compared (due to their relatively insignificant magnitude compared to the offline costs), the proposed approach circumvents the manifold interpolation step ~section \ref{ss:ROM_Interp}) as well as the iterative solution procedure of the projection-based approach making it relatively faster.

Future directions in this work include evaluation of the proposed approach with even higher-dimensional inputs and eventual incorporation into aerodynamic design optimization workflows. Furthermore, the authors plan to demonstrate the approach on probabilistic analysis problems such as estimating failure probabilities and propagating input uncertainties.

\section*{Acknowledgments}
This material is based upon work supported by the U.S. Department of Energy (DOE), Office of Science, Office of Advanced Scientific Computing Research, under Contract DE-AC02-06CH11357. This research was funded in part and used resources of the Argonne Leadership Computing Facility, which is a DOE Office of Science User Facility supported under Contract DE-AC02-06CH11357. SAR acknowledges Dimitri Mavris for the access to the high performance PACE cluster at GeorgiaTech to carry out the high-fidelity simulations. RM acknowledges support from the Margaret Butler Fellowship at the Argonne Leadership Computing Facility. This paper describes objective technical results and analysis. Any subjective views or opinions that might be expressed in the paper do not necessarily represent the views of the U.S. DOE or the United States Government. Declaration of Interests - None.

\begin{mdframed}
    The submitted manuscript has been created by UChicago Argonne, LLC, Operator of Argonne National Laboratory ("Argonne”). Argonne, a U.S. Department of Energy Office of Science laboratory, is operated under Contract No. DE-AC02-06CH11357. The U.S. Government retains for itself, and others acting on its behalf, a paid-up nonexclusive, irrevocable worldwide license in said article to reproduce, prepare derivative works, distribute copies to the public, and perform publicly and display publicly, by or on behalf of the Government. The Department of Energy will provide public access to these results of federally sponsored research in accordance with the DOE Public Access Plan (http://energy.gov/downloads/doe-public-access-plan).
\end{mdframed}
\appendix

\section{Appendix}
\subsection{Nonintrusive Model Reduction of the Euler Equations}
\label{a:Euler_MOR}
The following transformation is  performed  \[ [\rho u, \rho v, \rho uv, p, \rho u^2,\rho v^2, \rho uH, \rho vH]^\top  \rightarrow [y_1, y_2, y_3, y_4, y_5, y_6, y_7, y_8]^\top \] from the state variables to observables, leading to the lifted model, 
\begin{equation}
\begin{bmatrix}
\nabla_x & \nabla_y & & & & & &  \\
 &  & \nabla_y & \nabla_x & \nabla_x & & &  \\
 &  & \nabla_x & \nabla_y & &\nabla_y & &  \\
 &  & & & & & \nabla_x & \nabla_y  \\
\end{bmatrix} \begin{bmatrix}
y_1 \\
y_2 \\
y_3 \\
y_4 \\
y_5 \\
y_6 \\
y_7 \\
y_8,
\end{bmatrix} = \mathbf{0},
\end{equation}
where i empty spaces in the matrix denote zeros. The equation upon discretization leads to

\begin{equation}
\underbrace{ \begin{bmatrix}
 \mb{G}_x & \mb{G}_y &     &     &     &     &     &  \\
 	 &     & \mb{G}_y & \mb{G}_x & \mb{G}_x &     &     &  \\
 	 &     & \mb{G}_x & \mb{G}_y &     & \mb{G}_y &     &  \\
 	 &     &     &     &     &     & \mb{G}_x & \mb{G}_y  \\
\end{bmatrix}}_{\mb{A}} \underbrace{ \begin{bmatrix}
\mb{y}_1 \\
\mb{y}_2 \\
\mb{y}_3 \\
\mb{y}_4 \\
\mb{y}_5 \\
\mb{y}_6 \\
\mb{y}_7 \\
\mb{y}_8
\end{bmatrix}}_\mb{y}  = \underbrace{- \begin{bmatrix}
\mb{b}_{a1} \\
\mb{b}_{a2} \\
\mb{b}_{a3} \\
\mb{b}_{a4} \\
\mb{b}_{a5} \\
\mb{b}_{a6} \\
\mb{b}_{a7} \\
\mb{b}_{a8}
\end{bmatrix}}_{\mb{f}},
\label{e:Euler_FOM}
\end{equation}
where $\mb{G}_x$ and $\mb{G}_y$ represent the discrete version of the gradient operators $\nabla_x$ and $\nabla_y$, respectively, and the empty spaces  denote block matrices of zeros. The parameter-dependent $\mb{A}$ matrix is obtained directly by discretizing the linear terms $\mb{G}_x$ and $\mb{G}_y$ via the finite-volume method. The grid is exported in the CFD General Notation System (CGNS)~\cite{CGNS} for this purpose. The snapshots $\mb{y}$ are applied to the $\mb{A}$ matrix, and the right-hand side $\mb{f}$ is extracted for each parameter \textcolor{black}{snapshot, $\theta_i$, which in the present context represents aerodynamic shape parameters of the airfoils}. With the FOM reduced to the $\mb{A}\mb{y}=\mb{f}$ form and $\mb{A} \in \R^{4N\times 8N},~\mb{y}, \mb{f} \in \R^{8N \times 1}$, Equation (\ref{e:Euler_FOM}) represents an underdetermined system \textcolor{black}{due to the introduction of variables in excess of equations}. Therefore, they are closed by using nonlinear constraints presented in Equation (\ref{e:Euler_cont_Cons}). Note that the constraints express the relationship between the first $S=4$ observables ($y_1$ through $y_4$) and the remaining $O-S;~(O=8)$ observables ($y_5$ through $y_8$). As mentioned in the main text, the choice of the first $S$ observables and hence the $O-S$ constraints is not unique. From experience trying  different choices in this work, however, we found that the following heuristics ensure a stable transformation from the observables back to the state: (i) \textcolor{black}{the first $S$ observables must include all the primitive variables ($\rho, p, u, v$)},  (ii) the terms starting from the lowest order are picked as the first $S$ observables ($\rho u, \rho v, \rho uv, p$ in this case), and (iii) \textcolor{black}{at least} one among the first $S$ observables is set to be a state variable ($y_4 = p$ in this case). We  note that all the observables that are in excess of the number of equations can be expressed as some function of the rest because the number of independent observables is only as great. as the number of PDEs in the FOM ($S=4$); \textcolor{black}{additionally, heuristic (i) ensures this}. The constraints are expressed in terms of the continuous form of the state and observable as follows

\begin{equation}
\begin{aligned}
& h_1 = \rho u^2 - \frac{(\rho u)(\rho uv)}{\rho v}  \equiv y_5 - \frac{y_1 y_3}{y_2} = 0 \\
& h_2 = \rho v^2 - \frac{(\rho v)(\rho uv)}{\rho u}  \equiv y_6 - \frac{y_2 y_3}{y_1} = 0 \\
& h_3 = \rho uH - \rho u \left(E + \frac{p}{\rho} \right)  \equiv y_7 - y_1 \left(E + \frac{y_4y_3}{y_1y_2} \right)=0\\
& h_4 = \rho vH - \rho v \left(E + \frac{p}{\rho} \right) \equiv y_8 - y_2 \left(E + \frac{y_4y_3}{y_1y_2} \right)=0
\end{aligned}
\label{e:Euler_cont_Cons}
\end{equation}

\subsection{Discrete Empirical Interpolation Method (DEIM)}
\label{A:DEIM}

The Discrete Empirical Interpolation Method (DEIM) is briefly reviewed here, and as an illustration one of the nonlinear constraints used in Equation (\ref{e:Euler_cont_Cons}) is evaluated. For a nonlinear function $\mb{f}(\theta)\in \R^N$ the DEIM approximates $\mb{f}$ by projecting it onto a subspace spanned by $\lbrace \mb{x}_1,...,\mb{x}_q\rbrace\subset \R^N$ as
\begin{equation}
\mb{f}(\theta) \approx \mb{X} c(\theta),
\end{equation}
where $\mb{X}=[\mb{x_1},...,\mb{x_q}] \in \R^{N \times q},~~q<<N$ is determined via a POD of the snapshots of $\mb{f}$ and is assumed to be globally valid in the design space that bounds the design parameters $\theta$ and $\mb{c}(\theta)\in\R^q$ are the coefficients of the basis expansion. Then the approximation of $\mb{f}$ requires only the determination of $\mb{c}(\theta)$, which requires only $q$ equations. The DEIM gives a distinguished set of $q$ points from the overdetermined system $\mb{f}(\theta) = \mb{X}\mb{c}(\theta)$. Given a permutation matrix $\mb{P}$ that would give $q$ such distinguished rows of a matrix when premultiplied,  the $q\times q$ system necessary to solve for the coefficients is 

\begin{equation}
\mb{P}^\top \mb{f}(\theta) = (\mb{P}^\top\mb{X})\mb{c}(\theta) .
\end{equation}

The approximation of $\mb{f}(\theta)$ is then given by

\begin{equation}
\mb{f}(\theta) \approx \mb{X}(\mb{P}^\top\mb{X})^{-1} \mb{P}^\top \mb{f}(\theta).
\end{equation}

If the $q$ row-indices (that are extracted by premultiplying with $\mb{P}^\top$) are represented by a vector, $\mb{\varrho}$, then in the above equation $\mb{P}^\top \mb{f}(\theta)$ is equivalent to extracting the $\varrho$ rows of $\mb{f}$. Therefore the approximation of $\mb{f}(\theta)$ requires only $q$ computations, which is efficient because $q<<N$. Similarly, a nonlinear function that depends on the state $\mb{f}(\mb{u})$ can be approximated as

\begin{equation}
\mb{f}(\mb{u}) \approx \mb{X} (\mb{P}^\top\mb{X})^{-1} \mb{P}^\top \mb{f}(\mb{u}).
\end{equation}

Since $\mb{u} = \Phi_k^\top \tilde{\mb{u}}$ and setting $\tilde{\mb{f}}=\Phi_k^\top \mb{f}(\mb{u})$, $\tilde{\mb{f}}$ can be approximated as

\begin{equation}
\tilde{\mb{f}} = \Phi_k^\top \mb{X} (\mb{P^\top}\mb{X})^{-1} \mb{f}(\mb{P}^\top \Phi_k \tilde{\mb{u}}).
\end{equation}

In this equation, the term $\Phi_k^\top \mb{X} (\mb{P^\top}\mb{X})^{-1}$ is independent of the state and hence can be precomputed; and $\mb{P}^\top \Phi_k$ is just extraction of the $\varrho$ rows of $\Phi_k$. Therefore, by using the DEIM,  the nonlinear term can be expressed in terms of the reduced state $\tilde{\mb{u}}$ and hence can be efficiently computed.

Now DEIM is illustrated on evaluating the first constraint of Equation (\ref{e:Euler_cont_Cons}), which in discretized form is 

\begin{equation}
\mb{h}_1 = \mb{y}_5 - \frac{\mb{y}_1\mb{y}_3}{\mb{y}_2}.
\end{equation}

Let $\varrho_5$ be the vector containing the $q$ row-indices returned by DEIM via snapshots of the nonlinear term $\mb{y}_5$, and let $\Phi_1$, $\Phi_2$, $\Phi_3$, and $\Phi_5$ be the projection matrix of $\mb{y}_1$, $\mb{y}_2$, $\mb{y}_3$, and $\mb{y}_5$, respectively. Then

\begin{equation}
\tilde{\mb{h}}_1 = \tilde{\mb{y}_5} - \Phi_5^\top \mb{X} ~[\mb{X}(\varrho_5,:)]^{-1} \left\lbrace \frac{\Phi_1(\varrho_5,:)\tilde{\mb{y}_1}~~\Phi_3(\varrho_5,:)\tilde{\mb{y}_3}}{\Phi_2(\varrho_5,:)\tilde{\mb{y}_2}} 
\right\rbrace .
\end{equation}

In this equation, the term outside of the braces can be precomputed. Additionally, since $\mb{y}_5 = \frac{\mb{y}_1\mb{y}_3}{\mb{y}_2}$, then $\mb{X}=\Phi_5$; and hence the term reduces to $[\mb{X}(\varrho_5,:)]^{-1}$, which is $q \times q$ and  can be cheaply computed. Therefore, with  DEIM, the nonlinear constraints are evaluated in terms of the reduced-state variables, making the step computationally cheap.
\bibliographystyle{unsrt}  
\bibliography{References}  

\end{document}